\documentclass[twocolumn,showpacs,amsfonts,amssymb,amsmath,floats,superscriptaddress,aps]{revtex4-1}

\usepackage{graphicx}
\usepackage{multirow}
\usepackage{subfigure}

\usepackage{braket}

\def\e{\varepsilon}

\def\w{\omega}

\def\k{\vec{k}}
\def\non{\nonumber \\ }
\def\ul#1{\underline{#1}}
\def\mat#1{\ul{\ul{#1}}}
\def\myeqref#1{Eq.\ \eqref{#1}}

\usepackage{upgreek}
\usepackage{bbold}

\begin{document}

\title{Inelastic electron tunneling spectroscopy 
for probing strongly correlated many-body systems by scanning tunneling microscopy}

\author{Fabian Eickhoff}
\affiliation{Theoretische Physik II, Technische Universit\"at Dortmund, 44221 Dortmund, Germany}

\author{Elena Kolodzeiski}
\affiliation{Physikalisches Institut, Westf\"alische Wilhelms-Universit\"at M\"unster, 48149 M\"unster, Germany}

\author{Taner Esat}
\author{Norman Fournier}
\author{Christian Wagner}
\affiliation{Peter Gr\"unberg Institute (PGI-3), Forschungszentrum J\"ulich, 52425 J\"ulich, Germany}
\affiliation{J\"ulich Aachen Research Alliance (JARA), Fundamentals of Future Information Technology, J\"ulich, 52425 J\"ulich, Germany}

\author{Thorsten Deilmann}
\affiliation{Institut f\"ur Festk\"orpertheorie, Westf\"alische Wilhelms-Universit\"at  M\"unster, 48149 M\"unster, Germany}

\author{Ruslan Temirov}
\affiliation{Peter Gr\"unberg Institute (PGI-3), Forschungszentrum J\"ulich, 52425 J\"ulich, Germany}
\affiliation{J\"ulich Aachen Research Alliance (JARA), Fundamentals of Future Information Technology, J\"ulich, 52425 J\"ulich, Germany}
\affiliation{II. Physikalisches Institut, Universit\"at zu K\"oln, Z\"ulpicher Stra\ss e 77, 50937 K\"oln, Germany}

\author{Michael Rohlfing}
\affiliation{Institut f\"ur Festk\"orpertheorie, Westf\"alische Wilhelms-Universit\"at  M\"unster, 48149 M\"unster, Germany}

\author {F. Stefan Tautz}
\affiliation{Peter Gr\"unberg Institute (PGI-3), Forschungszentrum J\"ulich, 52425 J\"ulich, Germany}
\affiliation{J\"ulich Aachen Research Alliance (JARA), Fundamentals of Future Information Technology, J\"ulich, 52425 J\"ulich, Germany}
\affiliation{Experimentalphysik IV A, RWTH Aachen University, Otto-Blumenthal-Stra\ss e, 52074 Aachen, Germany}

\author{Frithjof B. Anders}
\affiliation{Theoretische Physik II, Technische Universit\"at Dortmund, 44221 Dortmund, Germany}

\date{\today}

\begin{abstract}

We present an extension of the tunneling theory for scanning tunneling microcopy (STM) to include different types of  vibrational-electronic  couplings responsible for inelastic contributions to the tunnel  current in the strong-coupling limit. It  allows for a better understanding of more complex scanning tunneling  spectra  of molecules on a metallic substrate in separating elastic and inelastic contributions.
The starting point is the exact solution of the spectral functions for the electronic active local  orbitals in the absence of the STM tip. This includes electron-phonon coupling in the coupled system comprising the molecule and the substrate to arbitrary order including the anti-adiabatic strong coupling regime as well as the Kondo effect on a free electron spin of the molecule.
The tunneling current is derived in second order of the tunneling matrix element which is expanded in powers of the relevant vibrational displacements. 
We use the results of an ab-initio calculation for the single-particle electronic properties as an adapted material-specific input for a numerical renormalization group approach for accurately determining the electronic properties of a NTCDA molecule on Ag(111) as a challenging sample system for our theory.
Our analysis shows that the mismatch between the ab-initio many-body calculation of the tunnel current in the absence of any electron-phonon coupling to the experiment scanning tunneling  spectra can be resolved by including two mechanisms: (i) a strong unconventional Holstein term on the local substrate orbital leads to reduction of the Kondo temperature and (ii) a different electron-vibrational coupling to the tunneling matrix element is responsible for inelastic steps in the $dI/dV$ curve at finite frequencies.

\end{abstract}

\maketitle

\section{Introduction}

The investigation of phonons and molecular vibrations by inelastic electron tunneling spectroscopy dates back more than 50 years \cite{JaklevicLambe1966,MolecularVibrationTunnel1968}.  For example, point contact spectroscopy \cite{PCSreview1989} has been successfully used to measure the electron-phonon coupling function that enters the Migdal-Eliashberg theory \cite{McMillanTc1968,AllenMitrovic} of superconductivity.  Recently, the increasing relevance of quantum nanoscience \cite{Khajetoorians2011,Baumann2015,Donati2016,Natterer2017,Esat2018,Cocker2016,Doppagne2018,Kimura2019,Wagner2019} revitalizes the interest in vibrational inelastic electron tunneling spectroscopy (IETS) of molecules adsorbed on solid surfaces \cite{Stipe1998, Guo2016, Wegner2013, Burema2013} or contacted in transport junctions \cite{Kim2011a, Vitali2010, Meierott2017, Bruot2012, Sukegawa2014}. While the fundamental mechanisms of the electron-phonon and electron-vibron interactions are well-understood (for simplicity, we will refer to both as electron-phonon interaction from now on), a quantitative theory with predicting power beyond a simplified picture comprising independent electronic degrees of freedoms and bosonic excitations is lacking. Even modern reviews \cite{REED2008} on this subject  present the inelastic tunnel process only on the original level of understanding   \cite{JaklevicLambe1966,MolecularVibrationTunnel1968}, i.e.~the emission or absorption of a single phonon when a single electron is tunneling, as depicted in Fig.~1 of Ref.\ \cite{MolecularVibrationTunnel1968} or Fig.~1(a) of Ref.~\cite{REED2008}. 

This commonly accepted picture is very adequate in the weak coupling limit \cite{MolecularVibrationTunnel1968} of the  adiabatic regime \cite{EntelGrewe1979,galperinNitzanRatner2006,EidelsteinSchiller2013,JovchevAnders2013}, whence the electron-phonon coupling is small on the energy scale of the hybridization between the relevant molecular orbital(s) and the surface (or electrode in a transport experiment), and provides a basic understanding of the relevant physical processes. However, it becomes problematic in systems dominated by polaron formation, or for systems in the crossover region between the adiabatic and the anti-adiabatic regimes  \cite{EntelGrewe1979,galperinNitzanRatner2006,EidelsteinSchiller2013}.

This calls for a more general treatment of the inelastic tunneling process. In this paper we provide such a theory, focussing in particular on the case of scanning tunneling spectroscopy (STS). We generalize the original picture \cite{JaklevicLambe1966,MolecularVibrationTunnel1968} to strongly correlated electron systems but maintain the notion that  inelastic contributions to the tunneling current require  absorption or emission of a phonon while the electron is crossing the tunnel  barrier. We treat the STM tip and the system of interest as initially decoupled and fully characterized by their exact Green's functions. After specifying  the tunneling Hamiltonian $\hat H_T$,  the tunnel current operator is derived from  the charge conservation. Then  the coupling between the system and the STM tip, $\hat H_T$, is switched on, and  the evolving steady-state  current is evaluated in second order of the tunneling matrix elements. All material-dependent spectral properties are encoded in the equilibrium spectral functions of the system. Combining an accurate determination of the molecular spectral function using Wilson's numerical renormalization group (NRG) approach \cite{Wilson75,BullaCostiPruschke2008} with a density functional approach \cite{RevModPhys.74.601} provides a theoretical approach to strongly coupled system with predicting power.

STS is an established technique and its theoretical background is well-understood \cite{TersoffHamann1983,TersoffHamann1985}.  Setting aside more challenging situations, commonly a featureless density of states in the STM tip is assumed,  and the STM is operated in the tunneling regime such that the measured $dI/dV$ curve  may be interpreted as being proportional to the local energy-dependent density of states (LDOS) of the sample at the given bias voltage.  Using spin-polarized tips \cite{SplittingRKKY2012} allows for the detection of the spin-dependent LDOS. Since electrons usually can tunnel from the STM tip to different orbitals in the target system, the quantum mechanical interference of  different paths \cite{SchillerHershfield2000a} may lead to Fano line shapes \cite{FanoResonance1961} in the tunneling spectra. 

The interpretation of electron tunneling becomes more complicated if the spectrum is dominated by the Kondo effect. The Kondo effect, originally discovered as resistance anomaly in metals containing magnetic impurities \cite{Kondo62,kondo_effect}, has been studied experimentally in quantum dots \cite{Kondo_QD0, Kondo_QD}, atoms and molecules on surfaces \cite{kondo_atom,LiSchneider1998,Manoharan2000,AgamSchiller2001,Kondo_molecule,kondo_molecule2}, and molecular junctions \cite{kondo_SM}. A comprehensive understanding has been developed \cite{Wilson75,kondo_anderson}: briefly, the at low temperatures logarithmically diverging antiferromagnetic exchange coupling between the unpaired spin and the itinerant electron states in the substrate (or leads) produces a singlet ground state with a low-energy single-particle excitation spectrum that is characterized by a resonance at zero energy. In such systems with their intrinsically highly non-linear LDOS in the vicinity of the chemical potential, it becomes very challenging to distinguish between elastic tunneling processes governed by the energy-dependent transfer matrix and additional inelastic contributions generated by the presence of an additional electron-phonon coupling. For example, in such systems so-called Kondo replica at vibrational frequencies have been observed \cite{Kondo_vib_SM,kondo_vib_bjunc, kondo_vib_bjunc2, kondo_vib_bjunc3,vib_kondo_stm, vib_kondo_stm2, vib_kondo_stm3}, whose precise nature is, however, not yet understood. The interplay between Kondo physics and electron-vibron coupling has also been studied theoretically \cite{PaakeFlensberg2005, vib_kondo_theo3,vib_kondo_theo}.

Since only the total tunneling current is accessible in experiments, its decomposition into individual processes requires guidance by a theory.  In this paper, we present  an approach providing this guidance. Specifically, we derive an extension to the comprehensive theory of the tunneling current in STM that was originally formulated by Schiller and Hershfield \cite{SchillerHershfield2000a} in the context of a magnetic adatom and generalized Fano's analysis \cite{FanoResonance1961} to inelastic contributions in the tunneling Hamiltonian which includes the calculation of the current operator from the local continuity equation. Notably, our theory accounts for two different types of electron-phonon interactions: (i) the intrinsic electron-phonon coupling in the system in the absence the STM tip and (ii) vibrationally induced fluctuations of the  distance between tip and molecule or substrate. The former is included in the system's Green's functions and only contribute to the elastic current. The latter enter the tunneling $H_T$ and, therefore,  are the origin of the inelastic current contributions.
 
Having developed said theory, we demonstrate its capabilities by applying the approach to
explain the experimental data. To this end, we have chosen the experimental system of naphthalene-tetracarboxylic-acid-dianhydride (NTCDA) molecules adsorbed on the Ag(111) surface. Similar systems, PTCDA/Ag(111) \cite{ptcda_kondo_cleavage,gated_molecular_wire,PTCDAAgMove,gated_wire_spectral} and PTCDA-Au complexes on Au(111) \cite{AU-PTCDA-monomer,AU-PTCDA-dimer}, have been investigated before but without the necessity of including phononic contributions. There, we applied a combination of density functional theory and many-body perturbation theory (DFT-MBPT)  and used the ensuing quasiparticle spectrum as input to a NRG calculation \cite{BullaCostiPruschke2008} to comprehensively understand the STS spectra. However, despite the similarity between NTCDA and PTCDA, STM experiments on NTCDA/Ag(111) cannot be explained using the same methodology. Specifically, the theory predicts a zero-bias resonance whose width is significantly  too large compared to the experiment. The origin of this deviation is not clear, as DFT-MBPT are expected to provide reliable input parameters for accurate NRG-calculated spectra \cite{PTCDAAgMove,gated_wire_spectral,AU-PTCDA-monomer,AU-PTCDA-dimer}. Moreover, the calculated spectra lack additional  features that are present in the experiment and hint towards inelastic electron-phonon contributions. The NTCDA/Ag(111) system, therefore, seems a good candidate as a first application case of our theory.

Indeed, we argue below that the NTCDA/Ag(111) experiments can be interpreted in a conclusive way by incorporating the very different effects of two vibrational modes into the description. One mode couples strongly to the local substrate electrons, thus dynamically modifying the hybridization function between the substrate and the molecule; this results in a substantial reduction of the Kondo temperature of the NTCDA molecule/substrate system. In contrast, the second mode couples only weakly to the electronic system. Both modes, however, cause modulations of the tunneling distance. While the second mode induces rather accentuated inelastic side peaks close to the vibrational frequency as consequence of a second-order phonon absorption/emission process, the polaronic entanglement of the first mode with the electronic system gives rise to two inelastic current contribution: A first order term involving only a single  phonon process is responsible for an asymmetric term while the second order contribution generates  only very weak inelastic features located in the spectrum at an renormalized phonon frequency. The strength of the theory presented here is the inclusion of both rather different mechanisms on an equal footing. It demonstrates how different vibrational modes with similar frequencies can nevertheless lead to distinctly different spectroscopic signatures. While, based on the commonly accepted level of understanding of electron-phonon effects in electron tunneling, the step-like structures are easily identified as vibration-related, the vibrational sharpening of the Kondo resonance in the presence of only marginal side peaks would be impossible to pinpoint without theoretical guidance.

Before going \textit{in medias res}, we briefly review the relevant literature regarding inelastic tunneling in STM and STS. Many papers in the literature focus on the theory of inelastic contributions to the tunneling current and, therefore, modifications to STS spectra. In particular, the influence of vibrational modes has been addressed \cite{MolecularVibrationTunnel1968,LorentePersson2000,REED2008,Leijnse2010}. Most of these theories \cite{Zawadowski1967,TersoffHamann1983,TersoffHamann1985} are based on the seminal many-body approach to tunneling by Bardeen \cite{Bardeen1961} that allowed to derive the Josephson current as a tunneling current between two superconductors \cite{Ambegaokar1963}. Higher-order electron-phonon processes in tunneling theories were investigated by Zawadwoski \cite{Zawadowski1967}, while  Caroli and collaborators \cite{Caroli71,Caroli72} employed the Keldysh approach to calculate the inelastic electron-phonon terms. Paaske and Flensberg investigated  the influence of vibrational effects onto the dynamics of a Kondo impurity \cite{PaakeFlensberg2005}. They combined a Schrieffer-Wolff transformation \cite{SchriefferWol66} with a third-order perturbation theory that is valid in the high-temperature regime well above the Kondo temperature and is limited to the anti-adiabatic regime. In their approximation, the atomic solution of a Holstein model \cite{LangFirsov1962} derived by Lang and Firsov  -- see also Mahan's text book \cite{Mahan81} -- modifies the Kondo coupling $J$ in the weak tunneling, large $U$ limit. This  Kondo coupling $J$ matrix becomes energy dependent  due to  polaron formation, inducing steps in the transmission matrix at multiples of the phonon frequency. Lorente and Persson \cite{LorentePersson2000} combined the Keldysh approach of Caroli et al.\ \cite{Caroli72} with density functional theory, both relying on the free-electron picture and decoupled vibrational modes. Such an approach is  only applicable in the adiabatic regime \cite{galperinNitzanRatner2006,EidelsteinSchiller2013}, but cannot address the anti-adiabatic regime that was considered by Paaske at al.\  \cite{PaakeFlensberg2005}. In two recent letters \cite{Wehling2008,PhysRevLett.110.026802}, electron-phonon effects included in the single-particle self-energy have been attributed to an inelastic electron tunneling contribution. If these self-energy corrections are only evaluated in the adiabatic regime, the effect on the current is so  small that it becomes visible only in the second derivative of the tunneling current $d^2I/dV^2$ \cite{WehlingTunnel2017}.

This  paper is organized as follows. In section \ref{sec:tunnel-current-general} we present our theory of the tunneling current. The theory is independent of the system Hamiltonian and  therefore of general nature. In particular, we discuss the inelastic and elastic contributions to the current, suggesting a partitioning that is based strictly on the question whether energy is transferred \textit{during} the tunneling process. As a necessary step towards the application of the tunneling theory to an actual physical system, we specify a system Hamiltonian in section \ref{sec:Modeling the system}. The choice of this Hamiltonian, consisting of a single impurity Anderson model (SIAM) and two distinct types of Holstein couplings, is motivated by the experimental system of NTCDA/Ag(111) which we introduce in section \ref{sec:experiment-NTCDA}. One of the Holstein couplings is unconventional in the sense that it couples vibrations of the adsorbed molecule to electronic states in the substrate. In section \ref{sec:application-to-NTCDA}, we apply the tunneling theory to the NTCDA/Ag(111) system. To this end, we present NRG calculations of differential conductance spectra  and compare them in detail to experimental scanning tunneling spectra (STS). As a result, we are able to present a  model of the NTCDA/Ag(111) system that provides a comprehensive understanding of all generic features in the STS spectra. In section \ref{sec:STS-anti-adiabatic-regime}, we consider the STS spectra that are to be expected for a Kondo impurity in the anti-adiabatic regime. In particular, we show that Kondo replica that are naively expected do not show up, at least in the parameter regime which we consider. The paper closes with a conclusion (section \ref{sec:conclusion}).   

\section{Theory of the tunnel current}
\label{sec:tunnel-current-general}

In this section of the paper, we derive a generalized tunneling theory for STS spectroscopy that incorporates previous approaches \cite{SchillerHershfield2000a,LorentePersson2000,PaakeFlensberg2005} as limiting cases in certain parameter regimes. We differentiate between, first, vibrational contributions that modify the electronic single-particle Green's function (GF) of the system even in the absence of the STM tip from, second, true inelastic contributions that are introduced during the electron tunneling process from the STM tip into the system, as illustrated in Ref.\ \cite{MolecularVibrationTunnel1968}. While the former enter the self-energy of the Green's function for arbitrary order in the electron-phonon coupling, the latter are included in the perturbative treatment of the tunneling Hamiltonian.

In essence, our approach is a generalization of the theory by Caroli et al.\ \cite{Caroli72} to arbitrary correlations in the system of interest, but with the limitation that it is strictly only correct up to second order in the tunnel matrix element. Higher-order corrections, as addressed by Zawadovski \cite{Zawadowski1967} for oxide interfaces, require a proper Keldysh theory that incorporates the feedback process from the system to the STM tip and vica versa. In such an approach, non-equilibrium distribution functions replace the Fermi functions that we use in our theory. In this so-called  quantum point contact regime \cite{PCSreview1989} the STM tip is not any more a weak probe, and the STS spectra would not only contain information about the system of interest, but also about its coupling to the tip. Therefore, we exclude these considerations here and restrict ourselves to the tunneling limit for the system-STM coupling.

\subsection{Tunneling Hamiltonian}

\begin{figure}[t]
\begin{center}
\includegraphics[width=0.4\textwidth]{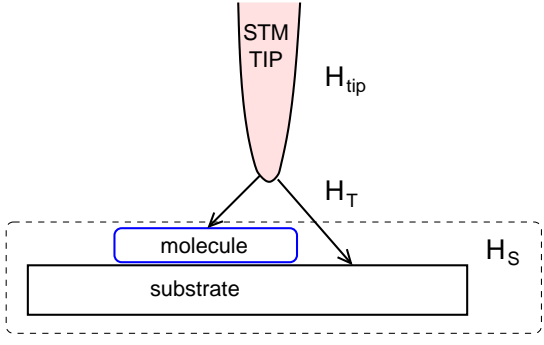}
\caption{Cartoon of the setup: a system comprising a molecule and a substrate is coupled to an STM tip. The arrows indicate the two transmission paths for electrons tunneling from the tip to the system S. The interference of such multi-orbital tunneling paths is responsible for Fano lineshapes in the tunneling spectra \cite{SchillerHershfield2000a}.
}
\label{fig:1-system-tip}
\end{center}
\end{figure}

We start from the most general situation for deriving the theory by dividing the total Hamiltonian of the coupled problem  as 
depicted in Fig.\ \ref{fig:1-system-tip} into three parts,
\begin{eqnarray}
\hat H &=& \hat H_{S} + \hat H_{\rm tip}  + \hat H_{T},
\end{eqnarray}
where $\hat H_{S}$ is the system Hamiltonian of the sample, comprising the adsorbed molecule and the substrate surface,  $\hat H_{\rm tip}$ denotes the Hamiltonian of the STM tip and $\hat H_{T}$ accounts for all tunneling processes between the tip and the sample system $\mathrm{S}$. We will keep the system Hamiltonian $\hat H_{S}$ unspecified without any restrictions.  In particular, we do not make any assumptions about its electronic, vibronic or even magnetic excitations. Therefore, the strong coupling limit, polaron formation or any other many-body effect, such as the Kondo effect or any kind of magnetism, may be included in the system $\mathrm{S}$. The STM tip, however, is modeled by a simple free electron gas
\begin{eqnarray}
\hat H_{\rm tip} &=& \sum_{\k \sigma}  \e_{\k\sigma} c^\dagger_{\k\sigma,\rm tip} c_{\k\sigma,\rm tip}, 
\end{eqnarray}
where $c^\dagger_{\k\sigma,\rm tip}$ creates a tip electron with spin $\sigma $ and energy $\e_{\k\sigma}$. If relevant, the Hamiltonian can be extended to a multi-band description, thus interpreting the index $\sigma$ as a combined spin and band label. 

Assuming some appropriately chosen orbital basis in the sample system S and a single-electron tunneling process, the most general bilinear tunneling Hamiltonian is given by \cite{Bardeen1961,TersoffHamann1985}
\begin{eqnarray}
\label{eq:general-tunneling-HT}
\hat H_{T} &=& \sum_{\mu \k \sigma \sigma'} 
\left(T^{\sigma\sigma'}_{\mu,\k}  d^\dagger_{\mu\sigma}  c_{\k\sigma',\rm tip} +h.c. 
\right)
\, \, \, ,
\end{eqnarray}
where $d^\dagger_{\mu\sigma}$ creates an electron in the as yet unspecified localized orbital $\mu$ of the system  $\mathrm{S}$. $\mu=0,\cdots, M-1$ labels different orbitals in the system. While only two tunneling paths are included in the cartoon Fig.\ \ref{fig:1-system-tip}, in general, electrons thus tunnel from the STM tip to $M$ different orthogonal orbitals of the system $\mathrm{S}$. These orbitals can be the orbitals of a molecule adsorbed on the substrate surface, or substrate Wannier orbitals in the vicinity of the STM tip. The shape of the STM tip has an influence which orbitals $\mu$ have to be included in Eq.\ \eqref{eq:general-tunneling-HT}. If $M\ge 2$, the quantum mechanical interference of the different tunneling paths includes the possibility of a Fano resonance \cite{SchillerHershfield2000a}. An additional capacitive coupling between the STM tip and the system \cite{Temirov2018} is ignored,  since we target the low bias regime of STM junctions. Such charging terms become relevant at large bias voltages which are not considered in this paper. 

A difficulty arises because even if $T^{\sigma\sigma'}_{\mu,\k}$ is spin-diagonal, it still depends on the details of the tip shape, which is unknown in experiment. Therefore, one usually makes several approximations that effectively absorb the details of the tip shape into an unknown ratio of tunneling matrix elements, but nevertheless turn out to be helpful for the understanding of spectroscopy data. For example, Tersoff and Hamann \cite{TersoffHamann1985} assume that the STM tip electrons are described by plane waves, leading essentially to a factorization of the matrix elements 
\begin{eqnarray}
\label{eq:Tersoff-Haman}
T^{\sigma\sigma'}_{\mu,\k} = a_{\k} t_{\mu}^{\sigma\sigma'}. 
\end{eqnarray}
Then a fictitious STM tip orbital can be introduced as
\begin{eqnarray}
\label{eq:stm-tip-c0}
c_{0\sigma,\rm tip} &=& \sum_{\k}  a_{\k} c_{\k\sigma,\rm tip} 
\end{eqnarray}
that we label with $i=0$. The local annihilation operator $c_{0\sigma,\rm tip} $ of an electron with spin $\sigma$ in this orbital is expanded in the free electron operators $c_{\k\sigma,\rm tip}$ with some expansion coefficients $a_{\k}$ whose details are not of interest and do not enter the theory, unless the STM tip is characterized by a strongly non-linear DOS in the relevant energy range. In the approximation Eq.\ \eqref{eq:Tersoff-Haman} and \eqref{eq:stm-tip-c0}, the STM tip shape has disappeared in some overall tunneling matrix elements $t_{\mu}^{\sigma\sigma'}$. However, we have to be aware that the STM tip breaks the local point group symmetry of the molecule. Therefore, one has to be careful when excluding tunneling channels purely on the basis of the symmetries of $\hat H_{S}$. 

For frozen nuclear positions $\{\vec{R}_i\}$, the above bilinear tunneling Hamiltonian is given by 
\begin{eqnarray}
\label{eq:H_T-bilinear}
\hat H_{T} &=& \sum_{\mu\sigma\sigma'}  t_{\mu}^{\sigma\sigma'}(\{\vec{R}_i\}) d^\dagger_{\mu\sigma}  c_{0\sigma',\rm tip} + h.c.
\end{eqnarray}
Here $t_{\mu}^{\sigma\sigma'}(\{\vec{R}_i\})$ denotes a tunnel matrix element that depends on some parameter set that is related to the atomic positions $\{\vec{R}_i\}$ in the system S and the tip, as well as on the spin. 

Since we are interested in the influence of molecular vibrations on the tunneling current, we must account for the change of the tunneling matrix element between the system S and the STM tip due to the vibrationally induced changes of the tip distance. To be more specific,  let us assume that $d^\dagger_{\mu\sigma}$ creates an electron in some extended molecular orbital spread over the entire surface-adsorbed molecule, or in a local Wannier state of the substrate in the vicinity of the STM tip. The molecule will have some vibrational eigenmodes, labelled by $\nu$, that deform the orbital. Imagining a perfectly rigid STM tip without any intrinsic vibrations, the tip-orbital distance will change as a function of this displacement.  Since the tunnel matrix elements are exponentially dependent on the distance, we model the tunneling matrix element by
\begin{eqnarray}
\label{eq:6}
  t^{\sigma \sigma'}_{\mu}(\{\vec{R}_i\}) &\to& t^{\sigma \sigma'}_{\mu}(\{\vec{R}^0_i\},\vec{R}_{\rm tip}) 
  e^{f_\mu(\{ \hat X_\nu\}) }
\end{eqnarray}
where $t^{\sigma \sigma'}_{\mu}(\{\vec{R}^0_i\},\vec{R}_{\rm tip})$ denotes the tunneling matrix element between the STM tip and the orbital $\mu$ if all atoms of the molecule are in their equilibrium positions $\{\vec{R}^0_i\}$ and the STM tip is located at position $\vec{R}_{\rm tip}$.  The unknown function $f_\mu$ depends on the superposition of all individual  dimensionless displacement operators  $\hat X _\nu = b_\nu +b^\dagger_\nu $ of each molecular eigenmode. Since we also allow for an electron-phonon coupling in the system (the system Hamiltonian $\hat H_{S}$ is as yet unspecified), the equilibrium position of atoms within the molecule might be shifted with respect to the equilibrium positions in the absence of this coupling \cite{galperinNitzanRatner2006,EidelsteinSchiller2013}. Therefore, it is useful to subtract the equilibrium position $x_{\nu 0}=\langle\hat X_{\nu}\rangle$ from $\hat X_{\nu}$ and define $\hat X_{\nu}'\equiv \hat X_{\nu} - x_{\nu0}$ in Eq.\ \eqref{eq:6}. We assume that vibration-induced changes in the tunneling matrix element are small and expand $\exp(f_\mu(\{ \hat X'_\nu\}))$ up to first order in the displacement. This leads to the simplification
\begin{eqnarray}
\label{eq:7}
 t^{\sigma \sigma'}_{\mu}(\{\vec{R}_i\}) & \approx 
& t^{\sigma \sigma'}_{\mu}(\{\vec{R}^0_i\})(1 +\sum_\nu \lambda^{\rm tip}_{\mu\nu} \hat X'_\nu)
 \label{eq:e-ph-tunnel}
\end{eqnarray}
where $\lambda^{\rm tip}_{\mu\nu}$ parameterizes the change of the tunnel coupling of the STM tip to the orbital $\mu$, induced by the excitation of the molecular vibration $\nu$. 
Similar terms have been considered in the context of Heavy Fermion superconductivity \cite{Grewe84} where the ionic breathing
mode couples to the lattice phonons.
Although this parameter $\lambda^{\rm tip}_{\mu\nu}$ can be spin-dependent in the case of a magnetically ordered surface, we have dropped this spin dependency. From now on we also drop the argument $\{\vec{R}^0_i\}$ on the right hand side of Eq.\ \eqref{eq:7} and use $t^{\sigma \sigma'}_{\mu}$ to refer to $t^{\sigma \sigma'}_{\mu}(\{\vec{R}^0_i\})$.

The matrix form of $t^{\sigma\sigma'}_{\mu}$ in spin space can be expressed in general as
\begin{eqnarray}
\mat{t}_{\mu}&=& t^0_{\mu} \mat{I} + \vec{t}_{\mu} \vec{\mat{\sigma}}
\label{spinmatrix}
\end{eqnarray}
by specifying four parameters $t^0_{\mu}$, $\vec{t}_{\mu}$ as spin-dependent tunnel matrix elements. The spinor matrix $\vec{\mat{\sigma}}$ could for example represent a free localized spin in the system S, or on the STM tip, that can couple to magnetic excitations such as magnons in a magnetic system S, which may cause additional magnetic inelastic contributions. In this paper, however, we assume that $\mat{t}_{\mu}$ is diagonal with diagonal elements 
\begin{eqnarray}
t_{\mu\sigma}= t^0_{\mu} +\sigma t^z_{\mu}.
\label{diagonalspinmatrixelements}
\end{eqnarray}
This allows spin-dependent tunneling matrix elements $t_{\mu\sigma}$, as they occur, e.g., for spin-polarized tips. If both the tip and the system S are paramagnetic, then $ \vec{t}_\mu=0$ in Eq.\ \ref{spinmatrix} and $t_{\mu,+1/2}=t_{\mu,-1/2}= t^0_{\mu}$. 
 
Note that a small $\lambda^{\rm tip}_{\mu\nu}$ in Eq.\ \ref{eq:e-ph-tunnel} does not imply that the electron-phonon coupling in the system S comprising substrate and molecule must be weak, because it is included in $\hat H_{S}$ and not connected to the parameters $\lambda^{\rm tip}_{\mu\nu}$ in the STM tunneling theory. In fact, the electron-phonon coupling in S can be arbitrarily strong  \cite{PaakeFlensberg2005} since the theory that we will present below only requires that $t^{\sigma \sigma'}_{\mu}$ is a very small energy scale and, therefore, the STM must be operated in the tunneling limit.

In the following we drop the prime in $\hat X'$ and demand $\langle{\hat X}\rangle=0$. We  discuss the generic case of $\langle{\hat X}\rangle=x_0\not = 0$ below in Sec.\ \ref{sec:e-phonon-displacement}, where we show that in leading order the total tunneling current is independent of $x_0$, as expected, although the partitioning between elastic and inelastic contributions is not unique.  This is not surprising, since the notion of an inelastic process requires the definition of the underlying  phonon basis sets.

\subsection{Tunnel current operator}

As the next step, we explicitly derive the analytic form of the tunnel current operator from charge conservation. This approach has the advantage that it allows the construction of the total current operator of the problem systematically and without adding terms by hand. We will show below that the derived total current operator contains all elastic and inelastic contributions. 

Since the tunnel current changes the number of electrons on the STM tip, the current operator $\hat j_{\rm STM}$ is related to the change of the charge $\hat Q_{\rm tip}=e \hat N_{\rm tip}$ on the tip, i.e.
\begin{eqnarray}
\label{eqn:current-operator-derivation}
\hat {j}_{\rm STM} &=&
 \frac{d\hat Q_{\rm tip}}{dt} = i \frac{e}{\hbar} [\hat H,\hat N_{\rm tip}] = i\frac{e}{\hbar} [\hat H_{T}, \hat  N_{\rm tip}]
\\
\nonumber
&=&  i\frac{e}{\hbar} 
\sum_{\mu\sigma}  t_{\mu \sigma}(\{\vec{R}_i\}) 
\left( 
d^\dagger_{\mu\sigma}  c_{0\sigma,\rm tip} 
-  c^\dagger _{0\sigma,\rm tip} d_{\mu\sigma} 
\right) ,
\end{eqnarray}
in order to enforce charge conservation in the total  system, consisting of the tip and the sample system S. Because the total particle number operator of the STM tip, $\hat N_{\rm tip}$, commutes with $\hat H_0=\hat H_{S}+ \hat H_{\rm tip}$, the current operator is generated by the tunneling Hamiltonian $\hat H_{T}$ only. Here we also have assumed that the tunneling matrix elements  are real, which can always be achieved by a local gauge transformation. Note that in Eq.\ \eqref{eqn:current-operator-derivation} we use $t_{\mu \sigma}(\{\vec{R}_i\})$ for an arbitrary but fixed set of atomic positions $\{\vec{R}_i\}$. The inelastic contributions to the tunnel current will become transparent once we substitute the linear expansion of $t_{\mu \sigma}(\{\vec{R}_i\}) $ in the displacements, Eq.\  \eqref{eq:e-ph-tunnel}, into Eq.~\eqref{eqn:current-operator-derivation}. Eq.\ \eqref{eqn:current-operator-derivation} demonstrates that the current operator depends only on the coupling between the two subsystems, $\hat H_{T}$, which is intuitively clear. A different $\hat H_{T}$, for example in the case of a magnetic interface, will modify the current operator derived in Eq.\ \eqref{eqn:current-operator-derivation}. Depending on the physics included in $\hat H_{T}$, this could include inelastic magnetic spin-flip contributions.

\subsection{STM tunnel current}
\label{sec:tunnel-current}

Since in the tunneling limit $\hat H_{T}$ defines the smallest energy scale of the system, we proceed in the interaction picture. We assume that the tunnel Hamiltonian $\hat H_{T}$ is switched on at time $t_0$. Then, the current evaluated at time $t>t_0$ is given by
\begin{eqnarray}
I(t) &=& \langle \hat j_{\rm STM}(t) \rangle = {\rm Tr}[\hat \rho_0 e^{i\hat H (t-t_0)} \hat{j}_{\rm STM} e^{-i\hat H (t-t_0)}]
\non
&=& \langle \hat U^\dagger(t,t_0) \hat j_{I}(t) \hat U(t,t_0) \rangle_0
\label{eq:current-1}
\end{eqnarray}
where $\hat j_{I}(t)= \exp[i \hat H_0 (t-t_0) ] \hat{j}_{\rm STM} \exp[-i \hat H_0 (t-t_0) ]$ is the STM current operator $ \hat j_{\rm STM}$ transformed into the interaction picture. 
Note that we absorb $\hbar$ in the time $t$, i.e.\ measure the time in units of inverse energy.

The time evolution operator $\hat U(t,t_0)$ obeys the standard equation of motion
\begin{eqnarray}
\partial_t  \hat U(t,t_0)  &=& - i\hat V(t) \hat U(t,t_0)
\end{eqnarray}
which is formally integrated to the time ordered operator
\begin{eqnarray}
 \hat U(t,t_0)  &=&T e^{  -i \int_{t_0}^t d t' \hat V(t') }.
\end{eqnarray}
$\hat V(t)$ denotes $\hat H_{T}$ in the interaction picture,
\begin{eqnarray}
\hat V(t) &=& e^{i \hat H_0 (t-t_0)} \hat H_{T} e^{-i \hat H_0 (t-t_0)} .
\end{eqnarray}
All expectation values have to be calculated with respect to the two decoupled system, $\langle  \hat A  \rangle_0 = {\rm Tr}\left[\hat \rho_0  \hat A \right]$, assuming thermodynamic equilibrium in each of the two uncoupled subsystems (S and the STM tip). Then, the density operator $\hat \rho_0$ factorizes into two independent contributions, 
\begin{eqnarray}
\label{eq:rho-0}
\hat \rho_0 &=& \hat \rho_{S}\hat \rho_{\rm tip}
\end{eqnarray}
with
\begin{eqnarray}
\hat \rho_{S} &=& \frac{1}{Z_{S}}e^{-\beta(\hat H_{S}  -\mu_{S} \hat N_{S})}
\non
\hat \rho_{\rm tip} &=& \frac{1}{Z_{\rm tip}}e^{-\beta(\hat H_{\rm tip}  -\mu_{\rm tip} \hat N_{\rm tip})}.
\end{eqnarray}
Here, we have introduced different chemical potentials for each subsystem: $\mu_{S}$ for the sample system S and $\mu_{\rm tip}$ for the STM tip. The bias voltage $V$ enters through the difference $\mu_{\rm tip}-\mu_S=eV$. For convenience, we define the chemical potential $\mu_{\rm S}$ of the system S as a reference energy and absorb it into the definition of the single particle energy. 


Evaluating the current up quadratic order in the tunneling matrix elements yields
\begin{eqnarray}
\label{eq:13}
I(t)  &=&  i \int_{t_0}^t d t'  \left[
\langle \hat V(t')  \hat j_{I}(t)   \rangle_0
-
\langle \hat j_{I}(t) \hat V(t')   \rangle_0
\right]
\non
&&
+ O(t^3_s),
\label{eqn:8}
\end{eqnarray}
where $t_s$ is a measure of the order of magnitude of the largest tunneling matrix element $t_{\mu\sigma}$. Note that $\langle \hat V(t') \hat j_{I}(t) \rangle_0=\langle  \hat j_{I}(t) \hat V(t') \rangle_0^*$, ensuring that the  current is real.

Substituting the linear expansion of tunneling matrix elements in the displacements of the vibrational modes $\nu$, Eq.\ \eqref{eq:e-ph-tunnel}, into $\langle \hat V(t')  \hat j_{I}(t)   \rangle_0$ yields
\begin{widetext}
\begin{eqnarray}
\label{eq:Vj}
\langle \hat V(t')  \hat j_{I}(t)    \rangle_0&=&
\sum_{\mu\sigma\mu'\sigma'}  
t_{\mu \sigma}t_{\mu' \sigma'}
\Big \langle (1 +\sum_\nu  \lambda^{\rm tip}_{\mu\nu}\hat X_\nu (t'))(1 +
\sum_{\nu'}  \lambda^{\rm tip}_{\mu'\nu'}\hat X_{\nu'} (t)  )
\left(
d^\dagger_{\mu\sigma}(t')  c_{0\sigma,\rm tip}(t') + 
c^\dagger_{0\sigma,\rm tip}(t')
d_{\mu\sigma}  (t')
\right)
\non
&& \times
\left(
d^\dagger_{\mu'\sigma'}(t)  c_{0\sigma',\rm tip}(t) - 
c^\dagger_{0\sigma',\rm tip}(t)
d_{\mu'\sigma'}  (t)
\right)
 \Big \rangle_0
\end{eqnarray}
\end{widetext}
Since $\hat H_{S}$ and $\hat H_{\rm tip}$ as well as the corresponding density operators commute, the expectation values factorize into products of the sample system S and the STM tip, and we arrive at
\begin{widetext}
\begin{eqnarray}
\label{eq:16}
\langle \hat V(t')  \hat j_I(t)   \rangle_0&=&
\sum_{\mu \mu'\sigma}  
t_{\mu \sigma}t_{\mu' \sigma}
\Big \langle 
(1 + \sum_\nu  \lambda^{\rm tip}_{\mu\nu}\hat X_\nu (t'))
(1 +\sum_{\nu'}  \lambda^{\rm tip}_{\mu\nu'}\hat X_{\nu'} (t) )
d_{\mu\sigma}  (t') d^\dagger_{\mu'\sigma}(t)
\Big \rangle_0
\Big \langle  
c^\dagger_{0\sigma,\rm tip}(t')  c_{0\sigma,\rm tip}(t)
\Big \rangle_0
\non
&& 
- 
\sum_{\mu\mu'\sigma}  
t_{\mu \sigma}t_{\mu' \sigma}
\Big \langle 
(1 + \sum_\nu  \lambda^{\rm tip}_{\mu\nu}\hat X_\nu (t'))
(1 +\sum_{\nu'}  \lambda^{\rm tip}_{\mu\nu'}\hat X_{\nu'} (t) )
d^\dagger_{\mu\sigma}(t')   d_{\mu'\sigma}(t)
\Big\rangle_0
\Big\langle  
c_{0\sigma,\rm tip}(t')  c^\dagger_{0\sigma,\rm tip}(t)
\Big\rangle_0,
\nonumber \\
\end{eqnarray}
\end{widetext}
under the assumption that the system S and the STM tip are in a normal conducting state. This factorization does not require a Wick's theorem, and, therefore, the Hamiltonian $\hat H_0$ remains fully general. Note  that the electronic correlation function of the STM tip is spin-diagonal, and hence the double sum over $\sigma \sigma'$ in Eq.\ \eqref{eq:Vj} collapses to a single sum over $\sigma$ in Eq.\ \eqref{eq:16}. It is clear that the displacements terms $\hat X_\nu$ and the electronic orbital operators $d_{\mu \sigma}$ do not factorize, because we explicitly allow for a strong electron-phonon coupling and thus polaron formation in the system S. Finally, the terms of type $\langle  d^\dagger_{\mu\sigma} d^\dagger_{\mu'\sigma }\rangle_0$ that we have neglected in Eq.\ \eqref{eq:16} must be included if either the STM tip or the sample system S are superconducting. In this case, our approach reproduces the well-known derivation of the Josephson current by Ambegaokar and Baratoff \cite{Ambegaokar1963}.

Eq.\ \eqref{eq:16} can be divided into elastic and inelastic contributions. The former are obtained by setting all $ \lambda^{\rm tip}_{\mu\nu}=0$, while the inelastic terms are given by the difference between Eq.\ \eqref{eq:16} for non-vanishing $\lambda^{\rm tip}_{\mu\nu}$ and for $\lambda^{\rm tip}_{\mu\nu}=0$. Similarly, the total current decomposes into the sum
\begin{eqnarray}
\label{eq:totalcurrent}
I_{\rm tot} &=&  I_{\rm el} +I_{\rm inel}, 
\end{eqnarray}
comprising an elastic and an inelastic current. This naturally defines the terminology used throughout the rest of the paper. 

In summary, we have related the total current to products involving a greater Green's function 
$G^>\sim 
\langle  C(\lambda^{\rm tip}) d^\dagger_{\mu\sigma}(t')  d_{\mu'\sigma}(t) \rangle_0$ of the system S and a lesser Green's function $G^< \sim \langle c_{0\sigma,\rm tip}(t')  c^\dagger_{0\sigma,\rm tip}(t)\rangle_0$  of the STM tip and vice versa \cite{Keldysh65}. Most importantly, the  Keldysh Green's functions of a fully interacting system entangling, in general,  vibrational and electronic operators, are employed for the system S. Therefore, the expressions derived and analyzed in the following sections go well beyond the standard literature.

\subsubsection{Elastic tunnel current}
\label{sec:elastic tunnel current}

Since we are interested in the asymptotic steady-state current, we perform the limit $t_0\to -\infty$ and calculate the current at the time $t=0$. Evaluating the greater and lesser Green functions with the equilibrium density operator in Eq.\ \eqref{eq:rho-0} and calculating the steady-state current for $\lambda^{\rm tip}_{\mu\nu}=0$ 
in Eq.\ \eqref{eq:13}, we obtain the well-known expression for the elastic tunnel current
\begin{eqnarray}
\label{eqn:el-current}
I_{\rm el}(t=0) &=&
\frac{2\pi e}{\hbar} 
\sum_{\sigma}  
\int_{-\infty}^{\infty} d\w  \rho_{\sigma,\rm tip}(\w) \tau^{(0)}_{\sigma}(\w)
\nonumber \\
&&\times
\left[
f_{\rm tip}(\w) - f_{S}(\w)
\right],
\label{eqn:elasticCurrent}
\end{eqnarray}
where $f_{\rm tip}(\w)=f(\w-eV)$ and $f_{S}(\w)=f(\w)$, with $f(\w)=[\exp(\beta \w)+1]^{-1}$ being the Fermi function.
\begin{eqnarray}
\tau^{(0)}_{\sigma}(\w) &=& \sum_{\mu\mu'}^M  t_{\mu \sigma}t_{\mu' \sigma}
\lim_{\delta\to 0^+} 
\frac{1}{\pi} \Im G_{d_{\mu\sigma},  d^\dagger_{\mu'\sigma}} (\w-i\delta),
\label{eqn:tauzero}
\end{eqnarray}
is the spin-dependent transmission function from the STM tip to the system S, 
and $\rho_{\sigma,\rm STM}(\w)$ denotes the spectral density of the STM tip,
\begin{eqnarray}
\rho_{\sigma,\rm tip}(\w) = \lim_{\delta\to 0^+} \frac{1}{\pi}  \Im G_{c_{0\sigma,\rm tip}, c^\dagger_{0\sigma,\rm tip}}
(\w-i\delta).
\end{eqnarray}
As usual, $G_{A,B}(z)$ refers to the equilibrium Green's function \cite{Rickayzen1980} defined in the complex frequency plane $z$ except on the real axis. Throughout the paper we moreover use the notation $\rho_{A,B}(\w)\equiv \Im G_{A,B}(\w-i0^+)/\pi$ to connect a Green's function of the two operators $A,B$ with its spectral function $\rho_{A,B}(\w)$. Note, however, that we have written Eq.\ \eqref{eqn:el-current} in terms of a transmission function $\tau^{(0)}_{\sigma}(\w)$ which includes the tunnel matrix elements $t_{\mu \sigma}$ as well as the spectral properties $\rho_{\mu \sigma, \mu' \sigma}(\w)$ of S, the advantage being that then all contributions to the current in Eq.\ \eqref{eq:totalcurrent}, i.e. Eqs.\ \eqref{eqn:el-current}, \eqref{eqn:inelasticCurrent} and \eqref{eq:second-order-inelastic-contribution}  have the same overall structure. In fact, the transmission function $\tau^{(0)}_{\sigma}(\w) $ can also be interpreted as the fermionic Green's function of the operator $A_\sigma= \sum_{\mu}^{M-1} t_{\mu \sigma}d_{\mu\sigma}$.

The usual assumption in STM experiments is that the density of states $\rho_{\sigma,\rm tip}(\w)$ is featureless in the energy (voltage) interval of interest. Then, it can be replaced by a constant $\rho_{\sigma,\rm STM}$ that only enters the prefactor in Eq.\ \eqref{eqn:el-current}. This confirms that a detailed knowledge of the expansion coefficients in Eq.\ \eqref{eq:stm-tip-c0} is not required, since these coefficients can be absorbed into this prefactor. 

We note that the result in Eq.\ \eqref{eqn:el-current} also allows for interferences among more than one elastic transport channels. For example, it is straightforward to show that Eq.\ \eqref{eqn:el-current} reproduces the result of Eq.\ (6) of Ref.\ \cite{SchillerHershfield2000a}, if we set $M=2$ and replace $d_{1\sigma}$ by the local surface conduction electron operator $\psi(\vec{R}_s)$. The Fano resonance \cite{FanoResonance1961} is generated by the quantum interference between the two or more elastic tunneling channels. 

\subsubsection{Inelastic tunnel current}
\label{sec:I-inelastic}

The contributions to the inelastic tunnel current are classified by the power of the electron-phonon coupling $ \lambda^{\rm tip}_{\mu\nu}$ in the tunneling Hamiltonian. We note again that the additional inelastic contributions arise from phonon absorption and emission \textit{during} the tunneling process, while all electron scattering processes within the system $S$ are included in $I_{\rm el}$. 

In first order in $ \lambda^{\rm tip}_{\mu\nu}$, we obtain an inelastic tunnel current
\begin{eqnarray}
I^{(1)}_{\rm inel} 
 &=&
 \frac{2\pi e}{\hbar} 
\sum_{\sigma}  
\int_{-\infty}^{\infty} d\w  \rho_{\sigma,\rm tip}(\w)
\tau^{(1)}_{\sigma}(\w)
\non
&& \times  \left[
f_{\rm tip}(\w) - f_{S}(\w)
\right]
\label{eqn:inelasticCurrent}
\end{eqnarray}
from Eq.\ \eqref{eq:16}, where following the general notation in this paper (see above) the transmission function $\tau^{(1)}_{\sigma}(\w)$ is defined as the spectral function of a composite Green's function $G^{(1)}_{d\sigma}$ which in turn involves tunneling matrix elements $t_{\mu \sigma}$ and correlation functions  $G_{X_\nu d_{\mu\sigma}, d^\dagger_{\mu'\sigma}}$ and $G_{d_{\mu\sigma},  X_\nu d^\dagger_{\mu'\sigma}}$ as 
\begin{eqnarray}
\label{eq:rho1explicit}
\tau^{(1)}_{\sigma}(\w)&=&
\frac{1}{\pi}\lim_{\delta\to 0^+} \Im G^{(1)}_{d\sigma} (\w-i\delta) \nonumber
\\
&=&\sum_{\mu\mu'}  t_{\mu \sigma}t_{\mu' \sigma}
\\
\times &\bigg(&\sum_{\nu}^{N_\nu}\lambda^{\rm tip}_{\mu\nu}\lim_{\delta\to 0^+} 
\frac{1}{\pi} \Im G_{\hat X_\nu d_{\mu\sigma},  d^\dagger_{\mu'\sigma}} (\w-i\delta) \nonumber
\\
&+& \sum_{\nu}^{N_\nu}\lambda^{\rm tip}_{\mu'\nu}\lim_{\delta\to 0^+} 
\frac{1}{\pi} \Im G_{d_{\mu\sigma},  \hat X_\nu d^\dagger_{\mu'\sigma}} (\w-i\delta)\bigg).
\nonumber
\end{eqnarray}
Since the expectation value of the anticommutator of a Green's function $G_{A,B}(z)$ equals the frequency integral of the corresponding spectrum $\rho_{A,B}(\w)$, we can conclude that the  spectra of $G_{d_{\mu \sigma}, \hat X_\nu d^\dagger_{\mu' \sigma}}$ and $G_{\hat X_\nu d_{\mu \sigma}, d^\dagger_{\mu' \sigma}}$ both individually integrate to $\langle \hat X_\nu\rangle \delta_{\mu,\mu'}$. Hence, for a vanishing displacement $\langle\hat X_\nu \rangle=0$, either the Green's function $G^{(1)}_{d\sigma}$ is identically zero, or its spectrum (i.e., the transmission function $\tau^{(1)}_{\sigma}(\w)$) has equal positive and negative spectral contributions. The first statement is true in the limit of vanishing electron-phonon coupling. For non-vanishing electron-phonon coupling, however, the correlation function $G^{(1)}_{d\sigma}$ does not vanish, since quantum fluctuations and hence non-zero correlators $G_{d_{\mu \sigma}, \hat X_\nu d^\dagger_{\mu' \sigma}}$ and $G_{\hat X_\nu d_{\mu \sigma}, d^\dagger_{\mu' \sigma}}(z)$ are allowed even if $\langle\hat X_\nu \rangle=0$.  

In second-order in $\lambda^{\rm tip}_{\mu \nu}$, the inelastic tunnel current
\begin{eqnarray}
I^{(2)}_{\rm inel} 
 &=&
 \frac{2\pi e}{\hbar} 
\sum_{\sigma}  
\int_{-\infty}^{\infty} d\w  \rho_{\sigma,\rm tip}(\w)
\tau^{(2)}_{\sigma}(\w)
\non
&& \times  \left[
f_{\rm tip}(\w) - f_{S}(\w)
\right],
\label{eq:second-order-inelastic-contribution}
\end{eqnarray}
involves the transmission function 
\begin{eqnarray}
\label{eq:rho2explicit}
\tau^{(2)}_{\sigma}(\w) &= & \frac{1}{\pi} \lim_{\delta\to 0^+} \Im G^{(2)}_{d\sigma} (\w-i\delta) \nonumber
\\
&=&\sum_{\mu\mu'}^M  t_{\mu \sigma}t_{\mu' \sigma}
\\
&&\times \sum_{\nu\nu'}^{N_\nu}\lambda^{\rm tip}_{\mu\nu}\lambda^{\rm tip}_{\mu'\nu'}\lim_{\delta\to 0^+} 
\frac{1}{\pi} \Im G_{\hat X_\nu d_{\mu\sigma},  \hat X_{\nu'}d^\dagger_{\mu'\sigma}} (\w-i\delta). \nonumber
\end{eqnarray}
Up to second-order, the total inelastic contribution to the tunneling current is thus given by $I_{\rm inel}  = I^{(1)}_{\rm inel} +I^{(2)}_{\rm inel}$. Again, the spectral sum rule of $G_{\hat X_\nu d_{\mu \sigma}, \hat X_{\nu'} d^\dagger_{\mu' \sigma}}(z)$ is related to the expectation value of the anticommutator,
i.\ e.\  $\langle \hat X_\nu \hat X_{\nu'} \rangle \delta_{\mu,\mu'}$.

\subsubsection{The limit of vanishing electron-phonon coupling in the system S}
\label{sec-lambda-0}

In order to make the connection to the literature and also to point out the major difference of our theory in comparison with earlier ones, we consider the limit of vanishing electron-phonon coupling in the system S, but maintain a small but non-zero $\lambda^{\rm tip}_{\mu\nu}$. Then, $\langle\hat X_\nu \rangle=0$. As argued in the previous section, as a consequence $G^{(1)}_{d\sigma}(z)=0$, $\tau^{(1)}_{\sigma}(\w)=0$ and $I^{(1)}_{\rm inel}=0$ hold, while $I^{(2)}_{\rm inel}$ is reduced to a simplified result \cite{MolecularVibrationTunnel1968}, because the correlation function $G_{\hat X_\nu d_{\mu \sigma}, \hat X_{\nu'}  d^\dagger_{\mu' \sigma}}(t)$ in Eq.\ \eqref{eq:16} factorizes in the time domain into the product of the electronic Green's function and the phonon propagator $G_{\hat X_\nu, \hat X_\nu}(t)$,
\begin{eqnarray}
\label{eqn:26}
G_{\hat  X_\nu d_{\mu \sigma}, \hat  X_{\nu'} d^\dagger_{\mu' \sigma}}(t) &=&  
G_{d_{\mu \sigma},  d^\dagger_{\mu' \sigma}}(t) G_{\hat X_\nu, \hat X_\nu}(t) \delta_{\nu\nu'}.
\end{eqnarray}
Therefore, the spectral function is given by a convolution in the frequency domain. 
Summing over all  free vibrational modes $\nu$ on a molecule with frequency $\w_{\nu}$, we obtain \cite{Mahan81}
\begin{widetext}
\begin{eqnarray}
\label{eqn:27}
\tau^{(2)}_{\sigma}(\w) 
 &=& \sum_{\mu\mu'\nu}  
t_{\mu \sigma}t_{\mu' \sigma}
\lambda^{\rm tip}_{\mu \nu} 
\lambda^{\rm tip}_{\mu' \nu} 
\left[
\rho_{d_{\mu \sigma},d^\dagger_{\mu'\sigma}}(\w-\w_\nu)(g(\w_\nu) +f_{S}(\w_\nu-\w)) 
+
\rho_{d_{\mu \sigma},d^\dagger_{\mu'\sigma}}(\w+\w_\nu)(g(\w_\nu) +f_{S}(\w_\nu+\w)) 
\right]
\nonumber \\
\label{eqn:freemode_inelastic}
\end{eqnarray}
\end{widetext}
where $g(\w)$ denotes the Bose function. Using the approximation \eqref{eqn:27} in Eq.\ \eqref{eq:second-order-inelastic-contribution} yields the identical inelastic current contribution as derived in Refs.\ \cite{Caroli72}. 

We briefly discuss two extreme cases for the electronic spectral function. For simplicity, we restrict ourselves to $M=1$ and a single vibrational mode with frequency $\w_0$. This excludes the possibility of a Fano resonance.  In the first extreme, we assume a featureless density of states in the vicinity of the Fermi energy over an interval  $[-2\w_0,2\w_0]$,  i.\ e.\ $\rho_{\mu,\mu'\sigma}(\w)\to {\rm const.}$ in \myeqref{eqn:27}, and  $\beta\w_0\gg 1$ so that the Bose function can be ignored. Then $\tau^{(2)}_{\sigma}(\w) $ will be dominated by the Fermi functions, introducing two threshold contributions in the overall differential conductance $dI/dV$ at $\pm \w_0$. These are the typical $dI/dV$ steps that are often encountered in inelastic tunnel spectroscopy as shown in Fig.~1 of Ref.~\cite{REED2008}.

In the second extreme, we consider an electronic DOS of the sample system S which possesses a sharp spectral peak located at $\w=0$  with a width $\Gamma\ll \w_0$. Then $\tau^{(2)}_{\sigma}(\w)$ exhibits again a sharp threshold behavior at $\pm \w_0$, but instead of a plateau the spectral function decreases with increasing $|\w|$ on a scale given by the peak broadening $\Gamma$. In this case, two "replicas" of the peak at $\w=0$ can be found at $\pm \w_0$ in the overall differential conductance $dI/dV$. However, the Fermi function $f_{\rm S}(\w_0\pm\w)$ cuts away the halves of the replicas on the low-$|\w|$ side and modifies them to a threshold behavior: a minimal energy transfer for $eV=\pm\w_0$ is required to generate an inelastic contribution replicating the standard picture \cite{MolecularVibrationTunnel1968}. These truncated replicas of the $\w=0$ peak are generated by the inelastic tunneling process due to the change of the distance between system S and STM tip. In case of a large electron-phonon coupling in S, the approximation \myeqref{eqn:27} is invalid and the proper Green's function $G_{\hat X_\nu d_{\mu \sigma}, \hat X_{\nu'}  d^\dagger_{\mu' \sigma}}(t)$ must be calculated, along with $G_{X_\nu d_{\mu\sigma}, d^\dagger_{\mu'\sigma}}$ and $G_{d_{\mu\sigma},  X_\nu d^\dagger_{\mu'\sigma}}$.

\subsubsection{Discussion}

The presented tunneling theory combines different limits \cite{MolecularVibrationTunnel1968,Mahan81,PerssonBaratoff1987,LorentePersson2000,PaakeFlensberg2005,REED2008} discussed the literature. We stress again that no assumption is needed with respect to the nature and dynamics governing the system S. On the contrary, the theory can address arbitrary strengths of both the electron-electron and electron-phonon interactions in the system. The only input that is required are the Green's functions of the participating orbitals and vibrational displacements in the absence of the STM tip. 

The theory is valid in the tunneling limit and we have restricted ourselves to the conventional single-particle electron transfer operator $\hat H_{T}$ \cite{Bardeen1961}. A further assumption that we have made concerns the relative distance changes between the tip and the system S that are induced by the relevant vibrational modes; they must be  small enough such that a linear expansion of the tunneling matrix elements in the displacements suffices and higher-order terms can be neglected. 

Apart from allowing quantitative calculations of tunneling spectra for realistic systems, one of the most important benefits of the theory is that it allows a systematic separation of elastic contributions to the tunneling current (charge transfer does not involve an energy transfer) from inelastic ones (arising from correlated tunneling processes involving a fermionic hopping and a displacement operator). This differentiation in some cases deviates from the one given in the literature. In fact, the terminology elastic vs.\ inelastic current is not unambiguous throughout the literature. 

In some cases, certain contributions to the elastic and inelastic tunnel currents may have even the same analytic structure. This can be illustrated for a single orbital in  the case of weak electron-phonon coupling in the system S, if moreover the free electron picture and the free phonon pictures are employed \cite{PerssonBaratoff1987,LorentePersson2000,Wehling2008}. We have seen in the previous section that in the case of vanishing electron-phonon coupling in S, but for finite coupling $\lambda^{\rm tip}$ in the tunneling matrix element, the general expression for $\tau^{(2)}_{\sigma}(\w)$ (Eq.\ \eqref{eq:rho2explicit}) in the inelastic current $I^{\rm (2)}_{\rm inel}$ takes on the shape as Eq.\ \eqref{eqn:freemode_inelastic}, leading to an inelastic current which for the special case of a flat DOS at the Fermi level leads to steps in the differential conductance at the vibrational energies $\pm \omega_0$.

\def\fmfL(#1,#2,#3)#4{\put(#1,#2){\makebox(0,0)[#3]{#4}}}
\begin{figure}[t]
\begin{picture}(100,100)(0,0)
\put(0,0){
\includegraphics{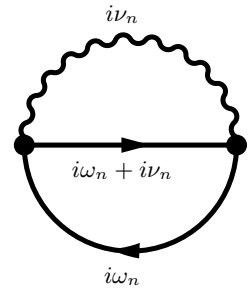}
}
\fmfL(60.22516,54.22514,t){$i\omega _n+i\nu _n$}%
\fmfL(60.22514,106.37524,b){$i\nu _n$}%
\fmfL(60.22514,14.07504,t){$i\omega _n$}%
\end{picture}
\caption{Second-order Feynman diagram of the generating Luttinger-Ward functional in the system S. The full line represents the full local electron Green's function $G_d(z)$, the wiggled line the full phonon propagator. $i\w_n = i\pi (2n+1)/\beta$ denote the fermionic Matsubara frequencies and $i\w_n = i2\pi n/\beta$ the bosonic Matsubara frequencies \cite{LuttingerWard1960}.
}
\label{fig:self-electron-phonon-diagram}
\end{figure}

We now compare this result to the perturbatively calculated elastic current in the same limit. Under these circumstances, the Green's function of the orbital with the single-particle energy $\e_a$ has the form $G_a(z)=[z- \e_a -\Sigma_{\rm el}(z) - \Sigma_{\rm el-ph}(z)]^{-1}$, where the self energy $\Sigma_{\rm el}(z)$ accounts for the purely electronic interactions and $\Sigma_{\rm el-ph}(z)$ arises from the additional electron-phonon coupling which is limited to the system S as assumed by Lorente et al.~\cite{LorentePersson2000}.  Introducing $G_{a}^{(0)} = [z- \e_a -\Sigma_{\rm el}(z)]^{-1}$ allows for a perturbation expansion in linear order of $\Sigma_{\rm el-ph}(z)$ in weak electron-phonon coupling,
\begin{eqnarray}
\label{equ:Gf-expansion}
G_a(z) &=& G_{a}^{(0)}(z) + G_{a}^{(0)}(z) \Sigma_{\rm el-ph}(z)G_{a}^{(0)}(z) + \cdots
\end{eqnarray}
If we substitute this expansion of $G_a(z)$ into the expression Eq.\ \eqref{eqn:el-current} for the elastic current, two contributions arise, first a purely electronic one generated by $ G_{a}^{(0)}(z)$, and second a contribution involving the self energy $\Sigma_{\rm el-ph}(z)$ in first order. This second term has been designated as an inelastic term in the literature \cite{PerssonBaratoff1987,LorentePersson2000,Wehling2008}, but we include it in the elastic part of the current, since the electron energy is conserved during the tunneling process and the scattering process occurs in the system S.

It is interesting to note that the second-order  contribution to  $\Sigma_{\rm el-ph}(z)$ in Eq.\ \eqref{equ:Gf-expansion} is proportional to $G_{\hat X_\nu d_{\mu \sigma}, \hat X_\nu d^\dagger_{\mu' \sigma}}$  in weak coupling. This can be seen from Fig.\ \ref{fig:self-electron-phonon-diagram}, which depicts the generating functional \cite{LuttingerWard1960} for the conserving approximation of $\Sigma_{\rm el-ph}(z)$: by differentiating with respect to the electronic Green's function $G_d(i\w_n)$ (equivalent to cutting the semi-circular full line)  we obtain the diagram of the electronic self energy $\Sigma_{\rm el-ph}(z)$ due to the electron-phonon interaction \cite{AllenMitrovic,JovchevAnders2013}. Evidently,  this diagram has the structure of $G_{\hat X_\nu d_{\mu \sigma}, \hat X_\nu d^\dagger_{\mu' \sigma}}$. Therefore, the analytic structure of the \textit{elastic} current calculated in second-order perturbation theory from Eq.\ \eqref{equ:Gf-expansion} is identical to that of the \textit{inelastic} current given by Eq.\ \eqref{eqn:freemode_inelastic}. But while the overall form of the two current contributions is identical, the physical mechanisms are different: In one case, an electron tunnels elastically from the tip into the system S, probing its density of states that includes the effects of vibration-induced electron scattering \textit{within the system S}. In the other case, the electron loses energy \textit{during the tunneling}, induced by the modulation of the distance between the tip and the system S, and transmits through the system S without further scattering on the vibration. Although both loss processes are governed by different coupling constants ($\lambda^{\rm tip}$ vs.\ $\lambda_c$ or $\lambda_d$, see section \ref{sec:modelling_the_system_vibrations}), in the limit of vanishing couplings these become of course indistinguishable. Thus, our approach incorporates the literature result in the limit of vanishing electron-phonon couplings. Once we leave the validity of the weak coupling limit, however, the two processes become distinguishable: On the one hand, the electron-phonon coupling in S leads to peaks in the density of states at the vibrational frequencies (see above), giving a distinct signature in the elastic current, and on the other hand the full composite Green's function $G_{\hat X_\nu d_{\mu \sigma}, \hat X_\nu d^\dagger_{\mu' \sigma}}$ must be calculated in the appropriate higher order to obtain the correct inelastic current as given by Eq.\ \eqref{eq:second-order-inelastic-contribution}; this also properly accounts for the renormalization of the phonon frequency in the adiabatic regime which will affect the inelastic current profoundly.

Returning to Fig.\ \ref{fig:self-electron-phonon-diagram}, we note that the phonon propagator obtains its self-energy by differentiation of the functional with respect to the phonon propagator (equivalent to cutting the wiggly phonon line that branches off from the particle-hole loop). In the spectral function and STS spectra, this correction to the phonon propagator has two consequences: First, it causes a renormalization of the phonon frequency itself \cite{HewsonMeyer02,EidelsteinSchiller2013} and second, it induces multi-phonon processes. In the literature, however, the self-energy of the phonon propagator is often neglected \cite{Wehling2008} and the phonon is treated as a free excitation with an infinite lifetime, such that only the bare phonon frequency enters the final expression \cite{PerssonBaratoff1987,LorentePersson2000}. Evidently, such approaches are limited to the case of a vanishing electron-phonon coupling in the system, i.\ e.\ the weak adiabatic limit, and cannot include renormalization effects stemming from multi-phonon processes. Already at moderate electron-phonon coupling corrections at $\w=2\w_0$ in the self-energy occur which also find their way into the tunneling spectra. Ref.\ \cite{JovchevAnders2013} discusses the deviations of the non-perturbatively calculated full electron-phonon self-energy from the lowest-order perturbative results.

In conclusion, we maintain the terminology of the elastic current for all current contributions where the electrons travel ballistically between the tip and sample system S. Internal many-body scattering processes within the system S are all included the spectral functions within $\tau^{(0)}_{\sigma}(\w)$ and no assumption of the strength of the internal interactions are required. Therefore, $I_{\rm el}$ describes the current for a static distance between the system S and the STM tip.

\section{Modeling the system}
\label{sec:Modeling the system}

In the previous section, we have presented a tunneling theory which relies on three spectral functions: one contains the information on the elastic tunneling current, the other two are connected linearly and quadratically to vibrational displacements. While this tunneling theory is completely general, for its application we need to specify the Hamiltonian of the system $\hat H_{S}$ and thus also the spectral functions which enter the tunneling theory. In the present section, we specify and discuss a $\hat H_{S}$ which turns out to be of sufficient generality to describe the physical sample system which we investigate experimentally in section \ref{sec:experiment-NTCDA}.  

\subsection{Electronic degrees of freedom}
We employ a single-orbital single impurity Anderson model (SIAM) for the electronic degrees of freedom
\begin{eqnarray}
\label{eq:H-e}
\hat H_{\rm e} &=& 
 \sum_{\k\sigma} \e_{\k\sigma} c^\dagger_{\k\sigma}  c_{\k\sigma} 
+
\sum_\sigma \e_{d\sigma} n^d_\sigma + U n^d_\uparrow n^d_\downarrow
\non
&& + \sum_{\k\sigma} V_{\k} ( c^\dagger_{\k\sigma} d_{0\sigma} + d^\dagger_{0\sigma}  c_{\k\sigma} )
\label{eqn:SIAM}
\end{eqnarray}
of the sample system S, thereby   assuming that only one single molecular orbital is relevant for the energy spectral properties accessed by the STM. $c^\dagger_{\k\sigma}$ creates an effective substrate electron of energy $\e_{\k\sigma}$,  momentum $\k$ and spin $\sigma$, while $ d^\dagger_{0\sigma}$ creates an electron in a local orbital, e.g.\ of a species adsorbed on the substrate surface, with the energy $\e_{d\sigma}$. The third term in the above equation specifies the Coulomb repulsion between electrons of opposite spin in the local orbital. The last term describes the hybridization between the substrate and the local orbital. We include the subscript 0 into the notation of the single active molecular orbital indicating that it will enter the tunnel Hamiltonian $H_T$, Eq.\  \eqref{eq:general-tunneling-HT}, as $\mu=0$ orbital.

For solving realistic systems with this ansatz, the SIAM needs to be mapped to the results of an atomistic simulation of the system in question. In this context, the projected density of states (PDOS) of the local orbital as calculated by a combination of DFT and many-body perturbation theory (MBPT) \cite{RevModPhys.74.601}, plays a crucial role, because the mean-field parametrization of the local orbital's Green's function
\begin{eqnarray}
G^{\rm mf}_d(z) = [ z- \e_{d\sigma} -Un_{-\sigma} - \Delta(z)]^{-1}
\label{eqn:MFGF}
\end{eqnarray}
can be employed to extract $\e_{d\sigma}$ as well as the hybridization function $\Delta(z)$ \cite{PTCDAAgMove,AU-PTCDA-monomer} as defined in the framework of the SIAM, 
\begin{eqnarray}
\Delta(z) &=& \sum_{\k} 
\frac{|V_{\k}|^2}{z-\e_{\k\sigma}}.
\label{eqn:HybFunc}
\end{eqnarray}
Both serve as the input for a NRG calculation \cite{BullaCostiPruschke2008}. We note that in the absence of an electron-phonon coupling the influence of the substrate on the  dynamics in the local orbital is completely determined by $\Delta(z)$, which justifies an effective single-band model \cite{BullaPruschkeHewson1997}. The spectral function required for the calculation of the elastic tunnel current through the system S as specified by the above Hamiltonian can be obtained by the standard approach \cite{PetersPruschkeAnders2006,WeichselbaumDelft2007} that is based on the complete basis set of the NRG \cite{AndersSchiller2005,AndersSchiller2006}. If the local orbital is close to integral filling, the exact solution of this model describes the Kondo effect \cite{Wilson75,KrishWilWilson80a,*KrishWilWilson80b}.

\subsection{Vibrational degrees of freedom and electron-phonon coupling}
\subsubsection{Hamilton operator}
\label{sec:modelling_the_system_vibrations}

Naturally, we need to include a vibrational component into $\hat H_{S}$ if we want to calculate the two inelastic transmission functions $\tau^{(1)}_{\sigma}(\w)$ and $\tau^{(2)}_{\sigma}(\w)$ that play a role in the tunneling theory of section \ref{sec:tunnel-current}. We divide the vibrational Hamiltonian into two parts, $\hat H_{\rm ph}$ and $\hat H_{\rm e-ph}$. We assume that there are $N_\nu$ phonon modes in the system S, and hence $\hat H_{\rm ph}$ is given by 
\begin{eqnarray}
\label{eq:H-ph}
\hat H_{\rm ph} &=& \sum_{\nu=0}^{N_\nu-1} \w_\nu b^\dagger_\nu b_\nu
\end{eqnarray}
where a phonon of mode $\nu$ is created by $b^\dagger_\nu$. Even in the absence of an electron-phonon coupling in the system S, this term must be included in $\hat H_{S}$ when we evaluate the tunneling current as presented in section \ref{sec:tunnel-current}, because each of the modes $\nu$ can in principle modulate the tunneling matrix element between the tip and the system S. 

The second term in the vibrational Hamiltonian, $\hat H_{\rm e-ph}$, describes the electron-phonon coupling in S. While in principle all $N_\nu$ phonon modes may couple to the electrons in the system S, for simplicity we restrict the electron-phonon coupling in the present section to a single mode $b^\dagger_0$ of frequency $\omega_0$. All NRG calculations were performed with the restriction to a single 
phonon mode in order to keep the number of  parameters in the model very small. Therefore $\omega_0$ always labels  the eigenfrequency of
the coupled vibrational mode that was included in the NRG, while $\omega_\nu (\nu>0)$ refers to  eigenfrequencies of modes with no electron-phonon coupling in $H_S$ but contribute to the tunneling Hamiltonian $H_T$. 
Of course, the number of phonon modes that couple to the electrons can straightforwardly be extended to whatever number is required to explain the experimental observations. For example, it turns out that $N_\nu=2$ phonon modes, one with a non-zero electron-phonon coupling within S, the other without, are sufficient to reproduce the experimental spectra of NTCDA/Ag(111) in section \ref{sec:experiment-NTCDA} with a minimal set of free parameters. We note that all molecular vibrations that do not have a finite or relevant electron-phonon coupling in the system S can be ignored in the calculation of the electronic properties of the system in absence of the STM tip. 
 
We assume that $\hat H_{\rm e-ph}$ is given by an extended Holstein Hamiltonian
\begin{eqnarray}
\label{eqn:AndersonHolstein}
H_{\rm e-ph} &=& \lambda_d \hat X_0 (\sum_\sigma d^\dagger_{0\sigma} d_{0\sigma} - n_{d0}) 
\\
&& \nonumber
+\lambda_c \hat X_0 (\sum_\sigma c^\dagger_{0\sigma} c_{0\sigma} -n_{c0}),
\end{eqnarray}
comprising \textit{two} Holstein couplings $\lambda_d$ and $\lambda_c$ to two distinct orbitals. One of these orbitals is the local orbital $d_0$, the other an effective local substrate electron $c_{0\sigma}$ that hybridizes with the local orbital $d$ as described in Eq.\ \eqref{eq:H-e}. The annihilation operator of the effective local substrate electron is defined by 
\begin{eqnarray}
c_{0\sigma} &=& \frac{1}{\bar V} \sum_{\k} V_{\k} c_{\k\sigma}
\label{eqn:c_0},
\\
\bar V^2 &=& \sum_{\k} |V_{\k}|^2,
\end{eqnarray}
and is entering the hybridization part in the SIAM,  Eq.\ \eqref{eqn:SIAM}. This operator and the corresponding $c_{0\sigma}^\dagger$ obey the fermionic anticommutation relation by construction. $\hat X_0=b^\dagger_0 + b_0$ denotes the dimensionless vibrational displacement operator of the phonon mode $\omega_0$. The unconventional Holstein coupling $\lambda_c$ is included in $\hat H_{\rm e-ph}$ since it captures the fact that a vibrational excitation of the adsorbed molecule may couple to electrons in the substrate when parts of the molecule periodically beat onto the substrate surface. In particular, we will show below that this unconventional Holstein coupling can reduce the Kondo temperature \cite{PAM-E-PHON2013} of the system S. 

The additional constants $n_{d0}$ and $n_{c0}$ in Eq.\ \eqref{eqn:AndersonHolstein} are often set to zero in the literature \cite{GalperinRatnerNitzan2007} when the polaronic energy shift in the single-particle energies is of primary interest, because they do not play a role then. Here, however, we focus on the quantum fluctuations with respect to some reference filling that are induced by the electron-phonon coupling \cite{EidelsteinSchiller2013,JovchevAnders2013} and use these constants to ensure $\langle \hat X_\nu \rangle=0$. Typical values are $n_{d0}= n_{c0}=1$ at half filling.

\subsubsection{Interaction-driven displacement of the harmonic oscillator}
\label{sec:e-phonon-displacement}

Away from particle-hole symmetry, an electron-phonon coupling as the one in Eq.\ \eqref{eqn:AndersonHolstein} generates a displacement of the equilibrium position of the corresponding harmonic oscillator. Since we are going to use an atomistic DFT calculation with relaxed atomic coordinates to generate the input parameters of the model Hamiltonian $\hat H_{\rm e}+ \hat H_{\rm ph} + \hat H_{\rm e-ph}$, such an additional displacement is not justified. We therefore include appropriately adjusted $n_{d0}$ and $n_{c0}$ added in Eq.\ \eqref{eqn:AndersonHolstein} to ensure $\langle\hat X_0\rangle=0$. However, the perturbative derivation of the tunnel current does not rely on explicitly vanishing displacements $\langle \hat X_\nu \rangle$, and therefore the absorption of the equilibrium displacement in Eq.\ \eqref{eq:7} is just a convention and must not alter the physics.

The equilibrium displacement generated by the electron-phonon coupling also touches upon a more fundamental issue: Evidently the physical observables such as the total STS spectra must not depend on the precise definition of the operators $b_\nu$. We therefore need to analyze our theory in this respect. Specifically, we show in this section that the total current STS spectra calculated in our theory do not depend on the choice of the basis for the operators $b_\nu$. Interestingly, however, this choice of basis does determine the partitioning between elastic and the inelastic contributions to the total current. Inelastic and elastic currents are therefore not physical observables, but an interpretation based on a model-dependent partitioning of the total current.  

Let us assume that we have made a particular choice $\hat X$ of the oscillator basis and find a non-zero $\langle \hat X_0 \rangle=x_0$ for the mode $\w_0$ (for which $\hat H_{\rm S}$ foresees an electron-phonon coupling). For simplicity, we assume that the other $N_\nu-1$ vibrational modes $\langle \hat X_\nu \rangle=0$ holds. Then we can define a new bosonic operator 
\begin{eqnarray}
\label{eq:XX}
\bar b_0 = b_0 - \frac{1}{2}x_0  
\end{eqnarray}
such that $\langle \hat{\bar{X_0}}=\bar b_0+\bar b_0^\dagger\rangle=0$. Substituting this expression into $\hat H_{\rm ph} + \hat H_{\rm e-ph}$ (Eqs.\ \eqref{eq:H-ph},\eqref{eqn:AndersonHolstein}) and ignoring all vibrational modes except $\omega_0$ yields
\begin{eqnarray}
\label{eq:35}
\hat H_{\rm ph} &+& \hat H_{\rm e-ph} =  \sum_{\nu=1}^{N_\nu-1} \w_\nu b^\dagger_\nu b_\nu \\
&+& \w_0 \bar b^\dagger_0 \bar b_0 +\lambda_d x_0  N_d + \lambda_c x_ 0\hat N_{c}  + E_0
\nonumber\\
&+&   \hat{\bar{X_0}} 
\left( \lambda_d  (\hat N_d  - n_{d0}) +\lambda_c(\hat N_c  -n_{c0}) +\frac{\w_0 x_0}{2}\right)\,
\nonumber
\end{eqnarray}
where we define $\hat N_d \equiv \sum_\sigma \hat n^d_\sigma$, $\hat N_c\equiv \sum_\sigma c^\dagger_{0\sigma} c_{0\sigma}$ and absorb all constants into $E_0$. To keep the Hamiltonian $\hat H_{S}=\hat H_{\rm e}+\hat H_{\rm ph} + \hat H_{\rm e-ph}$ invariant under the basis set change of the bosonic operator, we substitute $\e_d\to \e_d + \lambda_d x_0$ and define a single-particle energy $\e_c=\lambda_c x_0$ for the  orbital $c_{0\sigma}$.  In case of a non-zero $\lambda_c$, we also need to shift the constant $n_{c0}\to n_{c0} - \w_0 x_0/2\lambda_c$. Since under these conditions the Hamiltonian is unaltered, the dynamics of the fermion degrees of freedom remains identical and independent of this basis transformation $b_0\to \bar b_0$. In particular, this means that the Kondo temperature, if applicable, and other thermodynamic properties of the system S remain unchanged.

We now analyze the effect of a non-zero expectation value $\langle \hat X_0 \rangle=x_0$ on the tunnel current. For simplicity we set $N_\nu=1$ and assume identical  vibrational couplings in the tunneling Hamiltonian for all $M$ orbitals, i.e. $\lambda^{\rm tip}_{\mu \nu}=\lambda^{\rm tip}_{\mu' \nu}$. For the two inelastic density of states we require the Green's functions involving the vibrational displacements either linearly or quadratically. The relation between these in the two bases $\hat{X}$  and $\hat{\bar{X}}$ follows from \eqref{eq:XX} and is given by
\begin{eqnarray}
\label{eq:36}
G_{\hat X d_{\mu \sigma}, d^\dagger_{\mu' \sigma}}(z) &=& G_{\hat{ \bar X} d_{\mu \sigma}, d^\dagger_{\mu' \sigma}}(z)
+ x_0 G_{d_{\mu \sigma}, d^\dagger_{\mu' \sigma}}(z)\\
\label{eq:37}
 G_{\hat Xd_{\mu \sigma}, \hat X d^\dagger_{\mu' \sigma}}(z)
&=&
 G_{\hat {\bar X} d_{\mu \sigma}, \hat {\bar X} d^\dagger_{\mu' \sigma}}(z)
 \\
 &&
+ x_0\left( G_{ d_{\mu \sigma}, \hat {\bar X} d^\dagger_{\mu' \sigma}}(z)
+ G_{ \hat {\bar X} d_{\mu \sigma}, d^\dagger_{\mu' \sigma}}(z)\right)
\non
&& + x_0^2 G_{d_{\mu \sigma}, d^\dagger_{\mu' \sigma}}(z) .
\nonumber
\end{eqnarray}
Substituting these expression into the  formula for the total tunnel current and regrouping the different contributions, we obtain for the sum of the three transmission functions that enter the integral for the total tunnel current up to second order
\begin{eqnarray}
\label{basis change density of states}
\tau^{(0)}_\sigma + \tau^{(1)}_\sigma &+&\tau^{(2)}_\sigma = \\
	(&1& + \lambda^{\rm tip} x_0) ^2 \bar{\tau}^{(0)}_\sigma  + (1+\lambda^{\rm tip} x_0)\bar{\tau}^{(1)}_\sigma 
	+\bar{\tau}^{(2)}_\sigma
\nonumber
\end{eqnarray}
The purely fermionic Green's function $G_{d_{\mu \sigma}, d^\dagger_{\mu' \sigma}}(z)$ picks up the factor $(1+\lambda^{\rm tip} x_0)^2$ which therefore appears as a prefactor in the elastic density of states and in the corresponding tunnel current. Remembering that our theory is accurate to quadratic order in $\lambda^{\rm tip}$, we may add corrections of order $O([\lambda^{\rm tip}]^3)$ and higher to the right hand side of Eq.\ \eqref{basis change density of states}. Since $\bar{\tau}^{1}_\sigma$ and $\bar{\tau}^{(2)}_\sigma$ are of orders  $(\lambda^{\rm tip})$and $(\lambda^{\rm tip})^2$, respectively, we can thus write
\begin{eqnarray*}
\label{basis change density of states 2}
\tau^{(0)}_\sigma + \tau^{(1)}_\sigma + \tau^{(2)}_\sigma 
&\simeq&(1 + \lambda^{\rm tip} x_0)^2 (\bar{\tau}^{(0)}_\sigma + \bar{\tau}^{(1)}_\sigma +\bar{\tau}^{(2)}_\sigma)
\end{eqnarray*}
after adding the corresponding higher-order correction terms to the prefactors of $\bar{\tau}^{(1)}_\sigma $  and $\bar{\tau}^{(2)}_\sigma$. Therefore, a finite displacement $x_0$ generates an overall prefactor $(1+\lambda x_0)^2$ in the total tunnel current. This can be absorbed into the tunneling matrix element $t^2_{\mu \sigma}\to \bar t^2_{\mu \sigma}=  t^2_{\mu \sigma}( 1+\lambda x_0)^2 \approx [t_{\mu \sigma}\exp(\lambda x_0)]^2$, leading to an identical total tunnel current for the two bases $\hat{X}$ and $\hat{\bar{X}}$, up to $O([\lambda^{\rm tip}]^3)$ corrections. 

However, while the total current is invariant under the basis change of the harmonic oscillator, the attribution of elastic and inelastic contributions remains basis-dependent, which becomes immediately obvious from the Eqs.\ \eqref{eq:36} and \eqref{eq:37}: the inelastic current in the original oscillator basis contains an elastic part with respect to the shifted oscillator basis.

We adopt the following strategy in order to ensure that all properties are discussed in the framework of a harmonic oscillator basis with vanishing displacements in the presence of the electron-phonon coupling: First, we calculate the displacement for a given $\hat H_{S}$,  secondly we perform a basis set change of the harmonic oscillators to a basis $\hat{\bar{X}}_\nu$ with $\langle\hat{\bar{X}}_\nu\rangle=0$. This leads, thirdly, to a  renormalization of the model parameters in $\hat H_{S}$, as outlined in Eq.\ \eqref{eq:35}. Fourthly,  we calculate all spectral functions in the transformed basis. This implies that the effect of the displacement is absorbed into the definition of the prefactor via Eq.\ \eqref{eq:7} and is consistent with the notion that an additional electron phonon coupling does not change the atomic equilibrium positions as determined by the LDA.

\section{Experiments on NTCDA/Ag(111)}
\label{sec:experiment-NTCDA}

\subsection{Choice of system}
\label{sec:experiment-NTCDA choice of system}
To meaningfully test our tunneling theory of section \ref{sec:tunnel-current}, we need a system S that exhibits both strong electron-electron interaction and electron-phonon interaction and can be investigated in very clean conditions with STM and STS. Specifically, building on the Hamilton operator introduced in section \ref{sec:Modeling the system}, a quantum impurity system which shows the Kondo effect appears prospective. Molecular adsorbates on metals are a good starting point to realize a quantum impurity system \cite{Kondo_molecule,kondo_molecule2,ptcda_kondo_cleavage,vib_kondo_stm,AU-PTCDA-monomer,AU-PTCDA-dimer}, since they have localized orbitals that may interact with the electrons of the metal substrate.  At the same time, molecules display large number of vibrations, offering the possibility to find a sizable electron-phonon coupling at least for some of these modes. In fact, the combination of the Kondo effect and vibrational inelastic tunneling has been reported for a few molecule-on-metal systems \cite{kondo_vib_bjunc3, vib_kondo_stm, vib_kondo_stm2, kondo_vib_bjunc2, kondo_vib_bjunc}. For technical reasons, well-ordered, commensurate periodic layers have advantages, since in these layers the molecules are located at well-defined sites, enforced by both interactions with the substrate and interactions with the neighboring molecules. 

These considerations draw our attention to the system of 1,4,5,8-naphthalene-tetracarboxylic dianhydride (NTCDA) on Ag(111). For this system, the Kondo effect has been reported \cite{ziroff2012low}. An additional benefit is that NTCDA/Ag(111) bears similarity to PTCDA/Ag(111) and AuPTCDA/Au(111), for which the Hamiltonian in Eq.\ \eqref{eqn:SIAM} has allowed a quantitative modeling of the Kondo effect. However, unlike PTCDA/Ag(111), NTCDA/Ag(111) displays the Kondo effect even in the native adsorbed state \cite{ziroff2012low}, without artificially lifting the molecule from the surface, such that it can be studied in tunneling regime, a prerequisite for our theory. Moreover, it shows a rich vibrational signature \cite{C5CP06619K,braatz2016switching}. This makes NTCDA/Ag(111) ideally suited to the present purpose.

\subsection{Experimental details}
\label{sec:experiment-NTCDA experimental details}
The Ag(111) crystal was prepared by repeated cycles of $\mathrm{Ar}^+$ sputtering and annealing to $T \approx 800$\,K for $15$ minutes. A small coverage of NTCDA molecules (less than 15\% of a monolayer) was deposited from a home-built Knudsen cell onto the clean Ag(111) surface held at $T \approx 80$\,K. After deposition the sample was annealed at $T \approx 350$\,K for $5$ minutes and afterwards cooled down to $T \approx 80$\,K within $2-3$ minutes. In order to minimize contaminations the sample was transferred into the STM immediately after the preparation. 

The scanning tunneling microscopy (STM) and spectroscopy (STS) experiments on NTCDA/Ag(111) were carried out in ultra-high vacuum (UHV) in a Createc STM with a base temperature of $T \approx 9.5$\,K and JT-STM (SPECS) with a base temperature of $T \approx 4.3$\,K. The JT-STM offers magnetic fields up to $3$\,T in the out-of-plane direction. Differential conductance $dI/dV(V)$ spectra were recorded with the lock-in technique with the current feedback loop switched off. Typical parameters were a modulation amplitude of $0.6 - 2$\,mV and modulation frequency of $833$\,Hz. Before experiments on NTCDA, a featureless tip density of states was ensured by measuring the surface state of clean Ag(111). After changing the temperature of the STM we waited for $20$\,h to obtain equilibrium conditions before measuring $dI/dV$ spectra.

$dI/dV(V)$ spectra at different locations above the same molecule were measured as follows: First, the tip was positioned above the CH edge of a NTCDA molecule at tunneling current $I = 200$\,pA and bias voltage $V = 50$\,mV; then the feedback loop was switched off and the tip was moved at constant height to different locations above the molecule, followed by the measurement of $dI/dV(V)$ spectra at each of the desired positions.  

\begin{figure}
	\centering
		\includegraphics{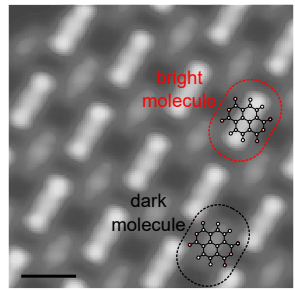}
		\caption{Constant-current STM image of the relaxed phase of NTCDA on Ag(111) ($I=200$\,pA, $V=50$\,mV). Graphical representations of (gas-phase) NTCDA molecules have been overlaid over the bright and dark molecules. The white, grey and red circles indicate hydrogen, carbon and oxygen atoms of NTCDA, respectively. The length of the scale bar is 10 \AA.}
	\label{fig:exp_fig1}
\end{figure}

\begin{figure}
	\centering
		\includegraphics{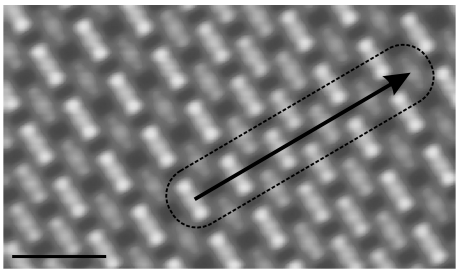}
		\caption{Constant-current STM image of the rippled phase of NTCDA on Ag(111) ($I=200$\,pA, $V=50$\,mV). The arrow indicates a line along which the character of the molecules changes gradually from bright to dark and vice versa. The length of the scale bar is 20 \AA.}
	\label{fig:fig4}

\end{figure}

\subsection{Structure}
\label{sec:experiment-NTCDA structure}
The geometric structure and the electronic properties of NTCDA on Ag(111) have  already  been studied in previous works \cite{stahl1998coverage,kilian2008commensurate,ziroff2012low, braatz2016switching, C5CP06619K}. There are two phases, commonly referred to as the relaxed and the compressed ones \cite{stahl1998coverage,kilian2008commensurate,braatz2016switching}. Here, we focus on the relaxed phase of NTCDA/Ag(111). In Fig.~\ref{fig:exp_fig1} an STM image of the relaxed phase of NTCDA is shown. The relaxed phase contains two molecules per unit cell, arranged in a brick-wall structure with a rectangular unit cell of area $11.57$\,\AA~$\times$ $15.04$\,\AA. The structure is commensurate \cite{braatz2016switching}. Because of their different appearance in the STM image the two molecules in the unit cell will from now on be referred to as bright and dark molecules, respectively. Both molecules are aligned with their long axis along the $[01\bar1]$ direction of the substrate \cite{braatz2016switching}. The difference between the two molecules most probably arises from different adsorption sites on the surface. Because the arrangement of the molecules in the unit cell is consistent with two distinct high-symmetry sites, on-top and bridge \cite{braatz2016switching}, it appears natural that the molecules are in fact located in these sites. However, it is not known whether bright molecules are in on-top and dark molecules in bridge sites or vice versa. From our PBE+vdW$^\text{surf}$ calculations (see below) we find that the NTCDA molecules at both sites are chemisorbed. The on-top molecule has an average distance of $\overline{z}=2.89$\,\AA\ and a corrugation of $\Delta z = 0.35$\,\AA\,, while for bridge molecule we observe $\overline{z}=2.86$\,\AA\ and $\Delta z=0.40$\,\AA.

We report here also a phase that to the best of our knowledge has not been reported before, the rippled phase. An STM image of the rippled phase, a variant of the relaxed phase, is shown in Fig.~\ref{fig:fig4} in which over a distance of approximately six unit cells along the $[01\bar1]$ direction of the substrate the bright molecule turns into a dark one and vice versa.

\subsection{Kondo effect}
\label{sec:experiment-NTCDA Kondo effect}
\begin{figure}
	\centering
		\includegraphics{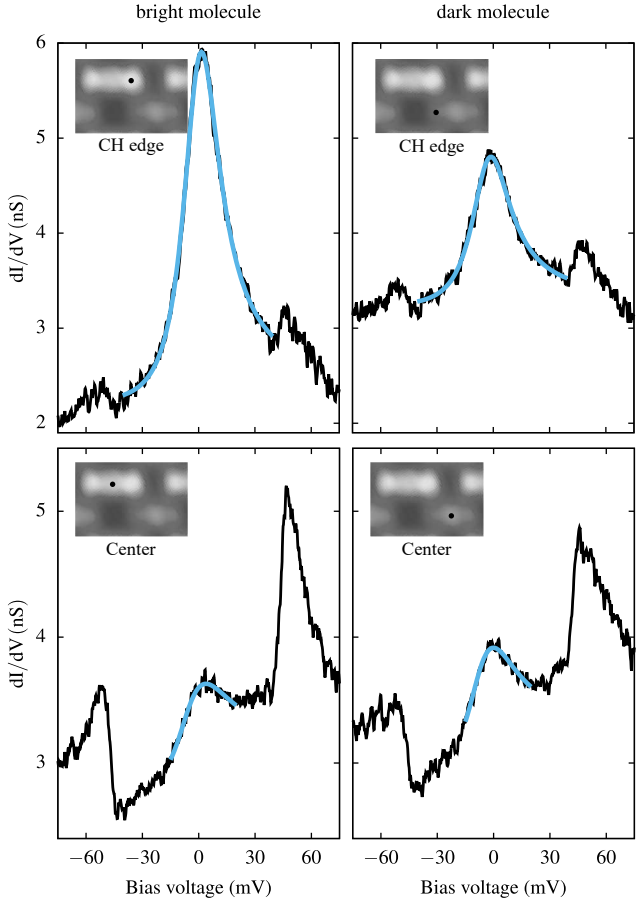}
		\caption{$dI/dV$ spectra of the bright and dark molecules acquired at the CH-edge (left) and at the center (right) of the bright (red line) and dark (black line) molecules, respectively. The spectra are plotted on the bias voltage axis as measured. A calibration of bias voltage scale to symmetrize the inelastic features would require shifting the spectra by $2.25$\,mV to the right. Fano fits according to Eq.\ \eqref{eq:Fano} are indicated by blue lines, the fit parameters are: bright molecule, CH-edge: $\delta = 24.04$, $q = 6.44$. Bright molecule, center:  $\delta = 32.50$, $q = 1.74$. Dark molecule, CH-edge:  $\delta = 25.42$, $q = 6.95$. Dark molecule, center: $\delta = 29.84$, $q = 2.08$. Average fit parameters are listed in table \ref{tab:fwhm_and_q}.}
	\label{fig:exp_fig2}
\end{figure}

Fig.~\ref{fig:exp_fig2} displays STS spectra recorded above NTCDA/Ag(111) in four different positions, namely in the vicinity of the CH edges of the NTCDA molecule and in the center of the molecule, each for both the bright and the dark molecules. These positions were chosen because at the CH edges the lowest unoccupied molecular orbital (LUMO) of NTCDA exhibits an intense lobe (for the bright molecule this lobe is directly seen in the STM image of Fig.~\ref{fig:exp_fig1}), while in the center of the molecule two nodal planes of the LUMO intersect. Note that the LUMO of NTCDA becomes partially occupied when the molecule adsorbs on the Ag(111)  \cite{ziroff2012low, braatz2016switching, C5CP06619K}. An image of the probability amplitude of  the LUMO is shown in Fig.~\ref{fig:exp_fig3}.

Fig.~\ref{fig:exp_fig2} shows that at the CH edge both molecules exhibit a peak at zero bias (the precise peak position for the bright molecule is $+1.9$\,meV, while for the dark molecule it is $-0.6$\,meV), although with very different intensities. For the bright molecule this peak is much more intense. Ziroff et al.\ suggested that a corresponding peak observed in photoelectron spectroscopy at the Fermi energy is a manifestation of the Kondo effect with a Kondo temperature of $T_K  \simeq 100$\,K ~\cite{ziroff2012low}, although its temperature evolution did not resemble the characteristic temperature dependence of a Kondo resonance. For the purpose of this paper we have to establish beyond doubt that the zero-bias peak for both molecules is indeed a Kondo resonance. To this end, we use a three-pronged approach, comprising measurements of the zero-bias peak as a function of temperature, magnetic field and hybridization, in each case looking for the dependence that is indicative of the Kondo effect.  

\begin{figure}
	\centering
		\includegraphics[width=85mm]{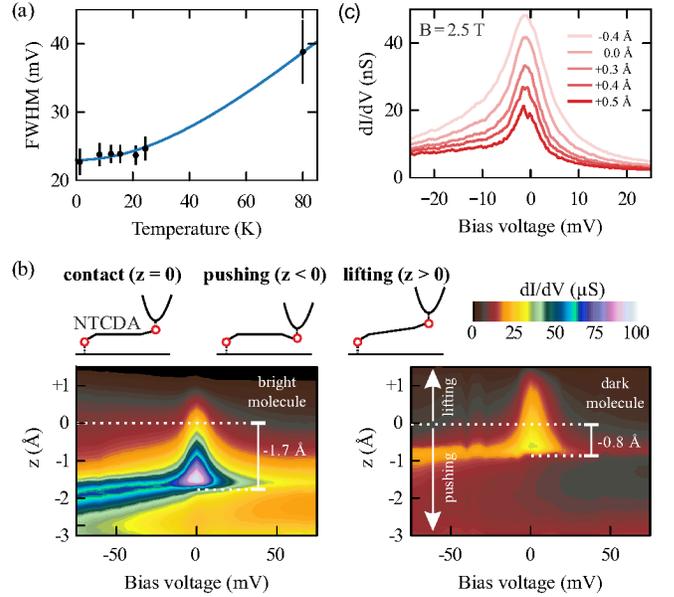}
		\caption{(a) Temperature evolution of the FWHM of the zero-bias peak of the bright molecule. (b) $dI/dV(V,z)$ maps for the bright (left) and dark (right) molecules, recorded after the formation of the tip-molecule bond.  The meaning of the $z$ coordinate is shown schematically in the illustrations. (c) $dI/dV$ spectra at various $z$ in a magnetic field of $B = 2.5$\,T, measured at $T = 4.3$\,K.}
	\label{fig:fig2}
\end{figure}

 We first analyze the temperature-dependence of the zero-bias peak for the bright molecule. In Fig.~\ref{fig:fig2}(a) its full width at half maximum (FWHM) is displayed. The FWHM was extracted by fitting with a Fano line shape \cite{FanoResonance1961, SchillerHershfield2000a}. Broadening effects due to temperature $T$ and modulation amplitude $V_\mathrm{mod}$ have been taken into account by subtracting them from the measured FWHM, using $\text{FWHM}= \sqrt{\text{FWHM}_\text{measured}^2-(1.7V_\mathrm{mod})^2 - (3.5k_{\mathrm{B}}T)^2}$
 \cite{Kroger2005}. 
 The such-determined intrinsic FWHM exhibits the expected temperature dependence of a Kondo resonance. Fitting the expression $\sqrt{(\alpha k_B T)^{2} + (2 k_{\mathrm{B}} T_K)^2}$ to the FWHM 
 \cite{Nagaoka2002}
 we find a Kondo temperature of $T_K^\mathrm{bright} = 133$\,K and $\alpha = 4.53$. It should be noted that this is only a rough estimate, because the FWHM is related to $T_K$ by a non-universal scaling constant \cite{AU-PTCDA-monomer}. A more accurate analysis in of the Kondo temperature will be presented in section \ref{sec:application-to-NTCDA}. 

Because of its low intensity and broad FWHM, the temperature dependence of the zero-bias peak of the dark molecule is difficult to study. A broad Kondo peak indicates that the system is in the weakly correlated regime, with a small  ratio $U/\Gamma$, where $U$ is the intra-orbital Coulomb repulsion (Eq.\ \eqref{eq:H-e}) and $\Gamma$ is an energy-averaged hybridization parameter (related to Eq.\ \eqref{eqn:HybFunc}), and a large Kondo temperature $T_{\rm K}$. To prove that the zero-bias peak of the dark molecule is also a Kondo resonance, we therefore apply a different strategy: Instead of decreasing the temperature to change its line shape, we decrease $\Gamma$, thus tuning the system further into the strong-coupling regime, in which the Kondo peak is sharper and more easy to pin down. The tuning of the hybridization is achieved by forming a contact (at $z \equiv 0$\,\AA, where $z$ is the vertical tip coordinate) between the tip apex and one of the corner oxygen atoms of NTCDA. The corresponding part of the molecule can then either be pushed towards the surface ($z < 0$\,\AA) or lifted up ($z > 0$\,\AA) \cite{ptcda_kondo_cleavage, gated_molecular_wire,PTCDAAgMove, gated_wire_spectral}. Fig.~\ref{fig:fig2}(c) displays $dI/dV$ spectra recorded at different $z$ for both molecules, plotted as color maps. Both maps exhibit very similar behavior, albeit shifted with respect to each other by $\Delta z=0.9$\,\AA\  along the vertical axis. Based on the similarity of the maps, and the fact that for the bright molecule we have already shown, employing the temperature-dependence, that the zero-bias peak is a Kondo peak, we can conclude that the same is also true for the dark molecule. In fact, both maps in Fig.~\ref{fig:fig2}(c) exhibit the expected dependence of a spin-$\frac{1}{2}$ Kondo effect, as the comparison with the well-studied case of lifting PTCDA molecules from Ag(111) shows \cite{ptcda_kondo_cleavage, gated_molecular_wire, PTCDAAgMove, gated_wire_spectral}.

\begin{figure}
	\centering
		\includegraphics[width=0.45\textwidth]{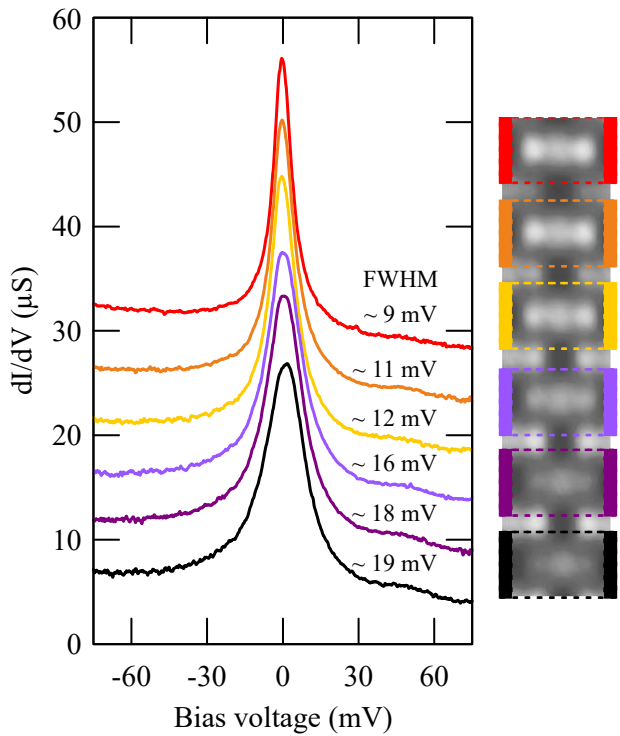}
		\caption{$dI/dV$ spectra of the different NTCDA molecules in the rippled phase, recorded directly after the formation of the bond of the tip to the molecules as indicated in the inset. The data was acquired at $T=4.3$\,K. Spectra are vertically offset by $5$\,$\mu$S for clarity.}
	\label{fig:fig7}
\end{figure}

\begin{figure}
	\centering
		\includegraphics[width=85mm]{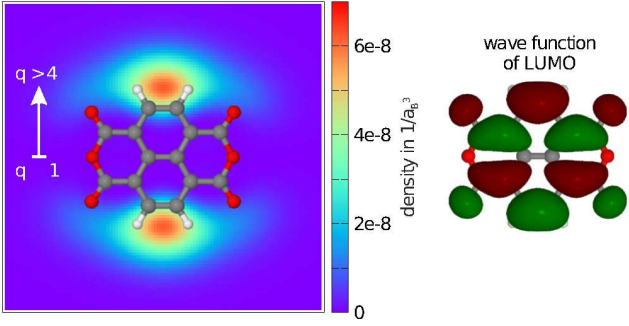}
		\caption{Local density of states (LDOS) of the LUMO of NTCDA calculated $4$\,\AA~above the gas-phase molecule (left panel). A graphical representation of the gas-phase NTCDA molecule has been overlaid for clarity. The right panel shows the top view of the LUMO of the gas-phase NTCDA molecule. The different colors indicate the positive ($\Psi (r) > 0$) and negative ($\Psi (r) < 0$) contributions of the wave function.}
	\label{fig:exp_fig3}
\end{figure}

The maps clearly show the sharpening of the Kondo resonance that is expected if the hybridization is reduced and the Kondo effect is tuned from the weak- towards the strong-coupling regime. Note that in addition to reducing $\Gamma$, lifting the molecule may reduce the charge transfer to the molecule and also lead to a smaller dielectric screening due to the larger molecule-surface distance, both resulting in a increased Coulomb interaction $U$, thus further increasing $U/\Gamma$ with increasing $z$. We note furthermore that the FWHM of the zero-bias peak decreases by a factor of $\approx 2.4$ for the bright and dark molecules when the molecule is contacted by the tip (from $22$\,mV for the non-contacted bright molecule in Fig.~\ref{fig:exp_fig2} to $9$\,mV for the contacted molecule at $z=0$, and from $45$\,mV to $19$\,mV for the dark molecule, see Fig.~\ref{fig:fig7}). This reduction of the FWHM can be explained by the partial dehybridization that occurs when the oxygen atom jumps into contact with the tip and lifts the surrounding parts of the molecule from the surface. It is well-known that the additional contact to the tip is electronically weak and does not lead to an appreciable hybridization with the LUMO \cite{Temirov2018,Esat2018}. Since the FWHM of the zero-bias peak decreases by approximately the same factor for the bright and dark molecules when the molecule is contacted by the tip, we can conclude that the initial $T_K^\mathrm{dark}$ of the dark molecule, without the contact to the tip, must also be larger than $T_K^\mathrm{bright}$. This explains the broader Kondo peak of the dark molecule in Fig.~\ref{fig:exp_fig2}.

In the strong-coupling regime moderate magnetic fields may split the Kondo resonance \cite{Zhang2013}. Applying a $B$ field of $2.5$\,T at an experimental temperature of $4.3$\,K, we indeed observe an incipient splitting of the Kondo resonance for the bright molecule at $z=0.5$\,\AA. Assuming a Land\'e factor of $g = 2$, the Zeeman energy at $2.5$\,T is $g \mu_B B \approx 0.29$\,mV, slightly smaller than the thermal fluctuations $k_B T \approx 0.37$\,mV at $T = 4.3$\,K (at $z=0.5$\,\AA, $T_K$ has dropped so far that $g \mu_B B\gg k_B T_K$). Therefore, the split is not well developed, but nevertheless clearly visible. Note that for the non-contacted bright molecule a $B_c \approx k_B T_K / (g \mu_B) \approx 49.5$\,T would be necessary to split the Kondo resonance, which is clearly out of reach.   

Another notable observation in Fig.~\ref{fig:exp_fig2} is the fact that in the center of the molecule step-like structures at zero bias are observed instead of a Lorentzian peak. The  spectra of the bright and dark molecules are almost identical and merely differ in the intensity of the step-like feature. Such features result from the quantum interference between two or more different tunneling paths \cite{FanoResonance1961, SchillerHershfield2000a}. This interference leads to a zero-bias feature with a so-called Fano line shape. In the simplest case of two interfering channels, the differential conductance is approximated by  
\begin{equation}
\label{eq:Fano}
\frac{dI}{dV} (V) \propto \rho_0 + \frac{(q+\epsilon)^2}{1+\epsilon^2},
\end{equation}
with  
\begin{equation}
\epsilon = \frac{eV-E_K}{(\delta /2)}
\end{equation}
and
\begin{equation} \label{NTCDA_q}
q = \frac{t_2}{\pi \rho_0 \Gamma t_1}.
\end{equation}
Here, $E_K$ describes the intrinsic position of the Kondo resonance, $\delta$ its FWHM, $\Gamma$ the hybridization between the local orbital and the substrate, $t_1$ and $t_2$ the tunneling probabilities from the tip directly into the substrate and into the local orbital, respectively, and $\rho_0$ the density of states. $q$ determines the line shape of the Kondo resonance. The blue lines in Fig.~\ref{fig:exp_fig2} display fits of Eq.\ \eqref{eq:Fano} to the experimental tunneling spectra. The fit parameters $q$ and $\delta$, averaged over fits for 10 data sets including the one shown in Fig.~\ref{fig:exp_fig2}, are summarized in Tab.~\ref{tab:fwhm_and_q}. As expected, $\delta$ is on average larger for the dark molecule than for the bright one (see above). However, it is noteworthy that the $\delta$ that is extracted from the spectra recorded at the CH edge is approximately the same as the one for the center of the molecule. This confirms that both the Lorentzian peak and the step-like feature indicate the same energy scale -- we can thus conclude that the step is also a manifestation of the Kondo effect that leads to the peak recorded at the CH edge. Table~\ref{tab:fwhm_and_q} also reveals that $q$ is significantly smaller in the center of the molecule, indicating that there the probability to tunnel directly from the tip into the substrate ($t_1$) is larger than at the CH edge. The reason for the larger tunneling probability directly into the substate when the tip is located in the center of the molecule is a direct consequence of the spatial distribution of the LUMO wave function, which has a node in the center of the molecule and a pronounced lobe at the CH edge (Fig.~\ref{fig:exp_fig3}b). This is reflected in Fig.~\ref{fig:exp_fig3}a, which displays the LDOS of the LUMO $4$\,\AA\, above the gas-phase NTCDA molecule. 

We note that the fits displayed in Fig.~\ref{fig:exp_fig2} and the derived parameters in Table~\ref{tab:fwhm_and_q} are merely heuristic and should only be used to ascertain that there are at least two tunneling paths present, and that the center of the molecule is more transparent to the tunneling current than the CH edge. More elaborate fits to the spectra, based on a more solid theoretical foundation, will be presented in section section \ref{sec:application-to-NTCDA}.
\begin{table}[htbp]
	\centering
		\begin{tabular}{|c|c|c|c|}
		\hline
			 molecule & location & $\delta$ & $q$ \\ \hline \hline
  \multirow{2}{*}{bright} & CH edge & $ (28.5\pm 2.3)$\,mV & $ 15.5\pm 6.9$ \\
	& center & $ (29.4\pm 6.8)$\,mV & $ 1.2\pm 0.2$ \\ \hline
	\multirow{2}{*}{dark} & CH edge & $ (52.1\pm 4.7)$\,mV & $ 21.4\pm 9.7$ \\
	& center & $ (48.5\pm 8.2)$\,mV & $ 0.9\pm 0.4$ \\
	\hline
		\end{tabular}
	\caption{$\delta$ and $q$ extracted from the fits of the Fano line shape (Eq.\ \eqref{eq:Fano}) to the $dI/dV(V)$ spectra of the bright and dark molecules. The values are averages over the fit parameters for ten different data sets. One data set is shown in Fig.\ \ref{fig:exp_fig2}.  }
	\label{tab:fwhm_and_q}
\end{table}

We conclude that there is overwhelming experimental evidence (from $T$-dependent data, junction stretching, magnetic field data and quantum interference) that both the dark and the bright molecule of the relaxed NTCDA/Ag(111) phase exhibit the Kondo effect. To explain the behavior of the present system quantitatively, it therefore appears natural to apply the theory that has been very successful for PTCDA/Ag(111) and AuPTCDA/Au(111) \cite{PTCDAAgMove,gated_wire_spectral,AU-PTCDA-monomer,AU-PTCDA-dimer}. This is done in section \ref{sec:application-to-NTCDA}.

\subsection{Vibrational features}
\label{sec:experiment-NTCDA vibrational side bands}

\begin{figure}
	\centering
		 \includegraphics[width=85mm]{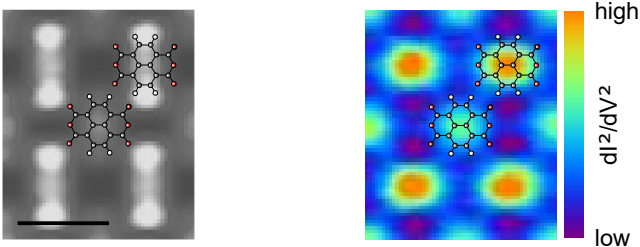}
		\caption{The panel on the left shows a STM topography image measured at constant height above the molecules of the rippled phase ($V=47$\,mV). The panel on the right shows the corresponding $d^2I/dV^2$ image. The length of the scale bar is 10 \AA.}
	\label{fig:fig10}
\end{figure}

\begin{figure}
	\centering
		\includegraphics[width=85mm]{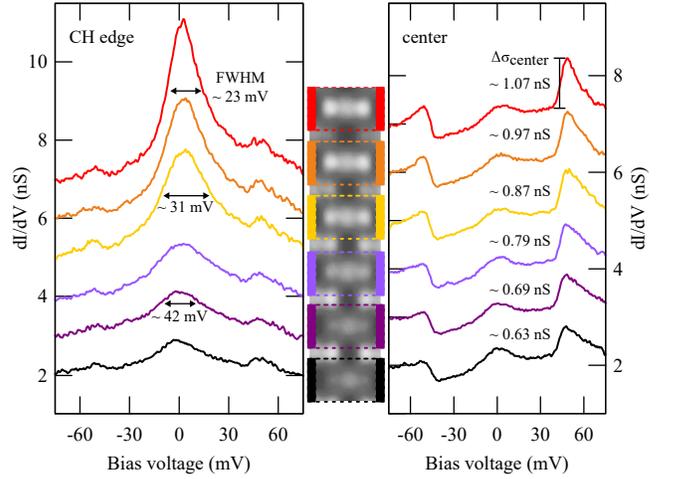}
		\caption{
		(left) $dI/dV$ spectra of the different NTCDA molecules in the rippled phase at $T=4.3$\,K. The spectra have been recorded at the CH edge of each molecule. Different colors correspond to the different molecules as indicated by the colored frames around the molecules. The color coding is the same as in Fig.\ \ref{fig:fig7}. (right) $dI/dV$ spectra of the different NTCDA molecules in the rippled phase at $T=4.3$\,K, measured at the center of each molecule. Colors as in panel b. Spectra are vertically offset by $1$\, nS for clarity.}
	\label{fig:fig9}
\end{figure}

%
%
%

In addition to the zero-bias features, the spectra in Fig.~\ref{fig:exp_fig2} also show features at finite bias voltages. Most notable are peaks at approximately $+(47.0 \pm 0.3)$\,mV and $-(51.5 \pm 0.3)$\,mV in the spectra recorded at the CH edges of both the bright and the dark molecules. However, we also observe weak features around $\pm 30$\,mV. The (nearly) symmetric location of in particular the stronger features around zero bias (up to a shift of $2.25$\,mV towards negative energies) is suggestive of inelastic excitations, either during the tunneling process or within the sample system. Such excitations can occur as a result of, e.g., vibrational or magnetic degrees of freedom.  We have not observed any change or shift of the side peaks in magnetic fields up to $3$\,T. It is therefore unlikely that the features are of magnetic origin and we conclude that they must be linked to vibrations. NTCDA indeed has a number of vibrational modes in the relevant frequency range \cite{C5CP06619K}. Some of them are listed in Table \ref{tab:vibrations}.  

\begin{table}[t]
	\centering
		\begin{tabular}{|c|c|c|}
		\hline
			 no. & symmetry & $\hbar\omega$   \\ \hline \hline
  1 & B$_{\rm 3g}$ & $ 41.6$\,meV ($335.35$\,cm$^{-1}$) \\
  2 & B$_{\rm 1u}$ & $ 46.2$\,meV ($372.41$\,cm$^{-1}$)\\
  3 & B$_{\rm 3g}$ & $ 50.4$\,meV ($406.62$\,cm$^{-1}$)\\
  4 & A$_{\rm g}$ & $ 50.7$\,meV ($408.96$\,cm$^{-1}$)\\
  5 & B$_{\rm 1g}$ & $ 52.8$\,meV ($525.53$\,cm$^{-1}$)\\ 
  \hline
		\end{tabular}
	\caption{Vibrational modes of gas-phase NTCDA in the energy range $40$ to $50$\,meV, calculated by DFT (taken from reference \cite{C5CP06619K}). Gas-phase NTCDA has the symmetry group D$_{\rm 2h}$.}
	\label{tab:vibrations}
\end{table}

When recorded in the center of the molecule, the features at $+(47.0 \pm 0.3)$\,mV and $-(51.5 \pm 0.3)$\,mV become much stronger. In Fig.~\ref{fig:fig10}, the spatial distribution of the vibrational features is displayed, recorded as a $d^2I/dV^2$ image at $\simeq 50$\,mV, in comparison with a  constant-height topographic image. One observes an image without nodal planes and a clear concentration of the intensity close to the center of the molecule. This is true for both molecules, although the bright molecule has a larger maximum intensity, which is consistent with Fig.~\ref{fig:exp_fig2}. Moreover, for the spectra measured in the center of the molecules there is a clear difference in the line shape between the bright and dark molecules. In both cases they are asymmetric with a steep rise at the low-bias side, but for the dark molecule the drop on the high-bias side is more moderate than for the bright molecule, giving the vibrational feature a more step-like appearance for the dark molecule, in contrast to a "half-peak" for the bright molecule.

\begin{figure}
	\centering
		\includegraphics[width=85mm]{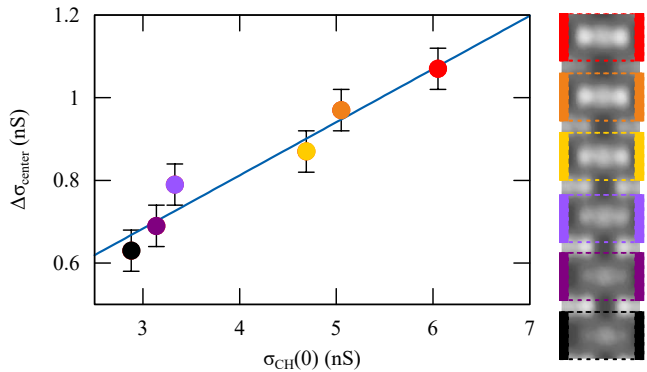}
		\caption{Plot of the step height $\Delta\sigma_{\rm center}$ measured at the center of the molecules in the rippled phase  as a function of the height $\sigma_{\rm CH}(0)$ of the Kondo resonance at the CH edge. The values are obtained from the spectra displayed in Fig.\ \ref{fig:fig9}. Color coding as in Fig.\ \ref{fig:fig9}. Data points have been fitted by a linear function of the form $\Delta\sigma_{\rm center}=c \times \sigma_{\rm CH}(0) +b$, fit parameters are $c = 0.13$, $b = 0.30$\,nS.}
	\label{fig:fig11}
\end{figure}

The dependence of line shape of the vibrational features at $+(47.0 \pm 0.3)$\,mV and $-(51.5 \pm 0.3)$\,mV on the shape of the spectrum at zero bias is also apparent in Fig.~\ref{fig:fig9}, which displays the evolution of the Kondo peak and the vibrational side bands in the transition from the dark to the bright molecule in the rippled phase of Fig.~\ref{fig:fig4}. As the intensity of the Kondo peak (measured at the CH edge) increases, the vibrational features measured in the center of the molecule at $+(47.0 \pm 0.3)$\,mV and $-(51.5 \pm 0.3)$\,mV turn from a step for the dark molecule into an asymmetric peak for the bright molecule, in agreement with Fig.~\ref{fig:exp_fig2}. For the feature at  $+(47.0 \pm 0.3)$\,mV  this behavior is also illustrated by Fig.~\ref{fig:fig11}, in which $\Delta\sigma_{\rm center}$ is plotted versus $\sigma_{\rm CH}(0)$ and a linear correlation between the peak height of the vibrational feature and the Kondo peak is observed \cite{Esat2017}.

\section{NRG results}
\label{sec:NRG-results}

\subsection{Application of the tunneling theory to NTCDA}
\label{sec:application-to-NTCDA}

\subsubsection{General approach}

In this section we specify step by step the theoretical framework needed to reproduce and explain the experimentally measured differential conductance $dI/dV$ spectra in Fig.\ \ref{fig:exp_fig2}, using Eqs.\ \eqref{eqn:elasticCurrent},  \eqref{eqn:inelasticCurrent} and \eqref{eq:second-order-inelastic-contribution} as a basis. We use an approach \cite{PTCDAAgMove,gated_wire_spectral,AU-PTCDA-monomer,AU-PTCDA-dimer} in which we map the results of density functional theory (DFT) calculations, combined with many body perturbation theory (MBPT) to include quasi-particle corrections, onto the Hamiltonian $\hat H_\text{SIAM}$ of a single-orbital Anderson model (SIAM), here also including Holstein terms, which is then solved by NRG calculations. In particular, the NRG is used \cite{Wilson75,BullaCostiPruschke2008} to exactly calculate all spectral functions that are required to calculate the transmission functions that enter Eqs.\ \eqref{eqn:elasticCurrent},  \eqref{eqn:inelasticCurrent} and \eqref{eq:second-order-inelastic-contribution}. As pointed out in Ref.~\cite{AU-PTCDA-monomer}, employing a fully energy-dependent hybridization function $\Gamma(\omega) = \Im \Delta(\w-i0^+)=\pi \Sigma |V_k|^2 \delta(\omega-\epsilon_k )$ in the NRG is crucial for an accurate description. Moreover, we do not impose particle-hole symmetry.  

Our theoretical framework is the same as discussed in 
Ref. \cite{PTCDAAgMove,AU-PTCDA-monomer,AU-PTCDA-dimer}.
Structural optimization is performed within density-functional theory (DFT), 
using the SIESTA package \footnote{We are using version 3.2 of the SIESTA which is available at 
http://departments.icmab.es/leem/siesta.
} \cite{Siesta1,Siesta2},
using ab-initio pseudopotentials and a double-zeta plus polarization basis (DZP).
We employ the PBE exchange-correlation functional \cite{PBE}.
Since the van der Waals interaction is crucial for weakly bound systems like 
organic molecules on metal surfaces, we include it in the formulation of 
Ruiz et al. \cite{vdWsurfPTCDA} (vdW$^\text{surf}$) for all structure optimizations.

In addition to the structural data, the electronic mean-field spectrum of the 
adsorbed molecule (in particular its LUMO state) is required as input for the NRG.
This cannot be calculated on the level of DFT, since DFT suffers from 
problems as long as electronic spectra are concerned 
\cite{PTCDAAgMove,AU-PTCDA-monomer,AU-PTCDA-dimer}.
Instead, many-body perturbation theory (MBPT) provides a systematic approach
to spectral features (except for the dynamical correlation to be treated by the NRG.)
Here we employ the same approach as discussed in 
Ref. \cite{PTCDAAgMove,AU-PTCDA-monomer,AU-PTCDA-dimer}.
Starting from a DFT-LDA calculation for a given geometry, we carry out a MBPT
calculation within our LDA+$GdW$ approach.
This yields realistic band-structure energies for all states of the adsorbate
system, fully including all screening and broadening effects resulting from
the metallic surface.
After projecting on the LUMO state of the bare molecule, we arrive at a projected
density of states (PDOS) as shown in Fig. \ref{NRG_fig4aa}.
From this PDOS one can deduce the level position $\e_{d\sigma}$ of the LUMO state 
when adsorbed on the surface, as well as  its hybridization function $\Delta(z)$
[see Eq. (\ref{eqn:MFGF})].
In addition, the internal Coulomb interaction $U$ of the LUMO state is also
obtained from MBPT \cite{PTCDAAgMove}.

\subsubsection{Input from ab-initio calculations and model without electron-phonon coupling}
\label{sec:Input from ab-initio calculations and model without electron-phonon coupling}
\begin{figure}[t]
\begin{center}
\includegraphics[width=0.5\textwidth]{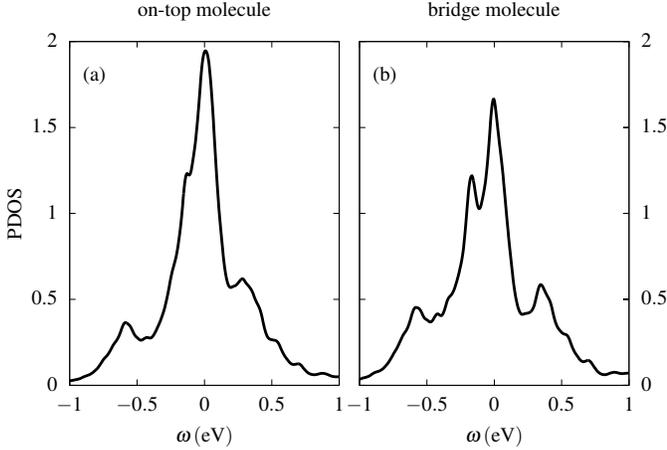}
\caption{PDOS of the LUMO of NTCDA/Ag(111) as calculated by a combination of DFT and MBPT. The latter includes quasiparticle corrections. (a) on-top molecule, (b) bridge molecule.
}
\label{NRG_fig4aa}
\end{center}
\end{figure}

As suggested by experimental data (see Fig.\ \ref{fig:exp_fig1} and discussion in section \ref{sec:experiment-NTCDA structure}), the molecules in the calculations are placed at on-top and bridge sites on the Ag(111) surface. According to the calculation, both molecules chemisorb stably at these sites. The \textit{ab initio} calculation predicts well-separated NTCDA molecular orbitals, an energy broadening of the orbitals due to their hybridization with the substrate, a partial occupation of the lowest unoccupied molecular orbital (LUMO) due to charge transfer from the substrate, and a substantial intramolecular Coulomb interaction $U=1.25$\,eV for the LUMO. The specific local environments of the two adsorption sites lead to slightly different positions of the on-top and bridge LUMOs, as can be seen in the projected densities of state in Fig.~\ref{NRG_fig4aa}, with the weight of the PDOS spectrum of the bridge molecule appearing further to the left. Similarly, the value of the hybridization functions at the chemical potential differ slightly for the two molecules, being $\Gamma^\mathrm{bridge}(0) = 190$\,meV and $\Gamma^\mathrm{top}(0) = 165$\,meV.

Because the Coulomb interaction is substantial compared to the hybridization strength, it enforces a single occupation of the LUMOs on both molecules, leading to a free spin of the radical that is ultimately screened by the Kondo effect for $T\to 0$. Since the \textit{ab initio} PDOS spectra only contain the Coulomb interaction on a mean-field level, the half-filled orbitals are spin-degenerate in a paramagnetic calculation and the effective mean-field orbital energy must be pinned close to the Fermi energy, as is indeed apparent in Fig.\ \ref{NRG_fig4aa}.

Both \textit{ab initio} spectra in Fig.\ \ref{NRG_fig4aa} are much too wide compared to the STS data which exhibits a peak width of the order of $30$\,meV in the differential conductance at zero bias. This indicates that many-body correlations play an important role and must be taken into account for matching theory with experiment. Interpreting the \textit{ab initio} PDOS as a mean-field solution \cite{PTCDAAgMove,gated_wire_spectral,AU-PTCDA-monomer,AU-PTCDA-dimer} allows us to extract the single particle energies $\epsilon_{d\sigma}^\text{bridge}=-0.77\,\text{eV}$ and $\epsilon_{d\sigma}^\text{top}=-0.67\,\text{eV}$ for both types of molecules as well as the full complex hybridization function $\Delta(z)$.

\begin{figure}[t]
\begin{center}
\includegraphics[width=0.5\textwidth]{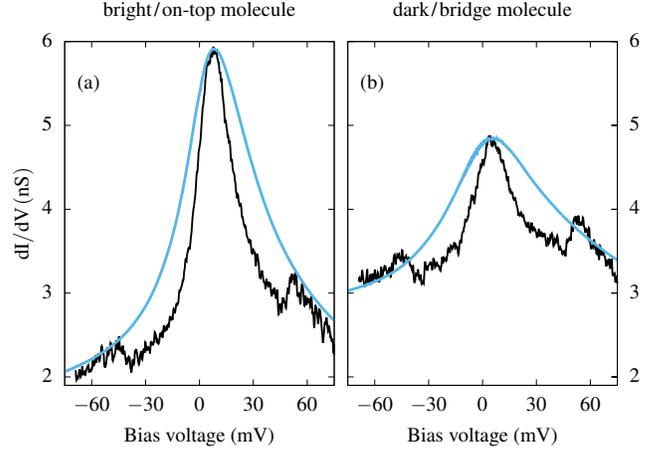}
\caption{Comparison between the experimental $dI/dV(V)$ spectra recorded on NTCDA/Ag(111) (black lines) and the results of NRG calculations for the SIAM with the PDOS of Fig.\ \ref{NRG_fig4aa} as input (blue lines). The NRG spectra have been adjusted with a constant offset ($\rho_\text{offset}=0.2$\,nS in panel (a), $\rho_\text{offset}=1.1$\,nS in panel (b) to account for an experimental background signal such that the zero-bias peak heights and the high-frequency tails are matched to the experimental $dI/dV$ curve. 
}
\label{NRG_fig4ab}
\end{center}
\end{figure}

Next, we use these sets of \textit{ab initio} parameters and functions as an input for NRG calculations to solve the SIAM for both the bridge and the on-top molecules in the absence of any electron-phonon coupling. For simplicity, we also set the tunneling matrix element $t_{c_{0\sigma}}$ from the tip to the local effective substrate orbital (defined in Eq.\ \eqref{eqn:c_0}) to zero and only include $\rho_{d_{0\sigma},d^\dagger_{0\sigma}}(\w)$ in the calculation of the theoretical $dI/dV$ curve (Eqs.\ \eqref{eqn:el-current} and \eqref{eqn:tauzero}), where $d_{0\sigma}$ is the annihilation operator for an electron in the LUMO. The result for the on-top and bridge molecules are displayed in Fig.\ \ref{NRG_fig4ab}, revealing a substantial narrowing of the zero-bias resonance relative to the \textit{ab initio} PDOS.

Thus, in the calculations as well as in experiments, both molecules exhibit a Kondo resonance. 
Driven by the differences in $\Gamma(0)$  the two molecules in the NTCDA/Ag(111) unit cell have different Kondo temperatures: The Kondo temperature of the bridge molecule is larger than that of  the on-top molecule. If we associate the larger measured to the larger calculated Kondo temperature, we can identify the dark molecule with the bridge site and the bright molecule with the on-top site. This identification is consistent with expectations based on structural arguments, namely that the  bridge molecule adsorbs slightly closer to the surface and exhibits a larger corrugation. This explains the larger $\Gamma$, the smaller coupling parameter $U/\Gamma$ and thus also the larger $T_K$. However, while the experimental trend $T_K^\mathrm{dark} > T_K^\mathrm{bright}$ is thus correctly predicted, the actual widths of the zero-bias anomalies, which is a measure  of  Kondo temperature, are too large compared to experiment, as indicated by the much narrower experimental curves in Fig.\ \ref{NRG_fig4ab}. 

\subsubsection{Identification of the problem. The unconventional Holstein model as a possible solution }
\label{sec:problem-and-strategy}

The SIAM  without electron-phonon coupling as discussed in the previous section has three shortcomings: (i) the Kondo temperature $T_{\rm K}$, determined from  the full-width half maximum (FWHM) of the calculated spectra, is too large, (ii) the calculated spectra naturally lack the additional vibrational features that are observed in experiment, and (iii) the exclusive coupling of the STM tip to the LUMO orbital NTCDA cannot explain the marked differences between the $dI/dV$ spectra measured in the center and at the CH-edge of the molecules.

While it is  clear how shortcomings (ii) and (iii) can be addressed, namely by reverting to the general tunneling theory of section \ref{sec:tunnel-current} that includes both inelastic tunneling and tunneling interference, we still need to identify a mechanism that is able to reduce the Kondo temperature below what is expected from standard theory without electron-phonon interaction as sketched out in the previous section.

In section \ref{sec:experiment-NTCDA vibrational side bands}, we have reported the observation of two sets of vibrational features in our STS spectra. At the same time, in our theoretical analysis of sections \ref{sec:tunnel-current-general} and \ref{sec:Modeling the system} we have identified three distinct mechanisms in which vibrations may influence the differential conductance spectra measured in STS: a vibration-induced change of the tunnel coupling of the STM tip to an orbital of the system S, as well as an electron-phonon coupling purely within the system S, the latter either of conventional of unconventional Holstein type. The coincidence of observing in experiment both a reduced Kondo temperature and vibrational features that possibly result from an electron-phonon coupling in the system S suggests that the two observations might be connected. 

We therefore briefly explore the possibility that the electron-phonon coupling within the system S influences, and in particular reduces, its Kondo temperature. A finite electron-phonon coupling $|\lambda_d|>0$ (see Eq. \ref{eqn:AndersonHolstein}) as in the conventional Holstein model \cite{HewsonMeyer02,GalperinRatnerNitzan2007} generates a reduction of $U\to U^d_{\rm eff}=U-\lambda^2_d/\w_0$  \cite{Mahan81,GalperinRatnerNitzan2007,EidelsteinSchiller2013,JovchevAnders2013} and would thus lead to an enhancement of the Kondo temperature. Only if $U^d_{\rm eff}<0$ the width of zero-frequency peak is rapidly reduced and the charge Kondo regime is entered \cite{HewsonMeyer02} -- see the discussion in the literature \cite{HewsonMeyer02} or in Sec.\ \ref{sec:anti-adiabatic-regime-spectrum}. However, for the case of NTCDA/Ag(111) the DFT calculation excludes a vibrational coupling of any local molecular vibrational mode to the molecular orbital, i.\ e.\ demands $\lambda_d\approx 0$ and, therefore, rules out the conventional Holstein model, both as a source of the sharpening of the Kondo peak and as a source of the vibrational features in the STS spectrum. Finally, a coupling of the phonon displacement to the hybridization $V_{\k}$ is known to lead to an enhancement of the effective hybridization \cite{Cornaglia2005},
as well as reduce the local Coulomb interaction \cite{Grewe84}
and thus an increase of the Kondo temperature, which can also be analytically derived by a employing a Lang-Firsov transformation \cite{LangFirsov1962}.
 
In contrast, a possibility to reduce the Kondo temperature is provided by an unconventional Holstein term which linearly connects the charging energy of the effective local substrate orbital $c_{0\sigma}$ to one of the molecular vibrational modes. In section \ref{sec:modelling_the_system_vibrations} above, we have included such a term, parameterized by the coupling constant $\lambda_c$, in Eq.\ \eqref{eqn:AndersonHolstein}. The physical idea behind this term is that a molecular vibration perpendicular to the substrate can induce a local potential change that shifts the single particle energy of local substrate orbitals as function of the displacement. Such an unusual Holstein coupling has been investigated in the context of the periodic Anderson model to provide a microscopic mechanism for the Kondo volume collapse \cite{PAM-E-PHON2013}, which is believed to be the origin of  structural $\ensuremath{\gamma}\ensuremath{\rightarrow}\ensuremath{\alpha}$ phase transition in Cerium \cite{KondoVolumKollapse1982}. It offers the possibility to reduce the width of the equilibrium Kondo resonance in a straightforward manner: In analogy to the discussion before, the Holstein coupling induces an attractive (negative) contribution to the local Coulomb interaction $U^c$ which is initially zero in an uncorrelated free conduction band. Consequently, the singly occupied spin-degenerate states are energetically separated from the lower-lying empty and doubly occupied states of $c_{0\sigma}$. This energy separation reduced the charge fluctuations with the LUMO. For a large $\lambda_c$, this reduces the hybridization $\Gamma \to \Gamma_{\rm eff}$ between the local orbital $d$ and the substrate $c_{0\sigma}$, thus providing a mechanism for the reduction of the Kondo temperature by bipolaron formation \cite{PAM-E-PHON2013}.

Before applying this model to the NTCDA/Ag(111) system, we investigate it in detail in the next two sections, first regarding its influence on the Kondo temperature and second regarding the spectral functions. To keep the analysis simple, we employ the assumption of particle-hole symmetry in the next two sections. 

\subsubsection{Analysis of the Holstein coupling $\lambda_c$: The influence on the Kondo temperature}
\label{sec:The influence on the Kondo temperature}

To set the stage for the realistic description the NTCDA/Ag(111) system, we investigate the influence of the unusual Holstein coupling $\lambda_c$ between a vibrational mode of the molecule and a local substrate orbital on the particle-hole symmetric single impurity Anderson Hamiltonian Eq.\ \eqref{eqn:SIAM} and its Kondo temperature. For clarity, we only include in Eq.~\eqref{eqn:AndersonHolstein} a featureless conduction band with a constant density of states $\rho_0=1/2D$, a spin-degenerate molecular orbital $d_{0\sigma}$ with single-particle energy $\epsilon_d=-U/2$ and a single vibrational mode $\omega_0$. The imaginary part of the hybridization function Eq.\ \eqref{eqn:HybFunc}, $\Gamma_0=\Im(\Delta(\w-i0^+))=\pi V^2\rho_0$, is a constant over the whole band width $\pm D$ and serves as the natural unit for all model parameters. We also set all $\lambda^{\rm tip}_{\mu\nu}=0$ and hence only include the elastic contributions to the tunnel current. By fitting the differential conductance that is calculated by the NRG to the empirical formula \cite{GoldhaberGordon} introduced by Goldhaber-Gordon et al.\ \cite{GoldhaberGordon}
\begin{align}
 \frac{d I}{d V}(V=0)=\frac{G_0}{\left[1+(2^{1/s}-1)(\frac{T}{T_{K}})^2\right]^s} ,
\label{eqn:Goldhaber}
\end{align}
with $s=0.22$ for a spin-$\frac{1}{2}$ system, we obtain the Kondo temperature $T_{K}$ as function of $\lambda_c$.

\begin{figure}[t]
\begin{center}
\includegraphics[width=0.45\textwidth]{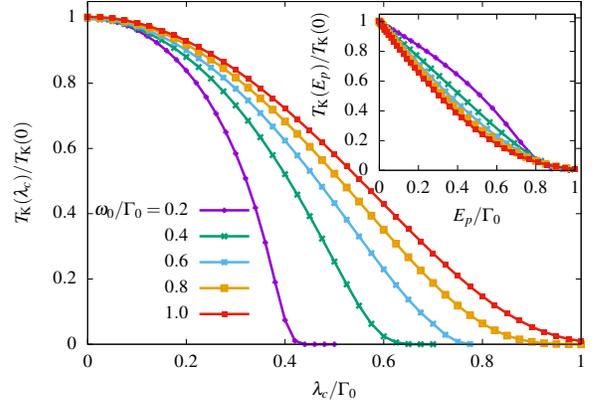}

\caption{Ratio $T_{K}(\lambda_c)/T_{K}(\lambda_c=0)$ calculated for the symmetric single impurity Anderson model plus unusual Holstein coupling $\lambda_c$ for a single vibrational mode $\omega_0$. The main panel shows the Kondo temperature as function of the coupling $\lambda_c$ for $U/\Gamma_0=10$ and different values for the vibrational frequency $\omega_0$, normalized at $\lambda_c=0$. The inset depicts the same quantity but plotted against the polaronic energy shift $E_p=\lambda_c^2/\omega_0$.
}
\label{fig:TK-Lc-a}

\end{center}
\end{figure}

In Fig.~\ref{fig:TK-Lc-a} we plot the ratio $T_{K}(\lambda_c)/T_{K}(\lambda_c=0)$ as function of the Holstein coupling $\lambda_c$ for different phonon frequencies $\w_0$. The additional coupling of the vibrational mode to the surface orbital indeed leads to a reduction of the Kondo temperature with increasing $\lambda_c$. For a fixed coupling strength $\lambda_c$ the reduction of $T_{\rm K}$  decreases with increasing  $\omega_0$. For a better understanding of the underlying mechanism, the inset shows the same data, but plotted as a function of the polaronic energy shift $E_{\rm p}= \lambda_c^2/\omega_0$. For phonon frequencies $\w_0 \simeq \Gamma_0$ and $\w_0>\Gamma_0$, the decrease of $T_{K}$ only depends on the polaron energy $E_{p}$. In this high-frequency or anti-adiabatic limit, the phonons can be integrated out and their main effect is to generate a negative effective $U$ in the local substrate orbital, $U^c_{\rm eff} \approx -2E_p$. For smaller frequencies, retardation effects play a role and we observe increasing deviations in the crossover regime to the adiabatic limit.

We have argued above in section \ref{sec:problem-and-strategy} that the attractive $U^c_{\rm eff}$ acting on the local substrate electrons primarily suppresses the hybridization between the molecule and the substrate to an effective value $\Gamma_{\rm eff}$, which as a consequence reduces the Kondo temperature due to the increasing correlation measured by the ratio $U/\Gamma_{\rm eff}$. However, a more careful analysis reveals that the notion of an effective hybridization $\Gamma_{\rm eff}$ may be misleading, since the absolute height of the Kondo peak is usually pinned at $1/\pi \Gamma_0$. Additional correlations  often only lead to a \textit{narrowing} of the peak width, parametrized by a reduction of $T_{K}$, but not to a change of the height of the peak. 

In order to distinguish between between a peak narrowing and $\Gamma_{\rm eff}$, we define the latter by the orbital 
spectral function at zero frequency for a particle-hole symmetric Hamiltonian,
\begin{eqnarray}
\rho_{d_{0\sigma},d^\dagger_{0\sigma}}(0) &\equiv & \frac{1}{\pi \Gamma_\text{eff}}.
\end{eqnarray}
Since the real part of the Green function must vanish at $\w=0$ in particle-hole symmetry,
\begin{eqnarray}
\label{equ:49-Gamma-eff}
\Gamma_\text{eff} &=&\Gamma_0 +  \Im[\Sigma_\sigma(-i0^+)] 
\end{eqnarray}
must hold using the general property $G_{d_{0\sigma}, d^\dagger_{0\sigma} }(z)=[ z-\e_{d \sigma} -\Delta(z) -\Sigma_\sigma(z)]^{-1}$, where we have divided the total self-energy of the molecular orbital $\Sigma_{\rm tot}(z)=\Delta_\sigma(z)+\Sigma_\sigma(z)$ into the hybridization-induced part $\Delta_\sigma(z)$ for the non-interacting problem and all correlation-induced and electron-phonon induced corrections $\Sigma_\sigma(z)$. Since the imaginary part of the self-energy $\Sigma_\sigma(z)$ vanishes for $T,\w\to 0$ in a local Fermi liquid in the standard case of a non-interacting conduction band, the Green's function is pinned to a fixed value $\rho_{d_{0\sigma},d^\dagger_{0\sigma}}(0)=\frac{1}{\pi} (\pi V^2\rho_0)^{-1}=(\pi \Gamma_0)^{-1}$ independent of the model parameters \cite{Langreth1966,YoshimoriZawadowski1982,Anders1991}.

The presence of the unusual Holstein coupling $\lambda_c$ however, leads to modifications of this picture. In appendix \ref{sec:EOM-GF}, we derive the exact analytic expression of the correlation-induced self-energy $\Sigma_\sigma(z)$ of the molecular orbital using the exact equation for motion (EOM) for the Green's functions \cite{BullaHewsonPruschke98}. The result is
\begin{eqnarray}
\label{eq:self-energy}
\Sigma_\sigma(z) &=& 
\frac{U F_\sigma(z) +\lambda_d M_\sigma(z)  + 
\frac{ \lambda_c}{V_0}\Delta(z)  N_\sigma(z) }{G_{d_{0\sigma},d^\dagger_{0\sigma}}(z)} ,
\end{eqnarray}
with the definitions
\begin{subequations}
\label{eq-def-corr-eom} 
\begin{eqnarray}
\label{equationa}
F_\sigma(z) &= &G_{d_{0\sigma} n^d_{-\sigma},d^\dagger_{0\sigma}}(z),\\
M_\sigma(z) &= & G_{\hat X_0 d_{0\sigma},d^\dagger_{0\sigma}}(z),
\label{equationb}
\\
N_\sigma(z) &= & G_{\hat X_{0} c_{0\sigma} ,d^\dagger_{0\sigma}}(z).
\label{equationc}
\end{eqnarray}
\end{subequations}
We explicitly use Eq.~\eqref{eq:self-energy} to obtain the Green's function of the molecular orbital  from the NRG solution which provides $F_\sigma(z)$, $M_\sigma(z)$, $N_\sigma(z)$ and $G_{d_{0\sigma},d^\dagger_{0\sigma}}(z)$ \cite{BullaHewsonPruschke98}.

As shown by Hewson and Meyer \cite{HewsonMeyer02}, the self-energy $\Sigma_\sigma(z)= [UF_\sigma(z) +\lambda_d M_\sigma(z)] /G_{d_{0\sigma},d^\dagger_{0\sigma}}(z)$ maintains Fermi liquid properties and its imaginary part vanishes for $T,\w\to 0$ for a coupling of the orbital to a free electron gas. This can be understood from the topology of a Feynman diagram expansion of these correlation functions independent of the analytic shape of $G_{d_{0\sigma},d^\dagger_{0\sigma}}(z)$. In the presence of a finite $\lambda_c$,   this statement does not hold any longer:  the imaginary part of the self-energy acquires a negative offset which we will quantify in the following.

Applying the EOM to $F_\sigma(z)$ reveals that this composite Green's function also contains additional self-energy corrections in the presence of a finite $\lambda_c$. Therefore, the self-energy contribution $\Sigma^U(z)= UF_\sigma(z) /G_{d_\sigma,d^\dagger_\sigma}(z)$ cannot be identified by the same skeleton expansion \cite{LuttingerWard1960} as for the $\lambda_c=0$ case.

Another modification stems from the third term in the nominator of Eq. \ \eqref{eq:self-energy},
\begin{eqnarray}
\Delta \Sigma(z) &=& \frac{ \lambda_c}{V_0 |G_{d_\sigma,d^\dagger_\sigma}(z)|^2}\Delta(z)  N_\sigma(z) G_{d_\sigma,d^\dagger_\sigma}^*(z) .
\end{eqnarray}
Assuming particle-hole symmetry and $T\to 0$, and using that the real part of $G_{d_\sigma,d^\dagger_\sigma}(z)$ as well as  the real part of $N_\sigma(-i0^+)$ vanish yields
\begin{eqnarray}
\label{eq:27}
\Delta \Sigma(-i0^+)  
&=& i \frac{  \lambda_c \Gamma_0}{V_0 \pi \rho_{d_{0\sigma},d^\dagger_{0\sigma}}(0)}
\Im  N_\sigma(-i0^+)  .
\end{eqnarray}
$N_\sigma(z)$ is an off-diagonal Green's function and its spectral integral is zero. Therefore, spectrum has equal positive and negative spectral weight in different frequency regions. The NRG calculation shows that $\Im  N_\sigma(-i0^+)<0$ for  $\lambda_c>0$ and $\Im  N_\sigma(-i0^+)\propto \lambda_c$ in leading order.

\begin{figure}[t]
\begin{center}
\includegraphics[width=0.5\textwidth]{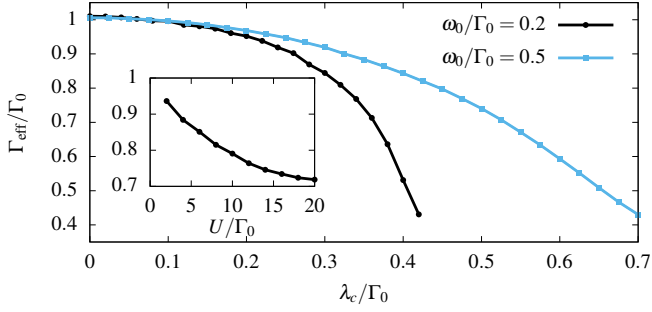}
\caption{Symmetric single impurity Anderson model plus unusual Holstein coupling $\lambda_c$ for a single vibrational mode $\omega_0$. Renormalized hybridization $\Gamma_0\rightarrow\Gamma_{\rm eff}(\lambda_c)$ as function of $\lambda_c$ for two different values for $\omega_0$ at $U/\Gamma_0=10$. The inset depicts the same quantity for $\omega_0/\Gamma_0=0.2$, $\lambda_c/\Gamma_0=0.35$ as function of the Coulomb interaction.}
\label{fig:TK-Lc-b}

\end{center}
\end{figure}
Particle-hole symmetric demands 
\begin{eqnarray}
G_{d_{0\sigma},d^\dagger_{0\sigma}}(-i0^+)&=& \frac{i}{\Gamma_\text{eff}} .
\end{eqnarray}
We substitute Eq.\ \eqref{eq:self-energy} into Eq.\ \eqref{equ:49-Gamma-eff}
with $\lambda_d=0$,
\begin{eqnarray}
\Gamma_\text{eff} &=&\Gamma_0 \\
&& \nonumber - \Gamma_\text{eff} \left[
U \Re F_\sigma(-i0^+) -\frac{ \lambda_c }{V_0} \Gamma_0\Im  N_\sigma(-i0^+) 
\right] ,
\end{eqnarray}
which we solve for the ratio 
\begin{eqnarray}
\label{eq:gamma-eff}
\frac{\Gamma_\text{eff}}{\Gamma_0} &=& 
\frac{1}
{
1 + U \Re F(-i0^+)- \frac{  \lambda_c}{V_0}\Gamma_0 \Im  N_\sigma(-i0^+) 
}
\end{eqnarray}
A negative  $\Im  N_\sigma(-i0^+)$ in combination with a positive $\Re F(-i0^+)$ leads to a reduction of $\Gamma_{\rm eff}$ which is quadratic in $\lambda_c$ for small $\lambda_c$, since then $\Im  N_\sigma(-i0^+)$ is proportional to $\lambda_c$. Clearly, the reduction $\Gamma_0 \to \Gamma_{\rm eff}$ is not only effected by $\lambda_c$, but also depends on $U$.  

Fig.\ \ref{fig:TK-Lc-b} shows the dependence, as calculated by NRG, of the effective hybridization on the coupling $\lambda_c$ for two different vibrational frequencies $\omega_0$ and a fixed $U/\Gamma_0=10$. For a fixed coupling strength $\lambda_c$, the reduction of $\Gamma_{\rm eff}$ decreases  with decreasing polaron energy $E_{p}$, as expected from the discussion in the context of Fig.\ \ref{fig:TK-Lc-a}, confirming the microscopic mechanism outlined above: the larger $E_{p}$, the more severe is the suppression of the hybridization and the stronger thus the reduction of the Kondo temperature. For weak electron-phonon coupling, we expect $\Im  N_\sigma(-i0^+)\propto \lambda_c$, as confirmed by NRG calculations. Furthermore, the change in real part of correlation function $F_\sigma(z)$ must also depend quadratically on $\lambda_c$, because it scales with the polaron energy. Hence Eq.\ \eqref{eq:gamma-eff} predicts an analytic form $1/(1+\alpha \lambda^2_c)$ for $\Gamma_{\rm eff}/\Gamma_0$, which agrees well with the data presented in Fig.\ \ref{fig:TK-Lc-b}. Moreover, $\Gamma_{\rm eff}/\Gamma_0$ should  decrease linearly with increasing $U$ for small $U$, as an expansion of Eq.\ \eqref{eq:gamma-eff} in powers of $U$ shows. The inset in Fig. \ref{fig:TK-Lc-b} confirms this prediction.

\begin{figure}[t]
\begin{center}
\includegraphics[width=0.5\textwidth]{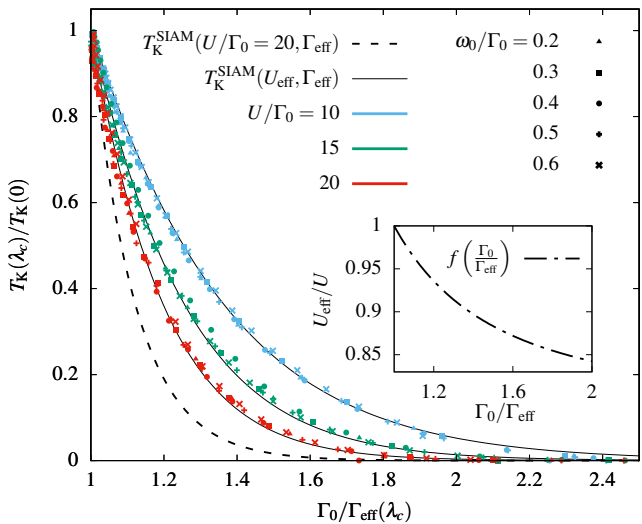}
\caption{Symmetric single impurity Anderson model plus unusual Holstein coupling $\lambda_c$ for a single vibrational mode $\omega_0$. $T_K$ plotted against the inverse of the renormalized hybridization. Different Coulomb interactions are indicated by different colors and different vibrational frequencies $\w_0$ by different points. For comparison we added the textbook expression for $T_K$ of the SIAM as the dashed line.}
\label{fig:TK-Lc-c}

\end{center}
\end{figure}

Finally, we address the question whether the change of $T_{K}$ could also be understood by using an effective SIAM \textit{without} explicitly including the phonons, whose effect would then be accounted for summarily by a renormalized $\Gamma_{\rm eff}$. As we will show below, the answer is no, one also needs a renormalization $U\to U_{\rm eff}$ \textit{in the molecular orbital}. Since it is possible to reproduce any $T_{\rm K}$ with an appropriate combination of $U$ and $\Gamma$, not much understanding would be gained if both parameters were left free. Therefore, we demand that  $U_{\rm eff}=Uf(x)$ depends only via a universal function $f(x)$ on the ratio $x=\Gamma_0/\Gamma_\text{eff}$. Assuming the validity of the standard expression for the Kondo temperature \cite{KrishWilWilson80a}, the ratio of the Kondo temperatures for fixed band widths but different hybridization strengths $\Gamma_\text{eff}$ is given by
\begin{equation}
\frac{T_K^{\rm SIAM}(U_{\rm eff}=Uf(x),\Gamma_\text{eff})}{T_K^{\rm SIAM}(U,\Gamma_0)}
=\frac{1}{\sqrt{x f(x)}} 
e^{ -  \frac{\pi U}{8\Gamma_0}( x f(x)-1)}
\label{eq:tk-ratio-gamma-eff}
\end{equation}
which for a fixed initial value  $U/\Gamma_0$ is only a function of $x$. 

In Fig.\ \ref{fig:TK-Lc-c} we plot the NRG data of Fig.\ \ref{fig:TK-Lc-a} as function of $x=\Gamma_0/\Gamma_\text{eff}$ and indeed observe universality: all data points for different phonon frequencies fall on top of a universal but $U$-dependent curve. However, a constant function $f(x)=1$, which would imply a $U_{\rm eff}=U$ that is not renormalized, yields a mismatch between the NRG results and Eq.\ \eqref{eq:tk-ratio-gamma-eff}, as shown for the case of $U/\Gamma=20$ by the dashed line in Fig.\ \ref{fig:TK-Lc-c}. We obtain an excellent fit of the numerical data with a phenomenological universal function $f(x) = 1 + 0.21(x^{-2}-1)$ (thin dotted lines for the different values of $U$ in Fig.\ \ref{fig:TK-Lc-c}.) The inset shows $f(x)$ on the same interval.  

In conclusion, our NRG solution of a physical model comprising an electron-phonon coupling $\lambda_c$ between the molecular vibration and a local effective substrate orbital reveals a reduction of the Kondo temperature of the spin-$\frac{1}{2}$ degree of freedom in the molecular orbital.  In the framework of a particle-hole symmetric single impurity Anderson model, this can be parametrized by a reduction of the hybridization between the molecular orbital and the substrate and a concurrent, but more moderate reduction of the intraorbital Coulomb repulsion $U$. The underlying mechanism is the generation of a negative $U^c$ in the local effective substrate orbital. In the antiadiabatic limit this negative $U^c$ is essentially given by the polaron energy. 

While it is intuitively clear that a negative $U^c$, by destabilizing the singly occupied state, will reduce the hybridization with the molecular orbital, the NRG shows that the screening of the intraorbital repulsion in the LUMO is also indirectly affected  by a coupling of the vibrational mode to the local substrate orbital. We have established the following consequences of $\lambda_c$: (i) a reduction of $\Gamma_{\rm eff}$, corresponding to an increase of the Kondo peak height relative to $\Gamma_0$, (ii) a reduction of $T_K$ leading to a narrowing of the Kondo resonance as well as (iii) a parametrization of the Kondo temperature by replacing $\Gamma\to \Gamma_{\rm eff}$ as well as $U\to U_{\rm eff}=Uf(\Gamma_{\rm eff}/\Gamma_0)$ in the standard analytic expression for $T_K$. Although Eq.\ \eqref{eq:gamma-eff} does not hold in the particle-hole asymmetric case, the qualitative features will remain valid.

\subsubsection{Analysis of the Holstein coupling $\lambda_c$: Spectral and transmission functions}

Next, we investigate the influence of the unconventional Holstein coupling $\lambda_c$ on the various spectral functions $\rho(\w)$ that make up the elastic and inelastic transmission functions $\tau^{(0)}_\sigma$, $\tau^{(1)}_\sigma$ and $\tau^{(2)}_\sigma$. This prepares the comparison of the calculated to the experimental differential conductance spectra for the NTCDA/Ag(111) system. As in the previous section, we will consider a particle-hole symmetric scenario for simplicity. Then, the Holstein coupling $\lambda_c$ does not lead to a displacement of the harmonic oscillator $\nu=0$ with energy $\omega_0$, and $\langle \hat{X_0} \rangle=0$ is always fulfilled. Furthermore, we allow for tunneling into $M=2$  states $\mu$, $\mu'$, namely the molecular orbital (tunneling matrix element $t_{0\sigma}=t_{d}$) and the effective local substrate orbital ($t_{1\sigma}=t_{c}$) where we have dropped the spin dependency of the tunneling matrix elements assuming  a non-magnetic tip. We thus explicitly include the possibility of a Fano interference in this section \cite{SchillerHershfield2000a}.

\begin{figure}[t]
\begin{center}
\includegraphics[width=0.5\textwidth]{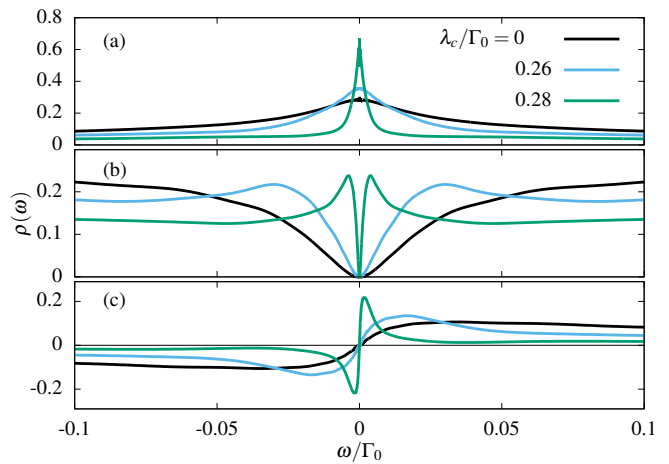}
\caption{Contributions to the elastic spectrum due to tunneling into the molecular orbital $d_{0\sigma}$ and the local surface orbital $c_{0\sigma}$,
leading to the constituents (a) $\rho_{d_{0\sigma},d^\dagger_{0\sigma}}(\w)$, (b) $\rho_{c_{0\sigma},c^\dagger_{0\sigma}}(\w)$ and (c)  $\rho_{d_{0\sigma},c^\dagger_{0\sigma}}(\w)$. The spectral functions have been calculated with NRG. We set $U/\Gamma_0=10$, $\omega_0/\Gamma_0=0.1$ and different colors indicate various unconventional Holstein couplings $\lambda_c$.
}
\label{NRG_fig2}
\end{center}
\end{figure}

The elastic part of the transmission function $\tau^{(0)}_\sigma(\omega)$ comprises three different contributions 
\begin{eqnarray}
\label{eq:transmission_elastic-two_channel_model}
 \tau_\sigma^{(0)}(\omega)& =&
 t_{d}^2\, \rho_{d_{0\sigma},d^\dagger_{0\sigma}}(\w)
  +t_{c}^2 \rho_{c_{0\sigma}, c_{0\sigma}^\dagger}(\w)
\nonumber\\
  &&
+t_{d} t_{c}\big[  \rho_{c_{0\sigma},d^\dagger_{0\sigma}}(\w)+ \rho_{d_{0\sigma},c_{0\sigma}^\dagger}(\w)\big],
\end{eqnarray}
stemming from the tunneling into the molecular orbital $d_{0}$ and into the effective local surface orbital $c_0$, introduced in Eq.\ \eqref{eqn:c_0} and implying $d_{1\sigma} =c_{0\sigma}$ in $H_T$, Eq.~\eqref{eq:general-tunneling-HT}. The three relevant spectral functions are plotted versus frequency for two different values of $\lambda_c$ and a fixed $U$ in Fig.\ \ref{NRG_fig2}. For $ \rho_{d_{0\sigma},d^\dagger_{0\sigma}}(\w)$, displayed in panel (a), the narrowing of the Kondo resonance with increasing $\lambda_c$ is illustrated. The increase of the peak height is connected to the reduction of $\Gamma_{\rm eff}$, as discussed extensively in the previous section. The corresponding anti-resonance in $\rho_{c_{0\sigma}, c_{0\sigma}^\dagger}(\w)$ is clearly visible in panel (b) of Fig.\ \ref{NRG_fig2}. This anti-resonance can be associated with the contribution to the Kondo screening by the electrons in local substrate orbital. The mixed contribution $ \rho_{c_{0\sigma},d^\dagger_{0\sigma}}(\w)= \rho_{d_{0\sigma},c_{0\sigma}^\dagger}(\w)$ in panel (c) is an antisymmetric function and thus its integrated spectral weight vanishes. This contribution captures the interference  between the two possible tunneling paths and generates Fano lineshapes in Eq.\ \ref{eq:transmission_elastic-two_channel_model}. We note that the low-frequency part of the all spectral functions in Fig.\ \ref{NRG_fig2} is governed by the same energy scale $T_{K}$ that is reduced with increasing $\lambda_c$.

\begin{figure}[t]
\begin{center}
\includegraphics[width=0.5\textwidth]{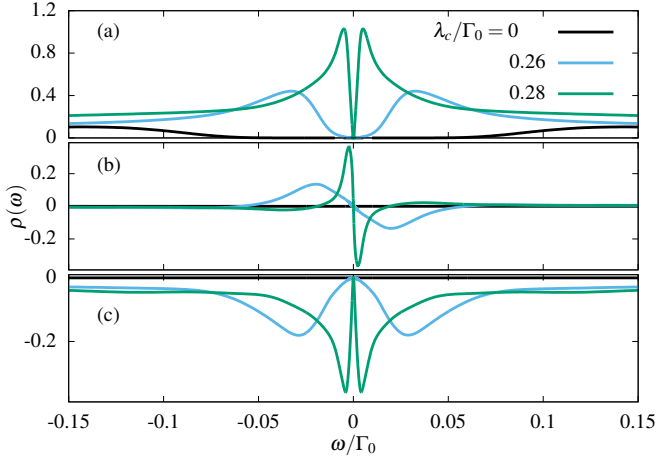}
\caption{Contributions to the inelastic spectrum due to tunneling into the molecular orbital $d_{0\sigma}$, the local surface orbital $c_{0\sigma}$ and a coupling $\lambda^\text{tip}_{\mu\nu}$ of the vibrational mode $\omega_0$ to the STM tip,
leading to the constituents (a) $\rho_{\hat{X}_0d_{0\sigma},\hat{X}_0d^\dagger_{0\sigma}}(\w)$, (b) $\rho_{\hat{X}_0d_{0\sigma},d^\dagger_{0\sigma}}(\w)$ and (c)  $\rho_{\hat{X}_0d_{0\sigma},c^\dagger_{0\sigma}}(\w)$. The spectral functions have been calculated with NRG. We set $U/\Gamma_0=10$, $\omega_0/\Gamma_0=0.1$ and different colors indicate various Holstein couplings $\lambda_c$.
}
\label{NRG_fig3}
\end{center}
\end{figure}

The constituents of the inelastic spectrum
\begin{eqnarray}
\label{eq:transmission_inelastic-two_channel_model}
 \tau^{(1)}_\sigma(\omega)&+&\tau^{(2)}_\sigma(\omega)= \lambda^{\rm tip} 
 \Big\{\lambda^{\rm tip}t^2_{d}
 \rho_{\hat{X}_0 d_{0\sigma},\hat{X}_0d^\dagger_{0\sigma}} (\w)\nonumber \\
&+&t^2_{d}
\big[ 
\rho_{ \hat{X}_0d_{0\sigma} ,d^\dagger_{0\sigma}}(\w)
+ \rho_{d_{0\sigma},\hat{X}_0 d^\dagger_{0\sigma}}(\w) 
\big]
\\
&+&t_{d} t_{c}\big[
\rho_{ \hat{X}_0 d_{0\sigma} ,c_{0\sigma}^\dagger}(\w) + \rho_{c_{0\sigma},\hat{X}_0 d^\dagger_{0\sigma}}(\w) \big]
\Big\}
\nonumber
\end{eqnarray}
are shown in Fig.\ \ref{NRG_fig3}. Here we have assumed that $\lambda^{\rm tip}_{\mu\nu}=0$ except for $\mu=0$ and $\nu=0$, i.e.\ only the molecular orbital $d_0$ (but not the local effective substrate orbital $c_0$) is coupled through the vibration $\hat X_0$ to the STM tip, with coupling constant $\lambda^{\rm tip}$. 

We note that $\rho_{\hat{X}_0d_{0\sigma},\hat{X}_0d^\dagger_{0\sigma}}(\w)$, displayed in panel (a) of Fig.\ \ref{NRG_fig3}, can in principle also be obtained from Eq.\ \eqref{eqn:freemode_inelastic} in the limit $\lambda_c=0$. It consists of two peaks at $\pm\omega_0$ that indicate the threshold for the excitation of a vibrational quantum by the tunneling electron. The smooth structure of the thresholds in the NRG spectrum is a consequence of broadening procedure in the NRG approach \cite{PetersPruschkeAnders2006,WeichselbaumDelft2007,BullaCostiPruschke2008}. 

Tracking the peak position of $\rho_{\hat{X}_0d_{0\sigma},\hat{X}_0d^\dagger_{0\sigma}}(\w)$ for increasing $\lambda_c$ reveals the well-understood renormalization of the phonon frequency $\omega_0^\prime(\lambda_c)$ in the adiabatic limit \cite{EidelsteinSchiller2013}. Furthermore, it is interesting to note that in the limit of a vanishing Holstein coupling $\lambda_c$, $\rho_{\hat{X}_0d_{0\sigma},\hat{X}_0d^\dagger_{0\sigma}} (\w)$ is the only non-zero contribution, because the inelastic terms that are linear in $\lambda^{\rm tip}$ require a non-zero electron-phonon coupling in the system S so the phonon number is not any longer a conserved quantity.

\subsubsection{Strategy for matching the experimental and theoretical spectra for NTCDA/Ag(111)}

We return to the NTCDA/Ag(111) system and show how the tunneling theory of section \ref{sec:tunnel-current-general}, the unconventional Holstein model (section \ref{sec:modelling_the_system_vibrations}) and the ab-initio input to the NRG (section \ref{sec:Input from ab-initio calculations and model without electron-phonon coupling}) can be combined to match the experimental spectra. 

In the first step (section \ref{Holstein coupling for NTCDA/Ag(111)}) we adjust the unconventional Holstein coupling $\lambda_c$ such that the zero-bias peak in the experimental differential conductance spectra is correctly reproduced, irrespective of the spectral signatures at finite voltages. This procedure can indeed reduce the simulated Kondo temperature sufficiently to achieve a match to the experimental Kondo temperature. In the second step (section \ref{sec:NRG-7}), we focus on the inelastic parameters $\lambda_{\mu\nu}^{\rm tip}$ which parameterize the change of the tunnel coupling of the STM tip to the orbital $\mu$, induced by vibration $\nu$, and govern the differential conductance spectra at higher energies. Both coupling mechanisms together allow modeling the differential conductance spectra in excellent agreement with experiment, as we will demonstrate in section \ref{NRG results for  NTCDA/Ag(111)}.

\subsubsection{Holstein coupling $\lambda_c$ for NTCDA/Ag(111)}
\label{Holstein coupling for NTCDA/Ag(111)}

First, we need to identify the vibrational mode(s) which may couple to the local effective orbital $c_{0\sigma}$. To this end we use the free vibrational modes of the gas phase in the absence of the substrate as guidance. The energies $\omega_\nu$ of molecular eigenmodes in the relevant energy range are given in Table \ref{tab:vibrations} of Sec.\ \ref{sec:experiment-NTCDA}, together with their irreducible representations. Two modes have B$_{\rm 3g}$ character, namely the ones at $\omega = 50.4\,\text{meV}$ and $\omega = 41.6\,\text{meV}$. The B$_{\rm 3g}$ eigenmodes describe specific out-of-molecular-plane vibrations, while the B$_{\rm 1u}$, A$_{\rm g}$ and B$_{\rm 1g}$ modes close in energy 
all correspond to in-plane vibrations. Since any mode that potentially contributes to the unconventional Holstein coupling $\lambda_c$ must change the local potential and thus shift the single particle energy of local substrate orbitals as function of the displacement, we can expect the relevant modes to involve displacements perpendicular to the surface. Thus, we can concentrate on the two B$_{\rm 3g}$ modes. Note that the symmetry considerations only provide guidance and are not strictly valid, since the $D_{2h}$ point group symmetry of the molecule in the gas phase is broken in the NTCDA/substrate system
as well as the presence of the STM tip.
  
\begin{figure}[t]
\begin{center}
\includegraphics[width=0.5\textwidth]{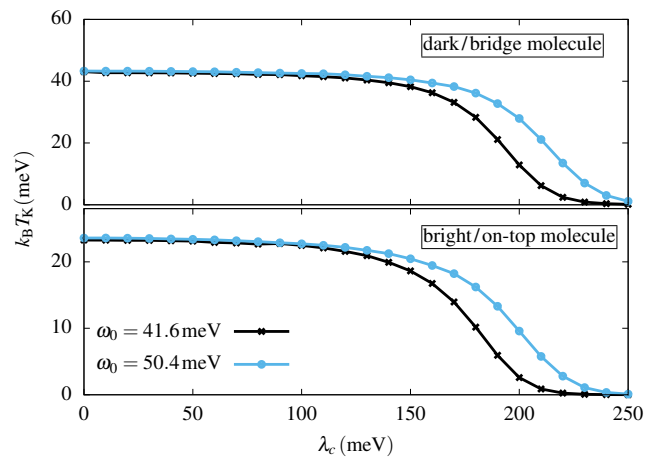}
\caption{NRG results for the Kondo temperature $T_K$ of the bright/on-top and dark/bridge molecules as a function of the unconventional Holstein coupling $\lambda_c$ for two different vibrational energies $\w_0=41.6,\text{meV}, 50.4,\text{meV}$.
}
\label{NRG_fig4a}

\end{center}
\end{figure}

Assuming that only one of the two possible B$_{\rm 3g}$ modes 
couples to the substrate, we calculate the Kondo temperature as function of the coupling strength $\lambda_c$ for both modes individually, using the electronic ab-initio parameters as input for the NRG calculations. The result is depicted in Fig.\ \ref{NRG_fig4a} for both types of molecules and both vibrational modes. A substantial narrowing is achieved for $\lambda_c>100\,\text{meV}$, whence the polaron energy $E_p$ exceeds the hybridization strength of $\Gamma^\text{dark}(0)=190\,\text{meV}$ and $\Gamma^\text{bright}(0)=165\,\text{meV}$ at the Fermi energy. Therefore, by fixing the appropriate value of $\lambda_c$ the NRG-calculated spectral width of the Kondo peak can be matched to the experimental findings, irrespective of which of the two B$_{\rm 3g}$ modes is used.

\subsubsection{Tip-system coupling $\lambda^{\rm tip}_{\mu\nu}$ for NTCDA/Ag(111)}
\label{sec:NRG-7}

The unconventional Holstein coupling $\lambda_c$ substantially reduces the Kondo temperature and thus improves the overall agreement of the experimental differential conductance spectrum with the combined ab-initio and NRG spectrum, but it does not explain the additional features in the $dI/dV$ spectra at finite frequencies, most notably the ones at  $+(47.0 \pm 0.3)$\,mV and $-(51.5 \pm 0.3)$\,mV which we have attributed to inelastic tunneling processes (section \ref{sec:experiment-NTCDA vibrational side bands}). Since in experiment these features are linked to the evolution of the zero-bias anomaly (Fig.\ \ref{fig:fig11}), we propose that they are incarnations of the so-called vibrational Kondo replica \cite{Kondo_vib_SM,kondo_vib_bjunc, kondo_vib_bjunc2, kondo_vib_bjunc3,vib_kondo_stm, vib_kondo_stm2, vib_kondo_stm3} and related to half of the Kondo peak shifted by $\pm \w_{\rm eff}$ as suggested by Eq.\ \eqref{eqn:freemode_inelastic} (see the discussion of the second limiting  case at the end of section \ref{sec-lambda-0}). The features at $+(47.0 \pm 0.3)$\,mV and $-(51.5 \pm 0.3)$\,mV thus very likely appear in the differential conductance spectrum as a result of inelastic tunneling in the limit of a free phonon mode. The relevant coupling is the vibration-induced change of the tunneling matrix element, i.e.\ the tip-system coupling $\lambda^{\rm tip}_{\mu\nu}$, and no electron-phonon coupling in the system S is required. 

This raises the question which electronic states $\mu$ and which vibrational modes $\nu$ may take part in the tip-system coupling $\lambda^{\rm tip}_{\mu\nu}$. We note that the presence of the tip at a general position above the system, as well as its possibly axially non-symmetric shape, breaks the symmetry of the molecule completely and principally allows the observation of any mode (no selection rules). The only condition is that the excitation of the vibrational mode modulates the tunnel matrix elements, as shown in Eq.\ \eqref{eq:7}. Most likely,  only the atomic displacements on the molecule will lead to a relevant change of the tunneling matrix elements and, therefore, $\lambda^{\rm tip}_{\mu\nu}$ is restricted to a tunneling into the molecular orbital $d_{0\sigma}$, i.e.\ $\lambda^{\rm tip}_{\mu\nu}=\lambda^{\rm tip}_{0\nu}$. Moreover, since electron densities vary exponentially with perpendicular distance from the surface of the system S, it is plausible that the most significant couplings $\lambda^{\rm tip}_{0\nu}$ will involve vibrational displacements perpendicular to the substrate --- among the modes in the relevant energy window listed in Table \ref{tab:vibrations} these are the B$_{\rm 3g}$ modes. Thus, it turns out that the \textit{same} modes which induce a local potential change on the surface and thereby provide the unconventional Holstein coupling $\lambda_c$ are those which also change the tunneling matrix elements from the tip into the system strongly.

Because a large $\lambda_c$ (which is needed for one vibrational mode to reduce the Kondo temperature, see section \ref{Holstein coupling for NTCDA/Ag(111)}) yields a sizable downward renormalization of the energy $\w_{\rm eff}$ at which the vibrational feature is observed, i.e.\ $\w_{\rm eff}< \w_0$ where $\w_0$ is the bare vibrational energy, and also because it  broadens the inelastic spectral functions substantially, a second mode which only couples weakly within the system S is needed to generate the sharp steps at  $+(47.0 \pm 0.3)$\,meV and $-(51.5 \pm 0.3)$\,meV in the total spectrum. Therefore, we require two distinct vibrational modes in the tunneling Hamiltonian in $\hat H_T$ ($N_\nu=2$). 

\begin{figure}
	\centering
\includegraphics[width=0.5\textwidth,, trim= 0 210  0 170,clip]{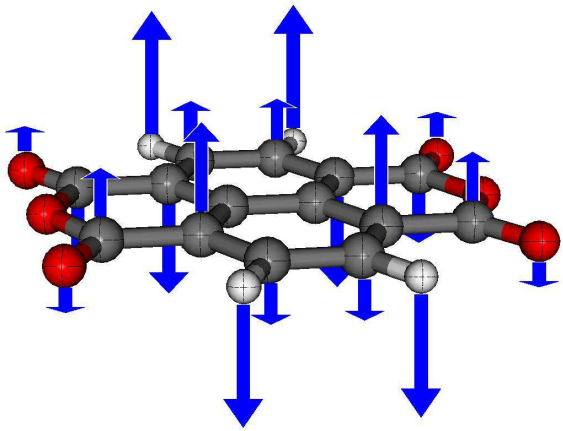}

		\caption{Elongation pattern of B$_{\rm 3g}$ vibrational mode No. 3 ($\omega_0=50.4\,\text{meV}$) from Table \ref{tab:vibrations}. }
	\label{Figure_Rohlfing_vibrational_mode}
\end{figure}

It is interesting to note that the mode at $\omega_0=50.4$\,meV in particular changes the tunneling matrix element very effectively, as a DFT analysis reveals. Fig.~\ref{Figure_Rohlfing_vibrational_mode} shows the elongation pattern of this mode, calculated for gas-phase NTCDA. The outer C atoms at the CH edge exhibit a small, the ones in line with the carboxylic C atoms the largest vibrational amplitude, while for symmetry reasons the C atoms located on the long axis of the molecule have zero amplitudes. If we estimate the contribution of each mode to the transmission functions in Eq.\ \eqref{eq:rho1explicit} and \eqref{eq:rho2explicit} separately by calculating in DFT the change of the LUMO local density of states (LDOS) on excitation of a single quantum of each vibration, we observe that the mode at $\omega_0=50.4$\,meV produces a substantial modulation of the LUMO LDOS \textit{above the center of the molecule}, although both the LUMO and the vibrational mode amplitude itself vanish there. The reason is that the positive and negative lobes of the LUMO (green and red in Fig.~\ref{fig:exp_fig3}), being of equal size for the non-vibrating molecule, are distorted differently on excitation of this mode. This breaks the symmetry in the center of the molecule, leading to a non-vanishing LUMO density of states in the \textit{center of the vibrationally distorted molecule}. Evidently, this gives rise to a large relative change of the LDOS on excitation of the vibration, and thus to a large $\lambda^{\rm tip}_{00}$. Apart from explaining the large value of $\lambda^{\rm tip}_{00}$, this also elucidates why the coupling is sharply focussed in the center of the molecule. 

\subsubsection{NRG results for  NTCDA/Ag(111)}
\label{NRG results for  NTCDA/Ag(111)}

In summary, we arrive at the following model to calculate differential conductance spectra for NTCDA/Ag(111): In accordance with DFT, we set $\lambda_d= 0$ for all vibrational modes of the NTCDA molecule, while one vibrational mode ($\omega_0=50.4\,\text{meV}$) exhibits a nonzero coupling $\lambda_c$ to the substrate orbital $c_{0\sigma}$ and two modes ($\omega_0=50.4\,\text{meV}$ and $\omega_1=41.6\,\text{meV}$) exhibit finite couplings $\lambda^{\rm tip}_{00}$ and  $\lambda^{\rm tip}_{01}$ to the tip. The inelastic contribution stemming from the mode $\omega_1$ is calculated via Eq.\ \eqref{eqn:freemode_inelastic}, using the NRG-calculated spectral function $\rho_{d_{0\sigma},d^\dagger_{0\sigma}}(\w)$, whereas the one stemming from the mode $\omega_0$ is calculated within the NRG
using the full formalism of Eq.\,\eqref{eq:rho1explicit} and \eqref{eq:rho2explicit}. For the bright (on-top) molecule we set $\lambda_c=200\text{meV}$, while $\lambda_c=220\text{meV}$ is selected for the dark (bridge) molecule. Since the DFT predicts a larger hybridization $\Gamma^\text{bridge}(0)=190\,\text{meV}$ compared to $\Gamma^\text{on-top}(0)=165\,\text{meV}$, a 10\% enhancement of the electron-phonon coupling for the bridge molecule appears justified. While the tunneling Hamiltonian may include an arbitrary number of orbitals in the system S, we focus on a minimal configuration $M=2$ to include (i) the Kondo effect, (ii) the feasibility of a Fano resonance, (iii) and the possibility to change the differential conductance spectra when moving from the CH edge to the center of the molecules by adjusting the tunneling matrix elements without altering the system S itself, i.e.\ using fixed spectral functions. For simplicity, we use the same two orbitals for the $M=2$ tunneling channels which have already been introduced in the framework of the two electron-phonon coupling mechanisms. As mentioned before, the hybridization functions $\Gamma^\text{on-top}(\omega)$, $\Gamma^\text{bridge}(\omega)$ and the intra-LUMO Coulomb repulsion $U$ are provided as input to the NRG calculation by a combination of DFT and MBPT, the latter in the shape of the GdW approximation. 

\begin{figure}[t]
\begin{center}
\includegraphics[width=0.5\textwidth]{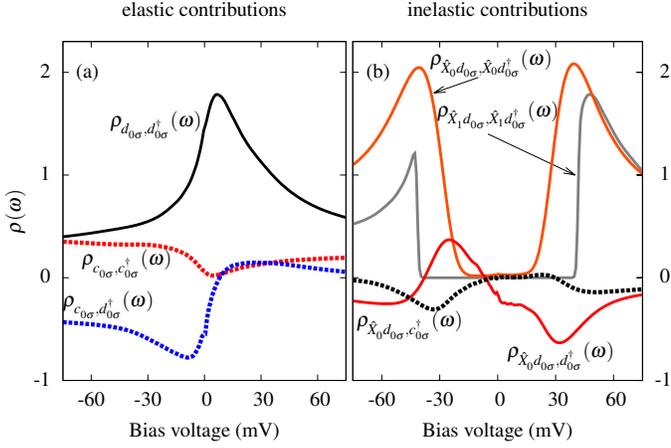}
\caption{Individual contributions to the (a) elastic and (b) inelastic spectrum for the bright (on-top) molecule that are combined in Fig. \ref{NRG_fig4} to fit the experimental $dI/dV$ spectra. Note that at the CH-edge we only include the spectra displayed with solid lines, whereas the Fano effect in the center of the molecule leads to additional contributions which are plotted as dashed lines in Fig \ref{NRG_fig4b}. Results for the dark (bridge) molecule are qualitatively the same. Parameters as in Fig. \ref{NRG_fig4}.
}
\label{NRG_fig4b}
\end{center}
\end{figure}

\begin{figure}[t]
\begin{center}
\includegraphics[width=0.5\textwidth]{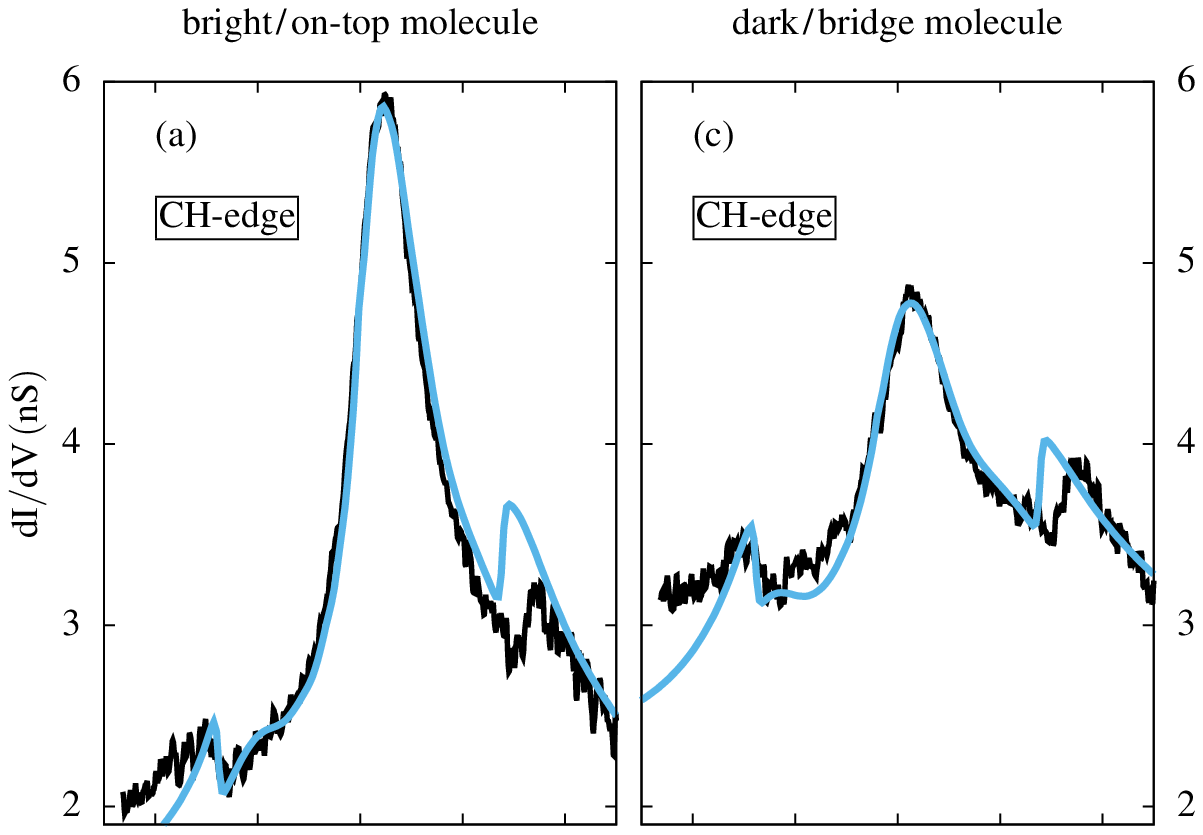}
\includegraphics[width=0.5\textwidth]{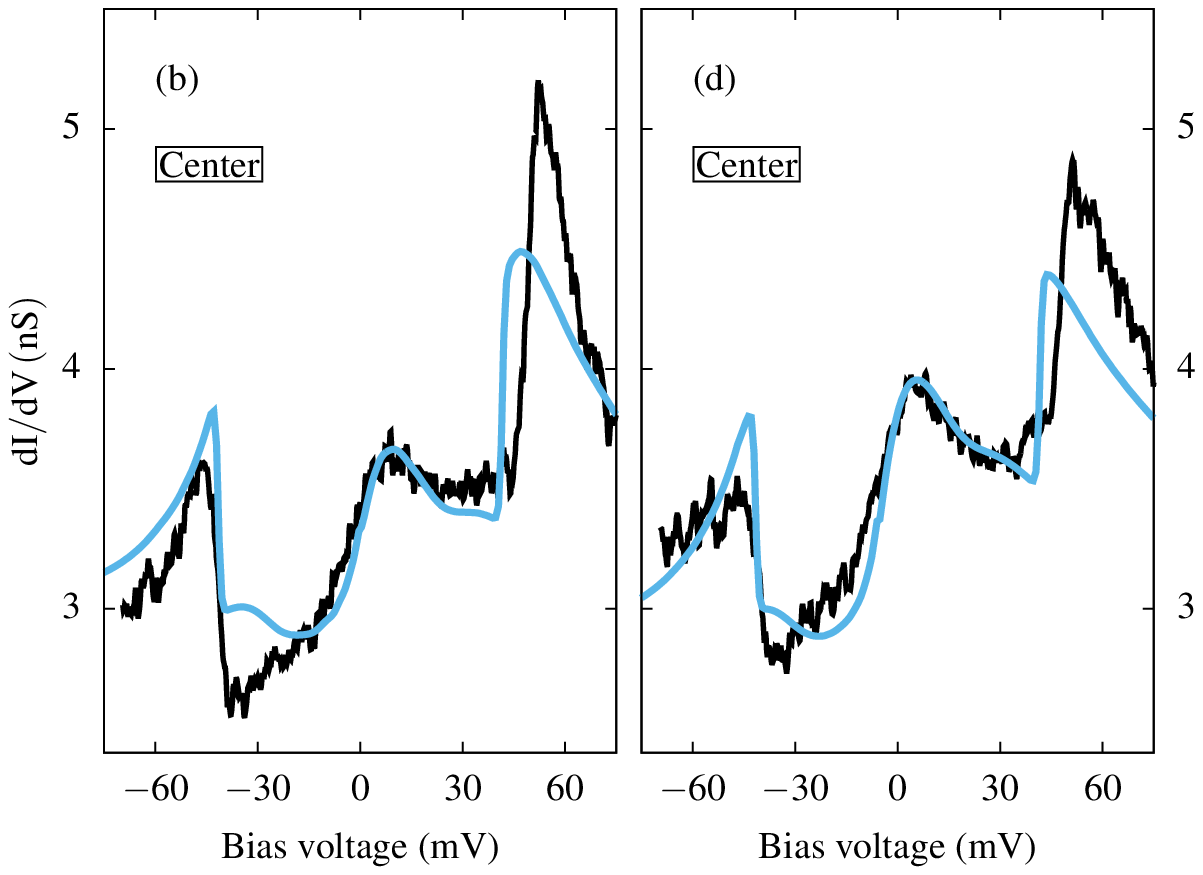}
\caption{Comparison between theoretical (blue) and experimental (black) $dI/dV$ spectra for both types of molecules (bright and dark), each measured at the CH-edge and at the center of a molecule. Two vibrational energies have been used, $\omega_0=50.4\,\text{meV}$ and $\omega_1=41.6\,\text{meV}$. Parameters are: (a) bright/on-top molecule, CH-edge: $\lambda_c= 200$\,meV  for $\omega_0$, $\lambda_c= 0$ for $\omega_1$, $\lambda^{\rm tip}_{00}= 0.25$, $\lambda^{\rm tip}_{01}= 0.35$, $t_c/t_d=0$, $\rho_{\rm offset}=0$. (b) bright/on-top molecule, center: $\lambda_c= 200$\,meV  for $\omega_0$, $\lambda_c= 0$ for $\omega_1$, $\lambda^{\rm tip}_{00}= 0.45$, $\lambda^{\rm tip}_{01}= 1.0$, $t_c/t_d=0.5$, $\rho_{\rm offset}=1.85$\,nS. (c) dark/bridge molecule, CH-edge: $\lambda_c= 220$\,meV  for $\omega_0$, $\lambda_c= 0$ for $\omega_1$, $\lambda^{\rm tip}_{00}= 0.28$, $\lambda^{\rm tip}_{01}= 0.40$, $t_c/t_d=0$, $\rho_{\rm offset}=0.52$\,nS. (d) dark/bridge molecule, center: $\lambda_c= 220$\,meV  for $\omega_0$, $\lambda_c= 0$ for $\omega_1$, $\lambda^{\rm tip}_{00}= 0.35$, $\lambda^{\rm tip}_{01}= 0.75$, $t_c/t_d=0.5$, $\rho_{\rm offset}=2.18$\,nS.
}
\label{NRG_fig4}
\end{center}
\end{figure}

Combining all elastic and inelastic contributions for the outlined SIAM-Holstein model of the NTCDA/Ag(111) system as displayed in Fig.\ \ref{NRG_fig4b}, the experiment/theory comparison of the $dI/dV$ curves is displayed in Fig.\ \ref{NRG_fig4} for two STM tip locations on both the dark (bridge) and bright (on-top) molecules. As the figure shows, our minimal model  yields a remarkable agreement between NRG and the experiment, with calculated Kondo temperatures $T_K^\text{bright}=103.6$\,K and $T_K^\text{dark}=140.4$\,K. In the comparison in Fig.\ \ref{NRG_fig4}, all other contributions beyond the $M=2$ tunnel paths that are explicitly contained in our model in section \ref{sec:tunnel-current-general} are included into a constant background $\rho_{\rm offset}$ that is added to the NRG-calculated $dI/dV$ curves. Its value is uniquely fixed by the condition that the maximum $dI/dV$ value of the zero-bias peak agrees between theory and experiment. It is reassuring that the experimental and theoretical values at larger bias ($\pm 70$mV) are also comparable.

We stress that the NRG curves in Fig.\ \ref{NRG_fig4} are not fits in the mathematical sense. Rather, we have chosen a set of electron-phonon input parameters for the NRG calculations (in addition to the electronic \textit{ab initio} parameters) which illustrate that our formalism of section \ref{sec:tunnel-current-general} and model of section \ref{sec:Modeling the system} are general enough to predict the generic features that are observed in the experimental differential conductance spectra of NTCDA/Ag(111). We have also tried a model in which the role of the two modes $\w_0$ and $\w_1$ is reversed. However, the resulting fit to the experimental data is significantly worse than the one in Fig.\ \ref{NRG_fig4}, the prime reason being that the vibrational mode renormalisation through $\lambda_c$ shifts the inelastic features substantially down to $\w_{\rm eff} < \omega_\nu$ and it is therefore preferential to start with a larger bare vibration energy. 

With regard to the comparison of the NRG-calculated spectra to the experiments in Fig.\ \ref{NRG_fig4} it should be noted that each experimental spectrum is inevitably measured with a slightly different tip, and different STM tips generally lead to different differential conductance spectra on the same molecule.  Within our theory this can be accounted for by the modification of the fictitious STM tip orbital in Eq.~\eqref{eq:stm-tip-c0} that also changes the individual matrix elements $t_{\mu \sigma\sigma'}(\{\vec{R}_i\})$ in the approximation Eq.~\eqref{eq:H_T-bilinear}. This modification of the tunneling matrix elements leads to a different background current and to a slightly different mixing of the different frequency components of the spectral functions. We therefore account for different tips by adjusting $\rho_{\rm offset}$. 

Experimental spectra may moreover contain a small offset in the voltage scale which is usually gauged away by a calibration. In the present case, the Kondo peak is a common feature in experiment and theory, and we have adjusted its precise location to coincide with the NRG calculation. To this end, we have shifted the experimental curve such that it coincides with the NRG curve at the Kondo peak. We have shifted the experimental curves rather than the theoretical ones because the precise calibration of the experimental energy axis has an uncertainty anyway 
and the location of NRG resonance is determined via the Friedel sum rule \cite{Langreth1966,YoshimoriZawadowski1982,Anders1991}
by LUMO orbital filling and the hybridization function,  both strongly constrained by the  DFT+MBPT  input.

The offset of the Kondo peak from zero bias is indicative of a particle-hole asymmetry. As such, the NRG-calculated offsets $+7.5$\,mV for the bright/on-top molecule and $+3.8$\,mV for the dark/bridge molecule in Fig.\ \ref{NRG_fig4} stem directly from the DFT+MBPT-calculated mean-field PDOS in Fig.\ \ref{NRG_fig4aa}. The as-measured experimental spectra exhibit Kondo peak positions of $+1.9$\,mV for the bright/on-top molecule and $-0.6$\,mV for the dark/bridge molecule. However, since we know that the inelastic features in the $dI/dV$ spectra should be located symmetrically around zero bias, we may use them for a calibration of the experimental bias voltage scale. This results in Kondo peak positions in the calibrated experimental spectra at $+4.2$\,mV for the bright/on-top molecule and $+1.7$\,mV for the dark/bridge molecule. This reveals that the NRG-calculation agrees with experiment regarding the \textit{direction} of the particle-hole asymmetry for both NTCDA molecules on Ag(111) (although the NRG predicts a larger particle-hole asymmetry than found in experiment), as well as regarding the fact that the particle-hole asymmetry is stronger for the bright/on-top molecule. In this respect, the absolute different  between experiment and NRG is only $1.2$\, mV (NRG predicts a difference of $3.7$\,mV between the on-top and bridge molecules, while in experiment the corresponding difference between the bright and dark molecules is $2.5$\,mV). 

In Fig.\ \ref{NRG_fig4} we have shifted the as measured CH-edge spectra $5.6$ / $4.4$\,mV (bright and dark molecules) to the right to achieve coincidence of the Kondo peaks with the NRG. A shift of $2.55$\,mV would have established a symmetric distribution of the inelastic features in the experimental curve. As a consequence, the inelastic features of the experimental spectrum appear off-center in Fig.\ \ref{NRG_fig4} with respect to their NRG counterparts (which are symmetric by construction). After what has been said it is clear that this difference is not an issue of the electron-phonon coupling in our model, but rather of the overestimated particle-hole asymmetry of the NRG calculation (and, more fundamentally of the DFT+MBPT calculation). In principle, a more correct comparison of the NRG and experimental curves in Fig.\ \ref{NRG_fig4} would require shifting the experimental Kondo peak by a larger value ($+5.6$ / $+4.4$\,meV for bright/dark molecules, to correct for the too large prediction of the particle-hole asymmetry) than the rest of the spectrum at the inelastic features ($+2.25$ \,meV for both molecules, to achieve the physically motivated symmetry of the inelastic features). 
Essentially, these  shifts are small and also reveals the overall uncertainties in our procedure matching theory and experiment.

In conclusion, Fig.\ \ref{NRG_fig4} shows that our model of the NTCDA/Ag(111) system explains all generic features of the experimental differential conductance spectra: (i) the different Kondo temperatures of the bright and dark molecules (by different adsorption heights and correspondingly different hybridizations with the substrate), (ii) the smaller-than-expected Kondo temperatures of both molecules including their absolute values (by electron-phonon coupling with an effective local substrate orbital through an unconventional Holstein term), (iii) the strong threshold features at approximately at $+(47.0 \pm 0.3)$\,mV and $-(51.5 \pm 0.3)$\,mV including their asymmetric peak shapes (by inelastic tunneling involving a free phonon including the replication of half of the Kondo peaks), (iv) the weak shoulders at lower bias (by inelastic tunneling involving the coupled vibration that is also responsible for the reduction of the Kondo temperatures), (v) the marked difference of the spectra at the CH edge and in the center of the molecules (by quantum interference between tunneling path into a molecular orbital and into the effective local substrate orbital which is also implicated in the unconventional Holstein coupling), (vi) the strong concentration of the inelastic tunneling in the center of the molecule (by the quantum interference and the symmetries of the involved modes), (vii) the offset of the Kondo peak to positive bias voltages in the calibrated spectra (by a particle-hole asymmetry in the PDOS of the NTCDA LUMO adsorbed on Ag(111)), (viii) the fact that the Kondo peak of the bright molecule appears at slightly larger bias voltages (by the stronger particle-hole asymmetry of the bright molecule).

\subsection{STS in the anti-adiabatic regime}
\label{sec:STS-anti-adiabatic-regime}

\begin{figure}[t]
\begin{center}
\includegraphics[width=0.5\textwidth]{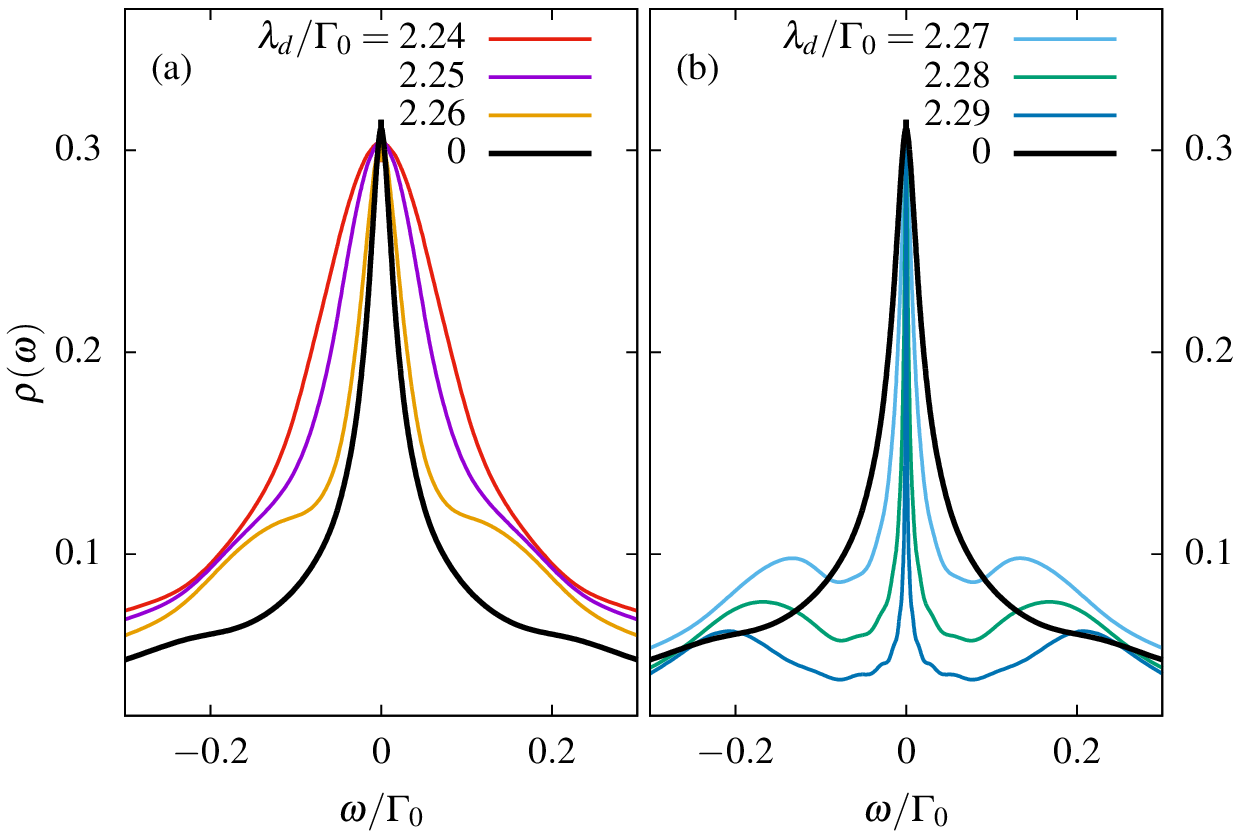}

\includegraphics[width=0.5\textwidth]{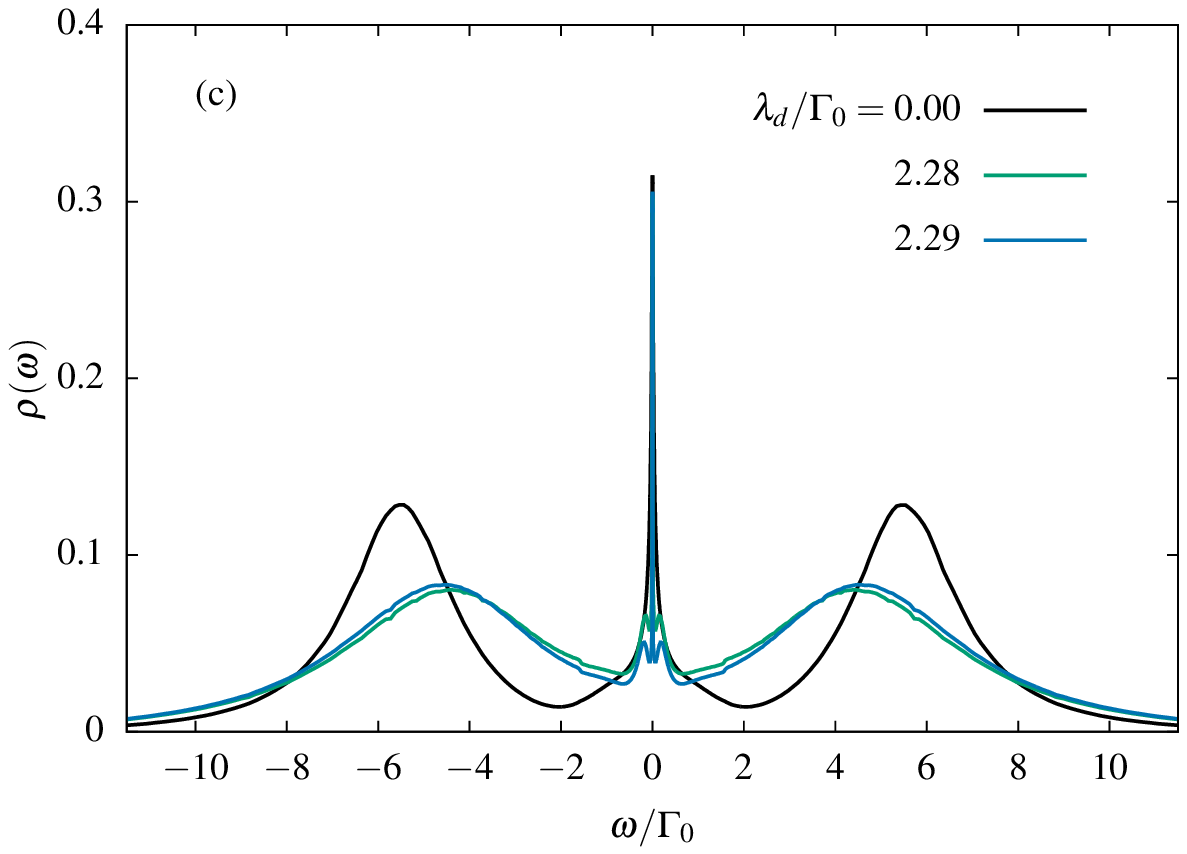}
\caption{Spectral function $\rho_{d_\sigma,d^\dagger_\sigma}(\w)$ 
of the molecular orbital for the particle-hole symmetric anti-adiabatic regime.
(a) and (b) spectral evolution for increasing $\lambda_d>  \lambda_{d,c}$. 
The $\lambda_d=0$ is added for comparison. 
(c) the spectral data of panel (b) plotted on a larger energy interval.
Parameters: $\rho=const$, $D/\Gamma_0=10$, $U/\Gamma_0=-2\e_{d}/\Gamma_0=10$.
}
\label{NRG_fig15}
\end{center}
\end{figure}

In this section we focus on the conventional setup ($\lambda_d>0$) 
\cite{galperinNitzanRatner2006,EidelsteinSchiller2013,JovchevAnders2013}
in anti-adiabatic regime and neglect the unconventional coupling of a local phonon mode to the substrate, 
i.~e.~$\lambda_c=0$. 
In the anti-adiabatic regime polaron energy $E_p$ exceeds 
the hybridization strength, i.e.~$E_p=\lambda^2_d/\w_0 >\Gamma(0)$.
For simplicity, we only consider a featureless symmetric conduction band with a 
constant density of states to separate the many-body effects from
single-particle energy shifts induced by particle-hole asymmetric  hybridization functions.

\subsubsection{Equilibrium electronic spectra}
\label{sec:anti-adiabatic-regime-spectrum}

We review the evolution of the molecular orbital equilibrium spectral properties
 \cite{HewsonMeyer02} with increasing electron-phonon coupling $\lambda_d$. The spectral function $\rho_{d_\sigma,d^\dagger_\sigma}(\w)$ for 
different values of $\lambda_d$ and a constant conduction band density of states
is shown  for a fixed $U/\Gamma_0=10$ and particle-hole symmetry in Fig.~\ref{NRG_fig15}. 
In order to obtain sharp spectral features, we averaged over $N_z=30$ z-values 
in the NRG calculation
and set  the NRG broadening parameter to $b=0.2$ -- see Ref.~\cite{AndersSchiller2006,
PetersPruschkeAnders2006,WeichselbaumDelft2007,BullaCostiPruschke2008} 
for the technical details.
The width of the zero-frequency resonance changes non-monotonically with  $\lambda_d$.
The initial width of the Kondo resonance
increases (not shown here) and, after reaching a maximum, it decreases again with increasing $\lambda_d$. Small
shoulders develop symmetrically around the zero-bias resonance  that evolves into two  separated peaks
as clearly seen in Fig.~\ref{NRG_fig15}(b). Simultaneously, the width of the zero-frequency peak rapidly declines.

The electron-phonon interaction generates an 
attractive contribution to electron-electron interaction \cite{Mahan81,LangFirsov1962} 
that is related to the polaron energy and renormalizes the bare value of $U\to U_{\rm eff} = U -2 E_p= U-2\lambda^2_d/\w_0$  \cite{HewsonMeyer02,Chowdhury2015,KleineAnders2015}.
$U_{\rm eff} $ vanishes at a critical value $\lambda_{d}^c= \sqrt{U\w_0/2}$ and changes its sign to an attractive
interaction upon further increase of  $\lambda_{d}$.

The spectral properties can be understood in terms of an  $U_{\rm eff}$
 \cite{HewsonMeyer02}. Starting from the purely electronic problem at $\lambda_d=0$,
 added as a black line to Fig.~\ref{NRG_fig15}  as a reference spectrum,
the decrease of $U_{\rm eff}$ with increasing $\lambda_d$ leads to an
increasing Kondo temperature up to $U_{\rm eff}\approx \pi\Gamma_0$.
The zero-frequency peak width monotonically grows up to this point.
The  zero-frequency peak width approached its largest values for $\lambda_d/\Gamma_0=2.24$ 
which is roughly a factor 2 larger than the value for $\lambda_d=0$ as shown in Fig.~\ref{NRG_fig15}(a).

Once $\lambda_d$ exceeds $\lambda_{d}^{c}$, the system  
entered the attractive $U$ regime at low frequencies \cite{HewsonMeyer02,Chowdhury2015,KleineAnders2015}
which is governed by a bi-polaron formation.
The spectral properties for this regime are shown in Fig.~\ref{NRG_fig15}(b) and (c).
There, the spin Kondo physics is replaced by a charge-Kondo effect with a rapidly decreasing low-temperature scale $T_K^c$ under further increasing of $\lambda_d>  \lambda_{d}^{c}$. 

The spectra develop two shoulders when increasing  $\lambda_d$ 
that are located approximately at $\pm U_{\rm eff}/2$. As depicted in Fig.~\ref{NRG_fig15}(b), 
this shoulders  grow into symmetric
side peaks once $|U_{\rm eff}|$ exceed the charge Kondo scale, i.\ e.\ $T_K^c < |U_{\rm eff}|$. 

Since the phonon frequency $\w_0$ is of the order of the charge fluctuation scale 
$\Gamma_0$ and
smaller than $U$, the concept of an effective $U_{\rm eff}$ is only useful at low energies.
In terms of the renormalization group approach \cite{Wilson75}, 
$U$ becomes frequency dependent in the presence of the electron-phonon
interaction and flows from its bare high energy value to $U_{\rm eff}$  for  $|\w|\ll \omega_0$.
Therefore the high energy features of the spectra depicted in Fig.~\ref{NRG_fig15}(c) 
are only moderately modified: the original charge excitations
around $\pm U/2$ are renormalized slightly to smaller values which indicate that the renormalization of 
$U\to U_{\rm eff}$ has set in very moderately at high frequencies $\w \approx U/2$. 
Once the flow of $U_{\rm eff}$ has converged,  the spectral developed
additional new peaks: in addition to the slightly shifted high-energy charge fluctuation peaks located at 
$\w\approx \pm U/2$
additional low-frequency
peaks located around  $\pm U_{\rm eff}$ develop leading to a much richer spectrum
as depicted in  Fig.~\ref{NRG_fig15}. 
As demonstrated in Fig.~\ref{NRG_fig15}(b),
the low-frequency side peaks evolves with  $U_{\rm eff}$.

In order to avoid entering the negative $U_{\rm eff}$ regime, 
one could fix  $U_{\rm eff} =const$ by adjusting the bare $U$
of the model upon increasing $\lambda_d$. However, with increasing $\lambda_d$ the renormalization of 
$\Gamma_0\to\Gamma_{\rm eff}\approx  \Gamma_0\exp[-\lambda_d^2/\w_0^2 f(\lambda_d/\w_0)]$  reduces rapidly
the charge fluctuation scale \cite{HewsonMeyer02} in the strong coupling regime. The reduction factor 
$\exp[-\lambda_d^2/\w_0^2]$ is generated by the local polaron formation and can be understood via 
the Lang-Firsov transformation \cite{Mahan81,LangFirsov1962}, while the scaling function 
 $f(\lambda_d/\w_0)$ accounts for additional reduction of  $\Gamma_{\rm eff}$ 
due to the softening of the phonon mode  \cite{HewsonMeyer02}. Consequently, the Kondo temperature
reduces rapidly once $E_p$ exceeds $\w_0$ suppressing the Kondo effect at finite temperature.
We investigated this limit but since the spectral functions qualitatively do not differ
much from those presented above, we spare the rather repetitive analysis.

\begin{figure}[t]
\begin{center}
\includegraphics[width=0.5\textwidth]{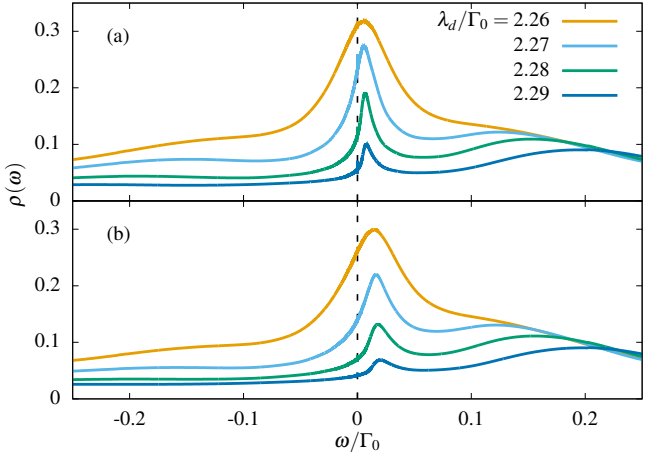}
\caption{Spectral functions for local  particle-hole asymmetry.
(a) small $\Delta\e/\Gamma_0=0.004$ and (b) $\Delta\e/\Gamma_0=0.01$.
NRG parameters as in Fig.~\ref{NRG_fig15}.
}
\label{NRG_fig-ph-asymmetry}
\end{center}
\end{figure}

A comment is in order with regards to particle-hole  asymmetry. While for particle-hole symmetry, the resonance in the spectral function remains
pinned to zero-frequency, a particle-hole asymmetry allows for 
a continuous change of the scattering phase \cite{Langreth1966,YoshimoriZawadowski1982}
of the low energy quasiparticles. In order to understand the spectra in this regime for $\lambda_d>  \lambda_{d}^{c}$,
we  can perform a particle-hole transformation of one spin species to convert an attractive U back to a repulsive
$U$ in the transformed model. Starting from the impurity Hamiltonian in the absence of an external magnetic field $ \e_{d\sigma} = \e_{d}$ 
and replacing $n_\uparrow = (1 - d_\uparrow d^\dagger_\uparrow) = 1 -\bar n_\uparrow$, where $\bar n_\uparrow$
is the number operator of the holes, 
we derive
\begin{eqnarray}
\sum_\sigma \e_{d\sigma} n^d_\sigma + U n^d_\uparrow n^d_\downarrow
= 
\non
\sum_\sigma \left(\frac{U}{2}-\sigma \Delta \e \right)  \bar n^d_\sigma  - U \bar n^d_\uparrow \bar n^d_\downarrow
+  \e_{d} \, ,
\end{eqnarray}
where $\Delta \e = \e_d + U/2$  serves as a measure of the particle-hole  asymmetry \cite{KrishWilWilson80b},
and $\bar n^d_\downarrow= n^d_\downarrow$.
A negative $U$ model describes  the same physics as the positive $U$ model after the 
particle-hole transformation but
with  $\Delta \e$ acting as effective magnetic field. The increasing $(-U_{\rm eff})$ 
leads  to a decreasing charge Kondo temperature $T^c_K$ since  the hybridization $\Gamma_0\to\Gamma_{\rm eff}$ is also reduced \cite{Mahan81,LangFirsov1962,EidelsteinSchiller2013}. 
Consequently the dimensionless 
magnetic field $\Delta\e/T_K(\lambda_d)$ increasing with further increasing of $\lambda_d$. 

Therefore, we can understand the evolution of the spectra  in Fig.\ \ref{NRG_fig-ph-asymmetry}
for two values of   $\Delta \e$ in terms of this analysis. 
The zero-bias Kondo resonance is shifted to a finite value $\Delta\e$, representing
the effective magnetic field in the transformed model. Furthermore its peak high is
increasingly reduced  due to the destruction of the  Kondo effect in a strong effective magnetic field. 
Therefore, the spectral properties shown in  Fig.~\ref{NRG_fig-ph-asymmetry}
are consistent with those of an effective Anderson model  in the  attractive $U$ regime.

\subsubsection{Inelastic contributions}

After reviewing the present understanding of  the electronic spectral function in 
the Anderson Holstein model  \cite{HewsonMeyer02} in its  anti-adiabatic,
particle-hole symmetric  as well as particle-hole anti-symmetric regime, 
we proceed by discussing the implications 
for a potential STS including elastic and inelastic contributions. We assume for simplicity
that the STM tip only couples to the molecular orbital excluding Fano physics.
In order to eliminate the coupling parameters that need to be adjusted for a specific experimental
setup, we define the following two spectral functions 
\begin{eqnarray}
\bar\rho^{(2)}(\w) &=& \frac{1}{(t_d \lambda^{\rm tip})^2} \tau^{(2)}(\w)\\
\bar\rho^{(1)}(\w)& = &\frac{1}{t^2_d \lambda^{\rm tip}}  \tau^{(1)}(\w)
\end{eqnarray}
that contain both inelastic terms. This eliminates the STM tip dependent prefactor and focuses 
only on the spectral features.

\begin{figure}[t]
\begin{center}
\includegraphics[width=0.5\textwidth]{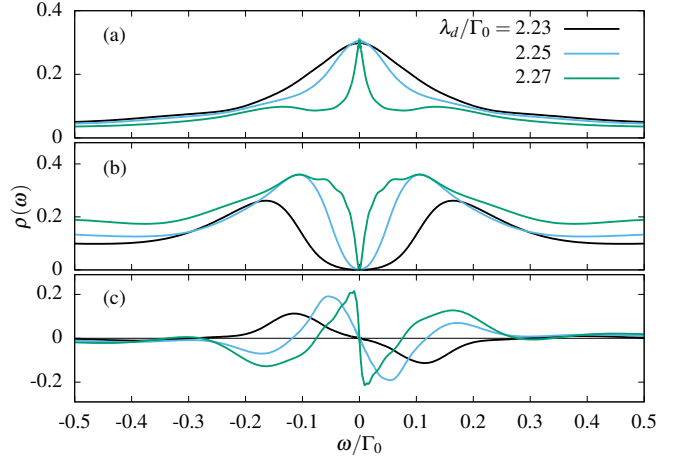}
\caption{All contributions to the STS spectra in the anti-adiabatic regime.
(a) spectral function taken from Fig.~\ref{NRG_fig15},
(b) $\bar\rho^{(2)}(\w)$ for $\lambda_d$ stated in panel (a) and
(c) $\bar\rho^{(1)}(\w)$ for $\lambda_d$ stated in panel (a).
Parameters as in Fig.~\ref{NRG_fig15}.
}
\label{NRG_fig16}

\end{center}
\end{figure}

The individual spectra contributing to the total STS are shown in Fig.~\ref{NRG_fig16}
for the different coupling constants $\lambda_d$.
Panel (a) includes some of the data contained already in Fig.~\ref{NRG_fig15} for comparison. 
Panel (b) of  Fig.~\ref{NRG_fig16} depicts the contribution to $\bar\rho^{(2)}(\w)$.
We observe the same narrowing of the distance between the two peaks
when increasing $\lambda_d$ as plotted in Fig.~\ref{NRG_fig3}. Note however,
that the data in  Fig.~\ref{NRG_fig3} were calculated for $\lambda_d=0$ and a finite $\lambda_c$
as well as a phonon frequency $\w_0 =0.1\Gamma_0$ which is ten times smaller than the
charge fluctuation scale. While the peaks are located at $\pm \w_0$ in the weak coupling
limit ($\lambda_d\to 0$), the energy difference between the two peak positions is 
significantly reduced in the anti-adiabatic regime.
Common to both cases, the previously discussed limit,
(i)  $\lambda_d>0$ and $\lambda_c=0$
and 
(ii) the focus in this section $\lambda_d=0$ and $\lambda_c>0$, 
is the renormalization of the phonon propagator
in the strong coupling limit. The charge susceptibility contributes to the phonon propagator
as can be understood  either in weak coupling derived from the Feynman diagram  in Fig.\ \ref{fig:self-electron-phonon-diagram} or in the atomic limit \cite{Mahan81,LangFirsov1962,HewsonMeyer02}. 
The softening of the phonon mode 
generates additional low frequency contributions to the correlated spectrum
which is the origin of the peak narrowing observed in  $\bar\rho^{(2)}(\w)$
as well as in the evolution of the  inelastic spectrum 
$\bar\rho^{(1)}(\w)$ 
shown in panel (c) of Fig.~\ref{NRG_fig16}.

\begin{figure}[t]
\begin{center}
\includegraphics[width=0.5\textwidth]{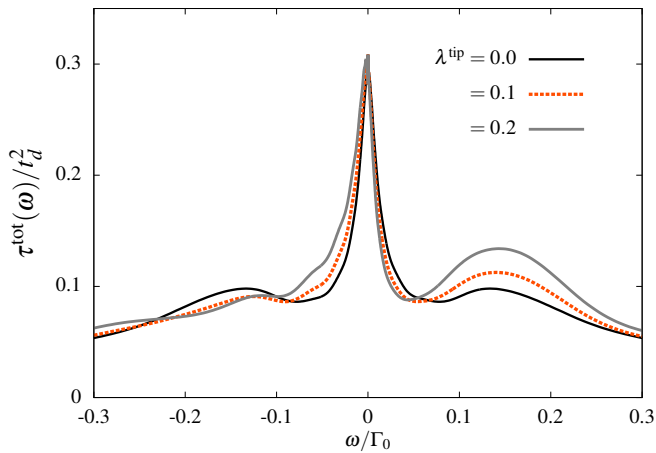}
\caption{Combination of all three contributions
for $\lambda_d/\Gamma_0=2.27$ and two values of $\lambda^{\rm tip}$.
NRG parameters as in Fig.~\ref{NRG_fig15}.
}
\label{NRG_fig18}
\end{center}
\end{figure}

After individually discussing the spectral contributions, we combine the results to
a total STS spectrum in Fig.~\ref{NRG_fig18}. We selected the spectra for the largest
$\lambda_d$ in Fig.~\ref{NRG_fig16}: $\lambda_d/\Gamma_0=2.27$.
For this value $\rho_{d_\sigma,d^\dagger_\sigma}(\w)$ clearly
shows side peaks associated with $U_{\rm eff}$ but not with $\pm \w_0$.
With increasing $\lambda^{\rm tip}$, the STS becomes increasingly asymmetric
due to the admixture of the odd function $\bar \rho^{(1)}(\w)$.
The correlated spectral function $\bar\rho^{(2)}(\w)$ 
of a electron removal or addition with 
a simultaneous displacement $X$ of the harmonic oscillator 
only provides an incoherent background
with a small gap at zero frequency remaining.  In this case the elastic contributions stemming from the side peaks
in  $\rho_{d_\sigma,d^\dagger_\sigma}(\w)$  could be mistakenly attributed to inelastic contributions stemming 
from a fictitious phonon at frequency $\w_0' = |U_{\rm eff}|$.

\section{Summary and conclusion}
\label{sec:conclusion}

We extended the  tunnel theory for STS  
that bridges between weak and strong electron phonon
coupling  in the system as well as includes the strong coupling limit of electronic degrees of freedom. It relies on an exact solution of all relevant system spectral functions of the system in the absence of the STM tip and requires only that the coupling to the tip remains weak so that the second order expansion in the tunneling matrix element
is sufficient.  Importantly, a Wick's theorem is not required.

One of the key incidence of our approach is the systematic derivation of the tunnel current operator from the charge conservation in the total system comprising the system S of interest and the STM tip.  The analytic form of the current operator is determined by the tunneling Hamiltonian $H_T$ connecting the two parts of the total system. The strength of the approach is the treatment of all tunneling processes, elastic  and inelastic contributions,  on equal footing. In particular, our approach includes linear as well as quadratic contributions to the inelastic tunneling current. Neglecting the linear term is only applicable in the limit of vanishing electron-phonon coupling  in the system. It becomes relevant in situations when the phonon mode not only enters $H_T$ but couples to the electronic degrees of freedom in the system S.

We presented experimental  STS for two different location of NTCDA molecules on Ag(111). Combining the LDA+MBPT with an NRG approach
clarifies that the so-called bright molecule corresponds to the top molecule and the dark molecule corresponds to the bridge molecule
in the ab-initio calculation. The projected LDA+MBPT  spectrum of the LUMO as well as the calculated screened intraorbital Coulomb interaction 
enters the NRG as ab-initio parameters. Guided from the inelastic features of the experimental  STS we added an extended  Holstein term to
the many-body calculation and were able to reproduce the  zero-bias peaks seen in the experimental  STS. 
The experimental differences in the peak width could be related to the different hybridzation strength of the LUMO orbital with the substrate
in the  bridge molecule and the  top molecule as calculated by the LDA.

In a second step, the different spectral functions calculated by the NRG were combined with sets of tunneling parameters to reproduce
the experimental STS. Consistent  with the influence of the LUMO electron density on the molecular motion of the B$_{3g}$ modes, the vibrational
couplings in the center of the molecule are a bit larger than at the CH site. While  at the CH the tunneling occurs mainly into the LUMO, a 
Fano mixing between the LUMO and the substrate orbital in the ratio of $2:1$ is found in accordance the phenomenological 
Fano fit of the experimental data.

We presented a generalized tunneling theory of STM spectra what include inelastic vibrations processes. 
Calculating accurate spectral functions with an combined  ab-inito DFT plus  many-body approach including the NGR
allows to reproduced the experimental STS in NTCDA molecules on Ag(111) and provide a deeper insight in this complex system.
Our approach can also be extended to  inelastic magnetic excitation processes 
during tunneling  and opens new doors for our understanding
of magnetic surfaces.

\begin{acknowledgments}
F.B.A.\ acknowledges support from the Deutsche Forschungsgemeinschaft via project AN-275/8-1. F.S.T.\ and M.R.\ and acknowledge support from the Deutsche Forschungsgemeinschaft via the Collaborative Research Center SFB 1083, projects A12 and A13, respectively. 
\end{acknowledgments}

\appendix

\section{Equation of motion for calculating the orbital Green function}
\label{sec:EOM-GF}

It is useful to derive a closed analytic expression for the self-energy of the molecular orbital
GF \cite{Kolodzeiski2017} 
which is used to increase the precision of the NRG GF \cite{BullaHewsonPruschke98} 
as well as
analyze the results. We consider the system Hamiltonian $H_S$
\begin{eqnarray}
H_S&=& \sum_{\k\sigma} \e_{\k\sigma} c^\dagger_{\k\sigma}  c_{\k\sigma} 
+\w_0 b^\dagger_0 b_0 
+
\sum_\sigma \e_{d\sigma} n^d_\sigma + U n^d_\uparrow n^d_\downarrow
\non
&& + \sum_{\k\sigma} V_{\k} ( c^\dagger_{\k\sigma} d_\sigma + d^\dagger_\sigma  c_{\k\sigma} )
\\
&& \nonumber
+ \lambda_d \hat X_0 (\sum_\sigma n^d_\sigma - n_{d0}) 
+\lambda_c \hat X_0 (\sum_\sigma c^\dagger_{0\sigma} c_{0\sigma} -n_{c0})
\end{eqnarray}
where we have defined
\begin{eqnarray}
c_{0\sigma} &=& \frac{1}{V_0}  \sum_{\k} V_{\k}  c_{\k\sigma} 
\\
V_0^2 &=&  \sum_{\k} |V_{\k}  |^2 \, .
\end{eqnarray}
We start from the commutators
\begin{eqnarray}
\, [d_\sigma, H_S] &=& \e_{d\sigma}d_\sigma + U n^d_{-\sigma} d_\sigma 
+V_0 c_{0\sigma} +\lambda_d \hat X_0 d_\sigma \\
\, [c_{k\sigma}, H_S]  &=& \e_{\k\sigma}  c_{\k\sigma} 
+\lambda_c \hat X_0 \frac{V_k}{V_0} c_{0\sigma} + V_k d_\sigma
\label{eq:commu-c0}
\end{eqnarray}
and obtain the  equation of motion (EOM)
\begin{eqnarray}
(z-\e_d) G_{d_\sigma,d^\dagger_\sigma}(z) &=& 1
+ U F_\sigma(z) +\lambda_d M_\sigma(z) \\
&& \nonumber
+ \sum_k V_k G_{c_{\k\sigma} ,d^\dagger_\sigma}(z)
\end{eqnarray}
after introducing the notation
\begin{eqnarray}
F_\sigma(z) &=& G_{d_\sigma n_{-\sigma},d^\dagger_\sigma}(z) \\
M_\sigma(z) &=& G_{\hat X_0 d_\sigma,d^\dagger_\sigma}(z) .
\end{eqnarray}
While the complex function $F_\sigma(z)$ contains the information about the local correlations between
the electrons of different spins $\sigma$,
the influence of the molecular vibration onto the equilibrium GF is account for by 
$M_\sigma(z)$ that also is relevant for the inelastic tunneling current 
-- see Sec.\ \ref{sec:I-inelastic}. In order to close the EOM, we use the commutator \eqref{eq:commu-c0}
to derive
\begin{eqnarray}
(z- \e_{\k\sigma}) G_{c_{\k\sigma} ,d^\dagger_\sigma}(z)
&=& V_k G_{d_\sigma,d^\dagger_\sigma}(z) \\
&& \nonumber
+\lambda_c \frac{V_k}{V_0} N_\sigma(z) 
.
\end{eqnarray}
The off-diagonal composite correlation function
\begin{eqnarray}
 N_\sigma(z)  &=& G_{\hat X_0 c_{0\sigma} ,d^\dagger_\sigma}(z)
\end{eqnarray}
accounts for  the correlations between hybridization process and the vibrational displacement $\hat X_0$.
We have seen its explicit importance for the renormalization of bare hybridization via Ref.\ \eqref{eq:27}.
Defining
\begin{eqnarray}
\Delta_\sigma(z)&=& \sum_k \frac{V_k^2}{z- \e_{\k\sigma}}
\end{eqnarray}
and using the standard parametrization of the GF in terms of self-energy corrections $\Sigma_\sigma$,
\begin{eqnarray}
G_{d_\sigma,d^\dagger_\sigma}(z)  &=& \frac{1}{z-\e_d -\Delta_\sigma(z) - \Sigma_\sigma(z)}
\end{eqnarray}
the self-energy can be expressed \cite{BullaHewsonPruschke98}
as
\begin{eqnarray}
\label{eq:APP-self-energy}
\Sigma_\sigma(z) &=& 
\frac{U F_\sigma(z) +\lambda_d M_\sigma(z)  + 
\frac{ \lambda_c}{V_0}\Delta(z)  N_\sigma(z) }{G_{d_\sigma,d^\dagger_\sigma}(z)}
\end{eqnarray}
Since the NRG can calculate each individual Green function $F_\sigma(z), M_\sigma(z),
N_\sigma(z)$ and $G_{d_\sigma,d^\dagger_\sigma}(z)$, Bulla 
et al.\ \cite{BullaHewsonPruschke98} 
have shown that replacing the GFs on the right side of \eqref{eq:APP-self-energy} by the NRG results yields
a self-energy that becomes almost independent of the NRG discretization parameters and, therefore,
is an accurate representation of the true self-energy for the continuum model.
Eq.\ \eqref{eq:self-energy} is analytically exact and is also used in the main text to present an better
analytical understanding of the numerical finding.


%

\end{document}